%% file: pod.tex
\documentclass[AMA,STIX1COL]{WileyNJD-v2}

\articletype{Research Article}

\received{\dots}
\revised{\dots}
\accepted{\dots}

\usepackage{etex}
\usepackage{graphicx}
\usepackage{amssymb}
\usepackage{amsmath}
\usepackage{mathtools}
\usepackage{hyperref}
\usepackage{xcolor}
\usepackage{pgfplots}
\usepackage{array}
\usepackage{float}
\usepackage{subfig}
\usepackage{setspace}

\usetikzlibrary{external}
\tikzset{every picture/.style={line width=0.9pt}}
\newlength\figureheight
\newlength\figurewidth
\newlength\hsep

\definecolor{std}{RGB}{0, 0, 128}%

\definecolor{fom}{RGB}{0,0,128}%
\definecolor{rom10}{rgb}{0.85000,0.32500,0.09800}%
\definecolor{rom50}{rgb}{0.92900,0.69400,0.12500}%
\definecolor{rom100}{rgb}{0.49400,0.18400,0.55600}%
\definecolor{rom200}{rgb}{0.46600,0.67400,0.18800}%
\definecolor{rom300}{rgb}{0.30100,0.74500,0.93300}%
\definecolor{rom400}{rgb}{0.63500,0.07800,0.18400}%
\definecolor{rom500}{rgb}{0.00000,0.44700,0.74100}%

\definecolor{rom_cos}{RGB}{255, 225, 25}%
\definecolor{rom_cob}{RGB}{60, 180, 75}%
\definecolor{rom_const}{RGB}{145, 30, 180}%
\definecolor{rom_direct}{RGB}{0, 128, 128}%
\definecolor{rom_grassmann}{RGB}{230, 25, 75}%
\definecolor{rom_direct_interp}{RGB}{128, 0, 0}%

\definecolor{t_other}{rgb}{0.00000,0.44700,0.74100}%
\definecolor{t_eval}{rgb}{0.85000,0.32500,0.09800}%
\definecolor{t_solve}{rgb}{0.92900,0.69400,0.12500}%

\definecolor{start}{RGB}{60, 180, 75}%
\definecolor{converged}{RGB}{230, 25, 75}%
\definecolor{p_sigma}{rgb}{0.00000,0.44700,0.74100}%
\definecolor{p_act_max}{rgb}{0.85000,0.32500,0.09800}%
\definecolor{p_act_min}{rgb}{0.92900,0.69400,0.12500}%
\definecolor{p_t_act}{rgb}{0.49400,0.18400,0.55600}%
\definecolor{p_t_deact}{rgb}{0.46600,0.67400,0.18800}%

\renewcommand{\vec}[1]{\boldsymbol{#1}}			% for vectors
\newcommand{\ten}[1]{\boldsymbol{#1}}			% for tensors
\newcommand{\vecrm}[1]{\boldsymbol{\mathrm{#1}}}	% for vectors in mathrm style
\newcommand{\matrm}[1]{\boldsymbol{\mathrm{#1}}}	% for matrices in mathrm style
\newcommand{\dd}{\mathrm{d}}					% differential d
\newcommand{\pd}{\partial}						% partial differentiation d
\newcommand{\pfrac}[2]{\frac{\pd #1}{\pd #2}}		% \pfrac{f(x,y)}{x} for partial derivative of f(x,y) with respect to x
\newcommand{\pfracl}[2]{\pd #1/\pd #2}
\newcommand{\norm}[2]{\left\| #2 \right\|_{#1}}				% \norm{L_2}{x} for the L2-norm of x, \norm{\infty}{x} for the Inf-Norm of x
\newcommand{\comment}[1]{}						% put around regions that are not to be compiled

\newcommand{\mor}[1]{\textit{MOR}}
\newcommand{\pmor}[1]{\textit{pMOR}}
\newcommand{\ppod}[1]{\textit{POD}}
\newcommand{\fom}[1]{\textit{FOM}}
\newcommand{\rom}[1]{\textit{ROM#1}}
\newcommand{\prom}[1]{\textit{pROM#1}}

\newcommand{\ds}{\vecrm{d}}
\newcommand{\rs}{\vecrm{R}^\mathrm{S}}
\newcommand{\ns}{n_\mathrm{s}}

\newcommand{\p}{\vecrm{p}}
\newcommand{\rd}{\vecrm{R}^\mathrm{0D}}

\newcommand{\T}[1]{#1^\mathrm{T}}
\newcommand{\proj}{\matrm{V}}
\newcommand{\projt}{\T{\proj}}
\newcommand{\dr}{\ds_\mathrm{r}}

\DeclareMathOperator*{\argmin}{argmin}
\DeclareMathOperator*{\argmax}{argmax}

\newcommand{\param}{\vecrm{\mu}}
\newcommand{\jac}{\vecrm{J}}

\newcommand{\defeq}{\vcentcolon=}

\begin{document}

\title{Parametric model order reduction and its application to inverse analysis of large nonlinear coupled cardiac problems}

\author[1]{M. R. Pfaller*}
\author[2]{M. Cruz Varona}
\author[1]{J. Lang}
\author[3]{C. Bertoglio}
\author[1]{W. A. Wall}

\authormark{M. R. PFALLER \textsc{et al}}

\address[1]{\orgdiv{Institute for Computational Mechanics}, \orgname{Technical University of Munich}, \orgaddress{\state{Boltzmannstr. 15, 85748 Garching b. M\"unchen}, \country{Germany}}}

\address[2]{\orgdiv{Chair of Automatic Control},\\ \orgname{Technical University of Munich}, \orgaddress{\state{Boltzmannstr. 15, 85748 Garching b. M\"unchen}, \country{Germany}}}

\address[3]{\orgdiv{Bernoulli Institute}, \orgname{University of Groningen}, \orgaddress{\state{Nijenborgh 9, 9747 AG Groningen}, \country{The Netherlands}}}

\corres{*M.~R.~Pfaller, Boltzmannstr. 15,\\ 85748 Garching b. M\"unchen, Germany. \email{martin.pfaller@tum.de}}

\abstract[Abstract]{Predictive high-fidelity finite element simulations of human cardiac mechanics co\-mmon\-ly require a large number of structural degrees of freedom. Additionally, these models are often coupled with lumped-parameter models of hemodynamics. High computational demands, however, slow down model calibration and therefore limit the use of cardiac simulations in clinical practice. As cardiac models rely on several patient-specific parameters, just one solution corresponding to one specific parameter set does not at all meet clinical demands. Moreover, while solving the nonlinear problem, 90\% of the computation time is spent solving linear systems of equations. We propose a novel approach to reduce only the structural dimension of the mo\-no\-li\-thi\-cally coupled structure-windkessel system by projection onto a lower-dimensional subspace. We obtain a good approximation of the displacement field as well as of key scalar cardiac outputs even with very few reduced degrees of freedom, while 
achieving considerable speedups. For subspace generation, we use proper orthogonal decomposition of displacement snapshots. To incorporate changes in the parameter set into our reduced order model, we provide a comparison of subspace interpolation methods. We further show how projection-based model order reduction can be easily integrated into a gradient-based optimization and demonstrate its performance in a real-world multivariate inverse analysis scenario. Using the presented projection-based model order reduction approach can significantly speed up model personalization and could be used for many-query tasks in a clinical setting.}


\maketitle

% why model order reduction is necessary
\section{Introduction}

% why cardiac mechanics
Cardiac solid mechanics simulations consist of solving large-deformation, materially nonlinear, elastodynamic coupled boundary-value problems. As the exact fluid dynamics of blood within the heart are usually not needed, the structural model is commonly coupled to lumped-parameter fluid models which provide the pressure to the endocardial wall \cite{westerhof08}. These so-called windkessel models are then coupled to cardiac solid mechanics \cite{kerckhoffs07}. For a comprehensive review of models currently utilized in cardiac mechanics the reader is referred to \cite{chabiniok16}.

\subsection{Model order reduction}

% why big problem -> mor
The needed huge number of degrees of freedom (\textit{DOFs}) and other challenges of solving coupled nonlinear problems make the solution of cardiac models computationally expensive and limit the models' use in clinical practice. For example, using a single node with 24 cores, a simulation of one heartbeat, which takes about one second in reality, takes about one day to compute with our high-fidelity four chamber model \cite{pfaller18}. The potential to reduce computation time motivates the use of reduced order models (\textit{ROMs}). In the following, different strategies in reduced order modeling are reviewed.

% bottom-up cardiac mor
An important category of cardiac \textit{ROMs} are "bottom up" \textit{ROMs}. For these models, the same system of differential equations as for the full order model (\fom{}) is solved, but on a simplified analytical geometry. The displacements are commonly parameterized by only one scalar degree of freedom (\textit{DOF}). These models are thus referred to as 0D models. Examples in this category include monoventricular cylindrical \cite{arts79}, spherical \cite{caruel14}, or prolate spheroid \cite{moulton17} or biventricular \cite{lumens09} geometries. These models allow extremely fast evaluation, with computation times less than one second. Their results are, however, only lumped quantities which usually need an extra correction step in order to predict the solution of a corresponding patient-specific 3D model.

% coarse discretization
Another approach of \mor{} in biomechanics is the use of coarsely discretized geometries, see e.g. \cite{biehler15,hirschvogel16}. Coarsely discretized models are easy to implement, since the computational framework is identical to the one of the \fom{}. The disadvantage of using coarsely discretized geometries is that there is no exact control over the approximation quality and important features of the \fom{} might not be preserved by the \rom{}.

% top-down cardiac mor
A third category of cardiac \textit{ROMs} are "top down" \textit{ROMs}, which are utilized in this work. These \textit{ROMs} make a model computationally less expensive by reducing the dimension of the problem, starting from the \fom{}. For example in cardiac electrophysiology, approximated lax pairs for propagating wave fronts were proposed in \cite{gerbeau14,gerbeau15}.  A local reduced basis method for parameterized cardiac electrophysiology was recently introduced in \cite{pagani18}. Reduced basis methods were proposed for general large deformation, material nonlinear finite element simulations \cite{krysl01,milani08}. A framework for linear coupled multiphysics problems was introduced in \cite{schilders14}.

% what we do
For large-scale finite element simulation, about 90~\% of the time is spent iteratively solving linear systems of equations. This proportion motivates the use of model order reduction by projection, where the full linear system is projected onto a much smaller dimensional subspace while preserving the model's most relevant features. The solution of the \fom{} is then approximated by a solution in the reduced space with a \rom{}. A popular method to generate such subspaces is proper orthogonal decomposition (\ppod{}), which is purely observation-based and independent of the underlying physics of the model. The snapshots, in our case transient observations of displacements, can be obtained from a \fom{} simulation of one heartbeat.

% review: cardiac mor
There are only few examples where \ppod{} has been applied to cardiac problems. The reduction of a patient-specific biventricular cardiac model using \ppod{} is described in \cite{chapelle2013galerkin}. A quasi-static cardiac model was reduced using \ppod{} in \cite{bonomi2017matrix} and combined with hyperreduction techniques. However, analysis was only carried out using an idealized ellipsoidal left ventricular geometry with few DOFs. While this is very instructive, results for speedup and accuracy of the \rom{} are not conclusive for real-world cardiac problems. Furthermore, both references used a purely structural cardiac model with prescribed blood pressures.

% why parametric -> pmor
Cardiac models rely on a large set of patient-specific parameters, describing constitutive behavior, hemodynamics, boundary conditions, or local fiber orientation. In order not to rely on a \fom{} simulation for each new \rom{} simulation, which would render the \rom{} simulation useless, the reduced subspace must be able to adapt to a changing parameter set. This adaption requires parametric model order reduction (\pmor{}). Among many global and local \pmor{} techniques, various subspace interpolation methods have been proposed in the past~\cite{benner2015survey}. Specifically, a popular method using a Grassmann manifold was proposed in \cite{amsallem2008interpolation} and illustrated with a large coupled aeroelastic model of a fighter jet. The method proposed in \cite{chapelle2013galerkin} uses direct interpolation of parameter-dependent solutions instead of interpolating subspaces. Furthermore, a global \pmor{} approach using a global basis over the whole parameter range is employed in \cite{
bonomi2017matrix}.

% what we do
The performance of \ppod{} in realistic coupled simulations of cardiac contraction is yet unknown. We demonstrate in this work the performance of \ppod{} applied to a patient-specific cardiac geometry with about $850\,000$ structural DOFs. To the best of the authors' knowledge, \ppod{} for large nonlinear models has only been applied to single fields, e.g. structural mechanics or fluid dynamics, separately \cite{farhat15}. In this work, we consider for the first time the case of a \ppod{}-reduced 3D structural model that is monolithically coupled to a 0D windkessel model, where we only reduce the structural dimension of the problem. Additionally, we compare several subspace interpolation methods for cardiac contraction. In these parametric simulations, we vary the contractility parameter controlling maximum active tension of the myofibers in our model, as it is the most influential parameter for cardiac function and commonly calibrated to experiments.

\subsection{Inverse analysis}

% why inverse analysis
Many of the cardiac model parameters depend on a patient's physiology and are \emph{a priori} unknown, as invasive experiments cannot be carried out on living human subjects. A predictive patient-specific cardiac model is thus subject to an iterative process termed \emph{inverse analysis}. In this context, the simulation of one heartbeat with given parameters can be regarded as the forward problem. The reverted task of matching the parameters to given observations from the patient-specific heartbeat is then the inverse problem. Common clinical measurements of cardiac kinematics are displacement data extracted from cine or tagged magnetic resonance imaging (MRI), representing an Eulerian and Lagrangian description of motion, respectively. Other measurements include blood pressure or electrocardiograms. As cardiac mechanics simulations pose an expensive forward problem, repeated evaluation during inverse analysis has incredible computational demands. Furthermore, algorithms for inverse analysis commonly scale 
linearly with the number of parameters. Inverse analysis is thus a promising application of reduced order modeling.%, often referred to as "curse of dimensionality".

% adjoint
During gradient-based optimization, the adjoint method offers computationally inexpensive gradient calculation. For example in \cite{sermesant06}, regional contractility was estimated from short axis cine MRI. Using the adjoint method, ischemic regions in cardiac electrophysiology were identified in \cite{chavez15}. Most recently in \cite{finsberg18}, passive material parameters and active fiber shortening was estimated for a biventricular geometry from ventricular volume and regional strain.

% gradient free
Gradient-free inverse analysis for cardiac problems was demonstrated in \cite{chabiniok12}, where regional cardiac contractility was estimated from cine MRI using the unscented Kalman filter (UKF). The reduced order UKF was further applied in \cite{bertoglio12a,moireau13} to estimate boundary condition parameters of the aorta for a fluid-structure-interaction problem. Other examples of gradient-free inverse analysis include \cite{asner16}, where left-ventricular active and passive material parameters were estimated from 3D tagged MRI using a parameter sweep.

% review: cardiovascular inverse analysis
There are some examples, where reduced order modeling has been combined with inverse analysis in biomechanics. For arterial hemodynamic fluid-structure-interaction problems, an inverse analysis with uncertainty quantification was performed in \cite{lassila13} using a reduced basis method. There are however few references for cardiac solid models. In \citep{chapelle2013galerkin}, the reduced order UKF was applied to estimate cardiac contractility in a healthy and an infarcted region. The forward simulations were carried out using \ppod{}, thus converging to a different solution than using the \fom{} only. In \cite{hirschvogel16} a multifidelity approach was proposed to calibrate hemodynamical and structural parameters of a cardiac model to ventricular pressure measurements. Here, a Levenberg-Marquardt-based optimization uses evaluations switching between a 3D \fom{}, a coarsly discretized version of the 3D \fom{}, and a 2D surrogate model. Another multifidelity approach was used in \cite{mollero18} between a 
3D \fom{} and a 0D surrogate model.

% what we do
Using coarsely discretized or surrogate models does however not guarantee that the most important features of the \fom{} are preserved. These surrogate models further lack the ability of \pmor{} to inherently "learn" from evaluations of the \fom{} to become more precise throughout the optimization. Instead, they require an additional mapping between \fom{} and surrogate model solutions. Most importantly, using 2D or 0D surrogate models during inverse analysis, the heart can only be tuned to scalar measurements. However, a calibration to spatial measurements from cine or tagged MRI might be desired in many applications, e.g. when detecting infarcted regions \cite{chabiniok12}. In this work, we thus propose a novel method of how a \rom{} can be integrated into any optimization-based inverse analysis leading to considerable savings in CPU time, and demonstrate its performance in a real-world multivariate inverse analysis scenario.

% structure of this paper
The remainder of this work is structured as follows. In section~\ref{sec_fom}, we introduce the full order elastodynamic and hemodynamic models. We derive a reduced formulation for the monolithically coupled system in section~\ref{sec_prom} and review several \prom{} subspace interpolation methods. In numerical experiments in section~\ref{sec_res}, we demonstrate the accuracy and speedup of our \rom{} and show its response to parametric variations. Furthermore, we propose in section~\ref{sec_inv_ana} a \pmor{}-based method for inverse analysis and analyze its performance with our four-chamber heart model. We close with a conclusion and future perspectives in section \ref{sec_conclusion}.

\section{3D-0D coupled cardiovascular modeling}
\label{sec_fom}
In this section we give a brief overview of our full order model (\fom{}) which is composed of a 3D elastodynamical model, see section~\ref{sec_fom_3D}, coupled to a 0D hemodynamical model, see section~\ref{sec_fom_0D}. We further provide insights into the numerical solution of the model in section~\ref{sec_fom_solution}. For a more detailed description, the reader is referred to \cite{pfaller18}.

\subsection{3D elastodynamical model}
\label{sec_fom_3D}
\begin{figure}
\centering
\subfloat[Posterior view.\label{geometry_back}]{
\includegraphics[width=.3\textwidth]{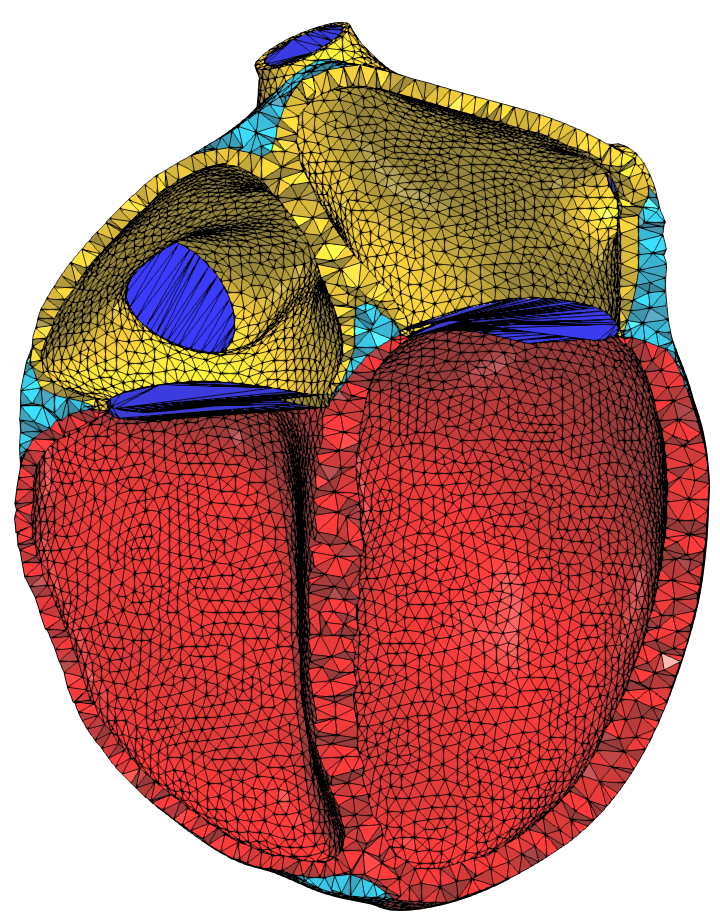}}\qquad
\subfloat[Anterior view.\label{geometry_front}]{
\includegraphics[width=.32\textwidth]{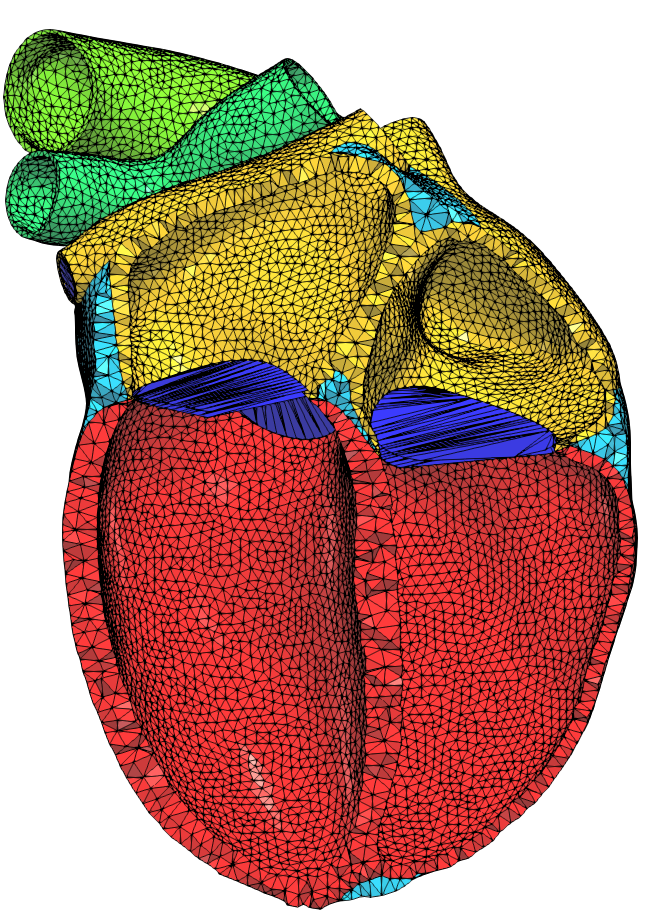}}
\caption{Computational mesh with quadratic tetrahedral elements cut in four-chamber view, colored by different materials.\label{geometry}}
\end{figure}

We follow the classic approach of nonlinear large deformation continuum mechanics to model the elastodynamic problem of 3D cardiac contraction. We define the reference configuration $\vec{X}$ and the current configuration $\vec{x}$, which are connected by the displacements $\vec{u} = \vec{x} - \vec{X}$. We calculate the deformation gradient $\ten{F}$, the right Cauchy-Green tensor $\ten{C}$, and the Green-Lagrange strain tensor $\ten{E}$
\begin{equation}
\ten{F} = \frac{\partial \vec{x}}{\partial \vec{X}}, \quad \ten{C}=\ten{F}^{\text{T}}\ten{F}, \quad \ten{E} = \frac{1}{2}(\ten{C} - \ten{I}).
\end{equation}
Balance of momentum, a Neumann windkessel coupling condition with left ventricular pressure~$p_v$, omni-directional  spring-dashpot boundary conditions, and pericardial boundary conditions yield the weak form of the 3D elastodynamic boundary value problem in the reference configuration
\begin{equation}
\begin{aligned}
\int_{\varOmega_0} \left[ \rho_0 \ddot{\vec{u}} \cdot \delta\vec{u}
+  \ten{S}:\delta \ten{E}\right]  \dd V
+ \int_{\varGamma_0^{\text{endo}}} p_v J \ten{F}^{-\text{T}} \cdot \ten{N} \cdot \delta\vec{u}\, \dd A
+ \int_{\varGamma_0^{\text{vess}}} \left[k_v\vec{u} + c_v\dot{\vec{u}}\right] \cdot \delta\vec{u}\, \dd A
&+ \int_{\varGamma_0^{\text{epi}}} \vec{N} \left[k_e\vec{u} \cdot \vec{N} + c_e\dot{\vec{u}} \cdot \vec{N}\right] \cdot \delta\vec{u}\, \dd A
= 0,
\label{eq_weak}
\end{aligned}
\end{equation}
with density $\rho_0$, accelerations $\ddot{\vec{u}}$, virtual displacements and strains $\delta\vec{u}$ and $\delta\vec{E}$, respectively, the Jacobian $J=\det\ten{F}$ of the deformation gradient, the second Piola-Kirchhoff stress tensor $\ten{S}$ , the reference surface normal $\vec{N}$, and spring stiffness $k_v,k_e$ and viscosity $c_v,c_e$ for vessel and epicardial surface, respectively. We define three surfaces for the imposition of boundary conditions $\varGamma_0^{\text{endo}}$, $\varGamma_0^{\text{vess}}$, and $\varGamma_0^{\text{epi}}$ at the left endocardium, the outside of the great vessels, and the epicardium respectively. At the cut-offs of the great vessels we apply homogeneous Dirichlet boundary conditions. We define different nonlinear materials for adipose tissue, ventricular myocardium, atria, aorta, and the pulmonary artery. The geometry and the different materials are shown in figure~\ref{geometry}. Each is composed of a hyperelastic and a viscous contribution depending on the rate of 
Green-Lagrange strains. Quasi-incompressibility is enforced by a penalty potential. The ventricular myocardium is modeled with an isotropic Mooney-Rivlin material and has an additive active stress component $\ten{S}_{\text{act}}$ modeling the contraction of myofiber bundles in reference fiber direction $\vec{f}_0$
\begin{equation}
\ten{S}_{\text{act}} = \tau(t) \cdot \vec{f}_0\otimes\vec{f}_0, \quad\text{with~} \dot{\tau}(t) = - |u(t)|\tau(t) + \sigma |u(t)|_+, \quad \tau(0)=0,
\label{eq_act_mat}
\end{equation}
with active stress $\tau\in~[0,\sigma[$ and the function $|u(t)|_+ = \max (u(t);0)$. The contractility parameter $\sigma$ controls the upper limit of the active stress component. The prescribed activation function $u(t)$ is
\begin{equation}
u(t) = \alpha_{\text{max}} \cdot f(t) + \alpha_{\text{min}} \cdot [1-f(t)], \quad 
f(t) = S^+(t-t_{\text{sys}}) \cdot S^-(t-t_{\text{dias}}), \quad S^\pm(\Delta t) = \frac{1}{2} \left[ 1 \pm \text{tanh} \left( \frac{\Delta t}{\gamma} \right) \right]
\label{eq_activation}
\end{equation}
with steepness $\gamma=0.005$~s and descending and ascending sigmoid functions $S^+$ and $S^-$, respectively. The indicator function $f\in~]0,1[$ indicates ventricular systole. The times $t_{\text{sys}}$ and $t_{\text{dias}}$ model the onset of systole and diastole, respectively. The times $t_{\text{sys}}$ and $t_{\text{dias}}$, and maximum and minimum myocyte activation rates $\alpha_{\text{max}}$ and $\alpha_{\text{min}}$, respectively, are calibrated to match the timing of ventricular systole as observed in cine MRI, as demonstrated in section~\ref{sec_inv_ana}.

We discretize displacements $\vec{u}$ and virtual displacements $\delta\vec{u}$ arising in the weak form \eqref{eq_weak} using quadratic basis functions on each tetrahedral finite element. Assembly of the discretized problem leads to the matrix notation of the spatially semi-discrete residual of the full order structural model
\begin{align}
\rs_{\mathrm{semi}} =\matrm{M} \ddot{\ds} + \vecrm{F} (\dot{\ds}, \ds, p_v) \overset{!}{=} \vec{0}
\end{align}
with mass matrix $\matrm{M}$, nonlinear force vector $\vecrm{F}$, and nodal displacements, velocities, and accelerations $\ds$, $\dot{\ds}$, and $\ddot{\ds}$ respectively. We discretize the boundary value problem in time using a combintation of Newmark's method \cite{newmark1959method} and the generalized-$\alpha$ scheme \cite{chung1993time} to obtain the time and space discrete structural residual $\rs(\ds_{j+1},p_{v,j+1})$ at time step $j+1$.

\subsection{0D hemodynamical model}
\label{sec_fom_0D}
We couple the left ventricular 3D structural model to a 0D lumped-parameter windkessel model of the circulatory system.  We utilize in this work a four element windkessel, using resistances~$R$, compliances~$C$, and an inertance~$L_p$. Pressures at different parts of the model are denoted by~$p_i$. We distinguish between a proximal part $p$ and a distal part $d$ of the aorta. The atrial pressure $p_{at}(t)$ is prescribed to simulate atrial systole. The venous pressure~$p_{\text{ref}}$ is kept constant. We model the atrioventricular and semilunar valves with a smooth diode-like behavior by nonlinear resistances $R_{av} \defeq R(p_{at},p_v)$ and $R_{sl} \defeq R(p_v,p_p)$ respectively, depending on a sigmoid function $R$. This yields the set of differential equations
\begin{equation}
\frac{p_v - p_{at}}{R_{av}} + \frac{p_v - p_p}{R_{sl}} + \dot{V}(\vec{u}) = 0,\quad q_p - \frac{p_v - p_p}{R_{sl}} + C_p \dot{p}_p = 0, \quad q_p + \frac{p_d - p_p}{R_p} + \frac{L_p}{R_p} \dot{q}_p = 0, \quad \frac{p_d - p_{\text{ref}}}{R_d} - q_p + C_d \dot{p}_d = 0
\label{eq_wk}
\end{equation}
which are coupled to the 3D structural model by the ventricular pressure $p_v$ and the change in ventricular volume $\dot{V}$, depending on the structural displacements $\vec{u}$. The vector of primary variables yields $\p=\T{[p_v, p_p, p_d, q_p]}$, including the flux $q_p$ through the inertance $L_p$. We discretize the set of windkessel equations~\eqref{eq_wk} in time with the one-step-$\theta$ scheme \cite{crank1947practical}. This yields the discrete windkessel residual $\rd(\ds_{j+1},\p_{j+1})$ at time step $j+1$.

\subsection{Solving the coupled problem}
\label{sec_fom_solution}
We solve the coupled 3D-0D model with the structural and windkessel residuals $\rs$ and $\rd$, respectively, for the displacements $\ds$ and windkessel variables $\p$ at time step $j+1$ with the Newton-Raphson method
\begin{align}
\renewcommand{\arraystretch}{1.5}
\left[
\begin{matrix}
\pfrac{\rs}{\ds} & \pfrac{\rs}{\p}\\
\pfrac{\rd}{\ds} & \pfrac{\rd}{\p}
\end{matrix}
\right]_{j+1}^i
\cdot
\left[
\begin{matrix}
\Delta \ds\\
\Delta \p
\end{matrix}
\right]_{j+1}^{i+1}
= -
\left[
\begin{matrix}
\rs\\
\rd
\end{matrix}
\right]_{j+1}^i,
\label{eq_fom}
\end{align}
linearizing the residuals in iteration $i$. The solution is converged if
\begin{align}
\norm{\infty}{\rs} < tol^\mathrm{S}_\text{res},
\quad
\norm{\infty}{\Delta \ds} < tol^\mathrm{S}_\text{inc},
\quad
\norm{2}{\rd} < tol^\mathrm{0D}_\text{res},
\quad
\norm{2}{\Delta \p} < tol^\mathrm{0D}_\text{inc},
\label{eq_conv_crit}
\end{align}
with the structural and windkessel residual and increment tolerances $tol^\mathrm{S}_\text{res}$,  $tol^\mathrm{0D}_\text{res}$, $tol^\mathrm{S}_\text{inc}$, and $tol^\mathrm{0D}_\text{inc}$, respectively. Note that the coupled model in \eqref{eq_fom} is independent of the concrete formulation of the structural and windkessel models from sections \ref{sec_fom_3D} and \ref{sec_fom_0D} respectively. It is valid for any arbitrary residuals $\rs$ and $\rd$.  For details of the model used here, again see \cite{pfaller18}.

\section{Nonlinear parametric model order reduction}
\label{sec_prom}
The 3D-0D cardiovascular model described above represents a large-scale, nonlinear, parametrized, and monolithically coupled model, featuring multiple sources of nonlinear system behavior and depending on several model parameters. Firstly, our model contains geometric nonlinearities, due to the use of the Green-Lagrange strain tensor $\ten{E}(\vec{u})$. Secondly, the utilized material laws for myocardial tissue induce material nonlinearity. The third and last source of nonlinearity is given by the nonlinear coupling between the structural and the hemodynamical model due to the Neumann windkessel boundary condition, acting in direction of the current normal vector of the endocardium.
%The nonlinear coupling between the structural and the hemodynamical model due to the Neumann windkessel boundary condition causes the third and last source of nonlinearity.  
Furthermore, our model depends on many parameters $\param = \T{\left[\mu_1,\ldots,\mu_{n_p}\right]} \in \Omega \subset \mathbb{R}^{n_p}$, classified in different categories. For instance, there exist parameters describing the constitutive behavior of the used materials (stiffness, viscosity, and incompressibility parameters), the additive active stress component $\ten{S}_{\text{act}}$ (e.g. the contractility $\sigma$, $\alpha_{\text{max}}$, $\alpha_{\text{min}}$, $t_{\text{sys}}$, $t_{\text{dias}}$), the hemodynamics (e.g. resistances $R$, compliances $C$, inertance $L_p$), as well as the boundary conditions for the outside of the great vessels and the epicardium (spring stiffnesses $k_v, k_e$ and dashpot viscosities $c_v, c_e$). Thus, our discrete nonlinear parametrized \fom{} reads
\begin{align}
\vecrm{R}(\ds, \p, \param) = 
\left[
\begin{matrix}
\rs(\ds, p_v, \param)\\[0.5em]
\rd(\ds, \p, \param)
\end{matrix}
\right]_{j+1} \overset{!}{=} \vec{0}.
\end{align}
The use of a 0D lumped-parameter windkessel model, instead of e.g. a 3D fluid dynamics model of the heart chambers and arteries, already simplifies the computational complexity of the coupled system. However, the numerical analysis of the present model still demands a high computational effort due to the large number of structural DOFs. While this is no problem for a few standard forward simulations, it is extremely challenging - and might even prohibit - fast model calibration, inverse analysis, and clinical applications. Therefore, our aim is to employ projection-based model order reduction to obtain a cardiovascular reduced order model (\rom{}) that accurately approximates the original model with substantially less DOFs and, consequently, less numerical effort.
%thus allowing for an efficient numerical simulation.
To this end, in section~\ref{sec_proj-ROM} we first apply the classical projection-based model order reduction framework to the structural component of our cardiovascular problem and further describe the numerical solution of the coupled \rom{}. A suitable strategy to compute the required projection matrix $\proj$ for a fixed parameter set is then explained in section~\ref{sec_subspace}. Afterwards, different subspace interpolation techniques are presented in section~\ref{sec_pmor}, in order to compute a parametric reduced order model (\prom{}) for any new parameter set. Finally, some implementation details are given in section~\ref{sec_implementation}.

%Then, the computation of the required projection matrix $\proj$ for a fixed parameter set is explained in section~\ref{sec_subspace}. 

%\begin{itemize}
%\item motivation
%\item short derivation
%\item generation of snapshots
%\item reduced coupled system
%\item reduced norm check
%\end{itemize}

\subsection{Cardiovascular reduced order model}
\label{sec_proj-ROM}
As mentioned before, the high dimension of the \fom{} comes from the discretization of the 3D elastodynamical model, whereas the 0D model only contributes few, in our case four, windkessel DOFs in $\p$. Consequently, we only apply model order reduction to the structural component while the 0D hemodynamical model remains unchanged.

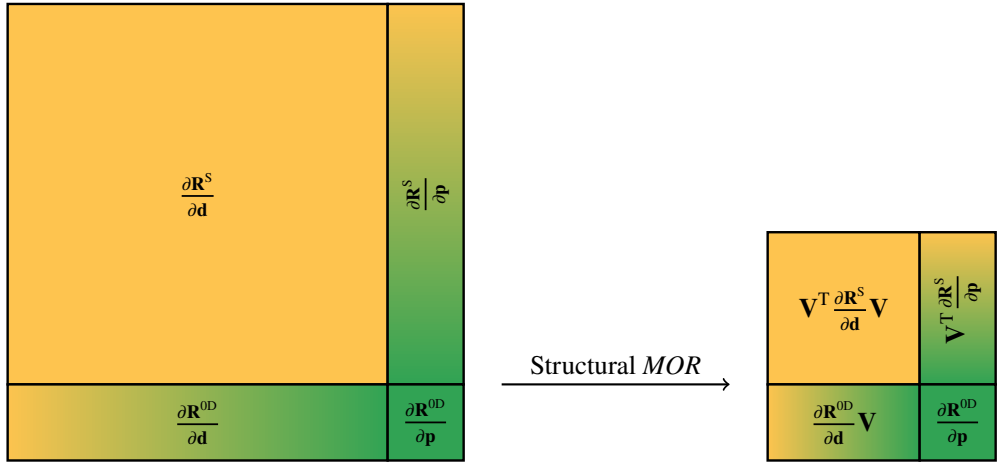
\begin{figure}
\centering
\input{tikz/coupled_mor.tex}
\caption{Visualization of the Jacobian during projection-based model order reduction of the structural dimension of the block matrix system in \eqref{eq_fom} to \eqref{eq:RedLinSysV}. The diagonal structural and windkessel blocks are colored yellow and green, respectively. Off-diagonal coupling blocks are shaded. Note that the dimension of the diagonal windkessel block remains unchanged.\label{coupled_mor}}
\end{figure}

Our aim is to approximate the high dimensional structural solution $\ds \in \mathbb{R}^n$ using a linear combination of $q \ll n$ basis vectors, with $n$ being the number of DOFs of the full elastodynamical model. With the set of basis vectors $ \vecrm{v}_i \in \mathbb{R}^n$ contained in a projection matrix $\proj = \left[ \vecrm{v}_1 \, ... \, \vecrm{v}_q \right] \in \mathbb{R}^{n \times q}$ and the vector of the reduced model's DOFs $\dr \in \mathbb{R}^q$, the approximation is
\begin{align}
\ds \approx \proj \dr. \label{eq:dApprox}
\end{align}
To obtain a square system, we project the structural residual onto the space spanned by $\projt$ yielding the spatially semi-discrete reduced residual $\rs_{\mathrm{semi,r}} \in \mathbb{R}^q$ 
\begin{align}
\rs_{\mathrm{semi,r}} = \projt \rs_{\mathrm{semi}}(\proj \ddot{\ds}_\mathrm{r}, \proj \dot{\ds}_\mathrm{r}, \proj \dr, p_v, \param).
\end{align}
This represents the Galerkin projection of the full structural residual $\rs_{\mathrm{semi}}$ onto the subspace $\mathcal{V}$ spanned by the vectors in $\proj$. After discretization in time, we obtain the Newton-Raphson update $i+1$ at time step $j+1$ of the coupled reduced order model
\begin{align}
\renewcommand{\arraystretch}{1.5}
\begin{bmatrix}
\projt \pfrac{\rs}{\ds} \proj & \projt \pfrac{\rs}{\p}\\
\pfrac{\rd}{\ds} \proj & \pfrac{\rd}{\p}
\end{bmatrix}_{j+1}^i 
\cdot
\begin{bmatrix}
\Delta \dr\\
\Delta \p
\end{bmatrix}_{j+1}^{i+1}
=-
\begin{bmatrix}
\projt\rs\\
\rd
\end{bmatrix}_{j+1}^i
\label{eq:RedLinSysV}
\end{align}
by linearizing the spatially and time discrete form of the residual. Through the chain rule of differentiation, the left column entries of the block Jacobian matrix are right-multiplied by $\proj$. The structural block $\projt \, \pfracl{\rs}{\ds} \, \proj$ is now of reduced dimension $\mathbb{R}^{q \times q}$. Note that the windkessel Jacobian $\pfracl{\rd}{\p}$ and windkessel degrees of freedom $\p$ remain unchanged. For the off-diagonal coupling blocks $\pfracl{\rd}{\ds}$ and $\pfracl{\rs}{\p}$ only the structural dimension is right and left multiplied with the projection matrix $\proj$ or its transpose, respectively. The original \eqref{eq_fom} and reduced \eqref{eq:RedLinSysV} block-Jacobian are visualized in figure~\ref{coupled_mor}.

The update step of a Newton iteration is carried out by extrapolating the reduced DOFs $\dr$ to the full order displacement vector $\ds$ with the projection matrix $\proj$ by
\begin{align}
\renewcommand{\arraystretch}{1.5}
\begin{bmatrix}
\ds\\
\p
\end{bmatrix}_{j+1}^{i+1}
=
\begin{bmatrix}
\ds \\
\p
\end{bmatrix}_{j+1}^i
+
\begin{bmatrix}
\proj \Delta \dr\\
\Delta \p
\end{bmatrix}_{j+1}^{i+1}\,.
\label{eq_rom}
\end{align}
Note that the full order structural residual $\rs$ and the full order block-Jacobian are always evaluated using the full order displacements $\ds$. It is only after their full evaluation and assembly that their dimensions are reduced by projection. The convergence check is carried out with the reduced residual and reduced displacement increment
\begin{align}
\norm{\infty}{\projt \rs} < tol^\mathrm{S}_\text{res}, \qquad % \left(\proj \ds^{i+1}_{r,n+1} \right)
\norm{\infty}{\Delta \dr} < tol^\mathrm{S}_\text{inc}.
\label{eq:ConvCritRedOrderSys} %\ds_{r,n+1}^{i+1}
\end{align}
The convergence criteria for the 0D windkessel model remain unchanged. As the coupled full order model \eqref{eq_fom} in section~\ref{sec_fom_solution}, the coupled \rom{} in \eqref{eq:RedLinSysV} is valid for any full order structural and windkessel residual $\rs$ and $\rd$, respectively.

\subsection{Subspace computation via POD}
\label{sec_subspace}
In this work, we use the method of Proper Orthogonal Decomposition (\ppod{}) to compute the reduced basis $\proj$ required for the projection-based reduction of our full problem. \ppod{} \cite{kunisch1999control,chatterjee2000introduction} is a straightforward and very well-known nonlinear model reduction approach, which relies on so-called \emph{snapshots}, i.e. discrete-time observations of the solution of our \fom{} for a fixed parameter set $\param$, to construct the basis $\proj$. Given $\ns \ll n$ snapshots $\ds_i$ gained from a numerical simulation of the \fom{} sample point, we obtain the snapshot matrix
\begin{align}
\matrm{D}=\left[ \vecrm{d}_1,\dots, \vecrm{d}_{\ns} \right] \in \mathbb{R}^{n \times \ns}.
\end{align}
The goal of \ppod{} is to construct a basis for an optimal approximation of the \emph{solution manifold} spanned by the snapshot matrix. In other words, the aim is to generate a basis that optimally approximates the information gathered in the snapshots. Therefore, we perform a singular value decomposition (SVD) of the snapshot matrix
\begin{align}
\matrm{D} = \matrm{U} \matrm{\Sigma} \T{\matrm{T}}
\end{align}
with the orthogonal matrices $\matrm{U} \in \mathbb{R}^{n \times n}$ and $\matrm{T} \in \mathbb{R}^{\ns \times \ns}$ containing the left and right singular vectors, respectively, stored column-wise. The diagonal matrix
\begin{align}
\matrm{\Sigma} = \text{diag} (\sigma_1, \dots, \sigma_{\ns}) \in \mathbb{R}^{n \times \ns}, \quad\text{where~} \sigma_1 \geq \dots \geq \sigma_{\ns} \geq 0,% \quad\text{with~} r = \mathrm{rank}(\matrm{D})
\end{align}
features all $\ns$ singular values $\sigma_i$ sorted in descending order on its main diagonal. We now select the first $q$ singular vectors $\vecrm{u}_i$ from the columns of the left singular matrix $\matrm{U}$ corresponding to the $q$ largest singular values $\sigma_i$ in $\matrm{\Sigma}$ to obtain the basis vectors
\begin{align}
\vecrm{v}_i = \vecrm{u}_i, \quad \forall i \in \{1,\dots,q\}
\end{align}
of the projection matrix $\proj$. The singular values $\sigma_i$ are frequently used to define the Relative Information Content (RIC)
\begin{align}
RIC(q) = \frac{\sum_{i=1}^q \sigma_i^2}{\sum_{i=1}^{\ns} \sigma_i^2} \quad \in \left[ 0, 1 \right].
\end{align}
This measure allows to select an appropriate basis dimension $q$ such that $RIC(q) \geq 1 - \varepsilon_{\text{POD}}$ for a given small tolerance $\varepsilon_{\text{POD}}$.\cite{bonomi2017matrix} The approximation error made by selecting $q<\ns$ basis vectors can be quantified by the sum of the squared truncated singular values 
\begin{align}
\mathrm{e}(q) = \sum_{i=q+1}^{\ns} \sigma_i^2.
\end{align}
Note that this technique provides an optimal basis for the approximation of the snapshot matrix in a least-squares sense.\cite{kunisch2002galerkin,liang2002proper,kerschen2005method} Thus, the efficiency of \ppod{} and the basis quality crucially depends on the selection of snapshots, which is required to represent the model's dynamics behavior sufficiently. Further note that \ppod{} requires the expensive numerical simulation of the full forward model, in general for many parameter sets, to collect representative snapshots. Nevertheless, this data-driven approach is very well applicable for the reduction of any nonlinear system.
%
%Note that no upper bound for the overall \ppod{} approximation error can be given by RIC or any other measure. Overall \ppod{} approximation error crucially depends on how well the selection of snapshots represents the model dynamics.

\subsection{Interpolation of subspaces}
\label{sec_pmor}
The cardiac model described in section~\ref{sec_fom} relies on many patient-specific parameters, describing e.g. constitutive behavior, hemodynamics, boundary conditions, or local fiber orientation. Consequently, a repeated model evaluation and simulation for many different values of the parameters is indispensable to personalize the model.
%
%One naive way would be to simulate the full order model for each parameter set and then perform a corresponding reduction via \ppod{}. Since this is pointless, ...
%
The aim of parametric model order reduction (\pmor{}) is to find a reduced cardiovascular model that preserves the parameter-dependency, thus allowing a variation of any of the parameters directly in the reduced model without having to repeat the whole reduction process each time. The parametric reduced model can be then used e.g. for patient-specific parameter estimation or uncertainty quantification purposes.

%The parameter dependency requires simulating the model multiple times for many different values of the parameters. 

%We seek a parameter-dependent reduced order model of much smaller dimension that captures the relevant dynamics of the original model over a range of parameters. Hence, the goal of parametric model order reduction (\pmor{}) is to preserve the parameter dependency in the ROM, so that the whole reduction process has not to be repeated for every new parameter value.

To efficiently reduce the parametric cardiovascular model, we decompose the \pmor{} procedure into an offline and online stage. In the \emph{offline phase}, the parametrized full order model with $n_p$ parameters $\param = \T{\left[\mu_1,\ldots,\mu_{n_p}\right]} \in \mathbb{R}^{n_p}$ is first simulated for several parameter sample points~$\param_k, k=1,\ldots,K$, and then corresponding local projection bases~$\proj(\param_k)$ are computed via \ppod{} from the data obtained. In the \emph{online phase}, the projection matrix~$\proj(\param^*)$ for a new parameter value~$\param^*$ is generated by interpolating between the precomputed subspaces.

In this paper, different subspace interpolation techniques are examined, which will be explained in the following. To do so, we suppose that local basis matrices $\proj_1 \!=\! \proj(\param_1), \ldots, \proj_{K} \!=\! \proj(\param_K) \in \mathbb{R}^{n \times q}$ spanning the subspaces $\mathcal{V}(\param_1), \ldots, \mathcal{V}(\param_K)$ have been computed in the offline phase from the snapshot matrices $\matrm{D}(\param_1), \ldots, \matrm{D}(\param_K) \in \mathbb{R}^{n \times \ns}$ at the sample points $\param_1, \ldots, \param_K$. Each basis matrix $\proj(\param_k)$ is composed of the vectors $\left\{\vecrm{v}_i(\param_k)\right\}_{i=1}^q$. For the interpolation, appropriate weighting functions $w_k(\param^*)$ should be selected to compute the interpolated basis $\proj(\param^*)$ in the online phase. Basically, any multivariate interpolation method could be used for this purpose: e.g. polynomial interpolation (Lagrange polynomials), piecewise polynomial interpolation (splines), radial basis functions (RBF),
 Kriging interpolation (Gaussian regression), inverse distance weighting (IDW) based on nearest-neighbor interpolation or even sparse grid interpolation.\cite{benner2015survey} For simplicity, in this paper we consider the special case of piecewise linear interpolation. We compare in this work four interpolation methods: weighted concatenation of bases, weighted concatenation of snapshots, adjusted direct basis interpolation, and basis interpolation on a Grassman manifold. A detailed mathematical description of the subspace interpolation methods is given in appendix~\ref{sec_app_interp}.

%where the weighting functions for a new parameter point $\param^*$ lying between two sample points $\param_1$ and $\param_2$ are given by
%\begin{align} \label{eq:lin-weight}
%w_k(\param^*) = \frac{\param^* - \param_2}{\param_1 - \param_2} \in [0,1] \quad \text{for } \param^* \in [\param_1,\param_2]\, .
%\end{align}
%
%For simplicity and a better insight into the interpolation methods, the case of a single varying parameter~$\mu$ ($n_p=1$) and two sample points ($K=2$) is considered, although the more.  
%
%$\matrm{D}(\mu_1)$, $\matrm{D}(\mu_2) \in \mathbb{R}^{n \times \ns}$ \\
%
%$\matrm{\Sigma}(\mu_1)$, $\matrm{\Sigma}(\mu_2)$ \\
%
%$\left\{\vecrm{v}_i(\mu_1)\right\}_{i=1}^q$, $\left\{\vecrm{v}_i(\mu_2)\right\}_{i=1}^q$ \\
%
%$\proj(\mu_1) = \left[\vecrm{v}_1(\mu_1), \ldots, \vecrm{v}_q(\mu_1) \right] \in \mathbb{R}^{n \times q}$, $\proj(\mu_2) = \left[\vecrm{v}_1(\mu_2), \ldots, \vecrm{v}_q(\mu_2) \right] \in \mathbb{R}^{n \times q}$ \\
%with the weight
%\begin{align} \label{eq:lin-weight}
%w := w(\param^*) = \frac{\param^* - \param_2}{\param_1 - \param_2} \in [0,1] \ \text{for } \param^* \in [\param_1,\param_2]\, .
%\end{align}

\subsection{Implementation details}
\label{sec_implementation}
The coupled \fom{} and \rom{} in \eqref{eq_fom} and \eqref{eq:RedLinSysV}, respectively, are solved using our in-house parallel high-performance finite element software package BACI \cite{wall10}. The code is implemented in C++ making use of the Trilinos library \cite{heroux05}. To the solve \fom{}'s large linear system in \eqref{eq_fom} we use a parallel iterative GMRES solver with $2\times2$ block SIMPLE-like preconditioning. For the \rom{}'s small linear system in \eqref{eq:RedLinSysV} we use a serial direct solver. All preliminary calculations, i.e. singular value decompositions and the interpolation of subspaces, are performed in MATLAB (Release 2017b, The MathWorks, Inc., Natick, MA, USA).

\section{Numerical results and discussion}
\label{sec_res}
In this section we present results for the approximation of the \fom{} simulation with \rom{} simulations. We distinguish here between model order reduction and parametric model order reduction. For a fixed parameter set, using the contractility $\sigma=280$~kPa, we explore the approximation qualities of \ppod{} in section~\ref{sec_res_mor}. Afterwards, we analyze the approximation quality with respect to a changing contractility in section~\ref{sec_res_pmor}. We employ a four chamber geometry obtained in vivo from a 33 year old healthy female volunteer. The imaging data was acquired at King's College London, UK using a Philips~Achieva~1.5T magnetic resonance imaging (MRI) scanner. The geometry was meshed using Gmsh \cite{geuzaine09} with a resolution of $2\text{ mm}$, yielding $282\,288$ nodes and $167\,232$ quadratic tetrahedral elements, totalling $n=846\,864$ structural degrees of freedom (DOFs). Additionally, we have four windkessel DOFs. The cut geometry is displayed in figure~\ref{fig_modes_ref}.

\subsection{Model order reduction}
\label{sec_res_mor}
In this section we demonstrate the reducibility of our coupled hemodynamical-structural simulation model of a cardiac cycle and compare computational costs. Following analysis of the heart's eigenmodes in section~\ref{sec_modes}, we investigate the approximation quality of \ppod{} using a varying number of modes~$q\in\{10,50,100,200,300,400,500\}$ in section~\ref{sec_mor_quality}. The model \rom{10} was the model with the smallest mode number, where the cardiac simulation converged to a result in all time steps. The highest mode number for \rom{500} was chosen, since there is a plateau around $q=500$ in the decay of singular values in figure~\ref{fig_PODResult}. We further demonstrate the computational speedup achieved by using \ppod{} in section~\ref{sec_speedup} using again varying mode numbers.

\subsubsection{POD-modes of the heart}
\label{sec_modes}
To study the reducibility of our cardiac model, we analyze the decay of the singular values compared to the first one. This gives a measure of relative importance of the modes selected by \ppod{}. In figure~\ref{fig_PODResult} we show the normalized singular value $\sigma_i/\sigma_1$ of mode $i$. For modes $i<50$ there is a fast decay in relative importance, indicating good reducibility. There is a plateau for $250<i<700$, indicating that not much new information is gained by including those modes in the \rom{}.

\begin{figure}
\centering
\newcommand{\hscale}{0.14}
\subfloat[Reference configuration.\label{fig_modes_ref}]{
\includegraphics[scale=\hscale,trim=5cm 2cm 39cm 3cm,clip]{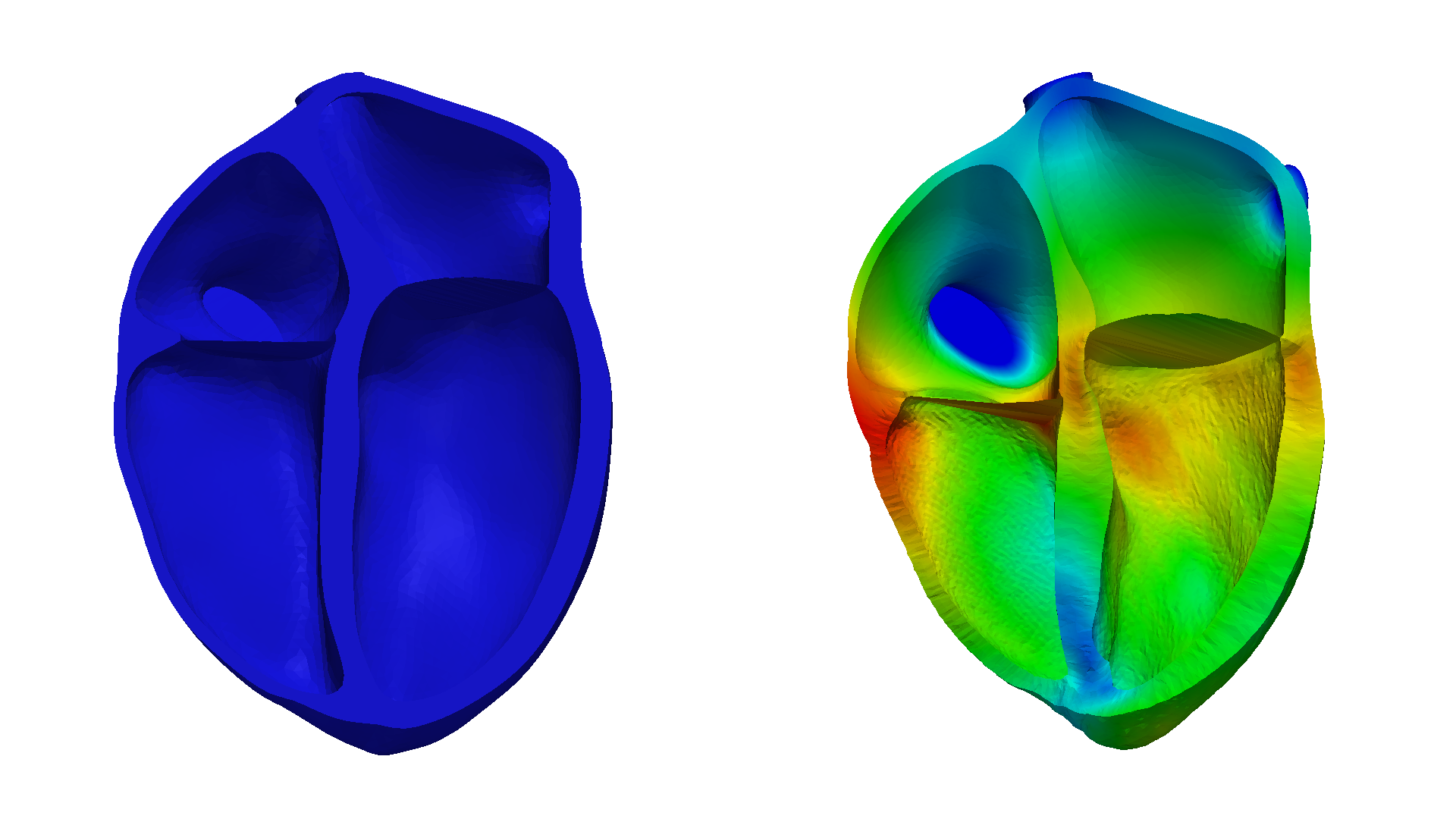}}~
\subfloat[End-systole.\label{fig_modes_es}]{
\includegraphics[scale=\hscale,trim=39cm 2cm 5cm 3cm,clip]{fig/Undeformed_ts505PeakSystole.png}}~
\subfloat[Mode $i=1$.\label{fig_mode1}]{
\includegraphics[scale=\hscale,trim=39cm 2cm 5cm 3cm,clip]{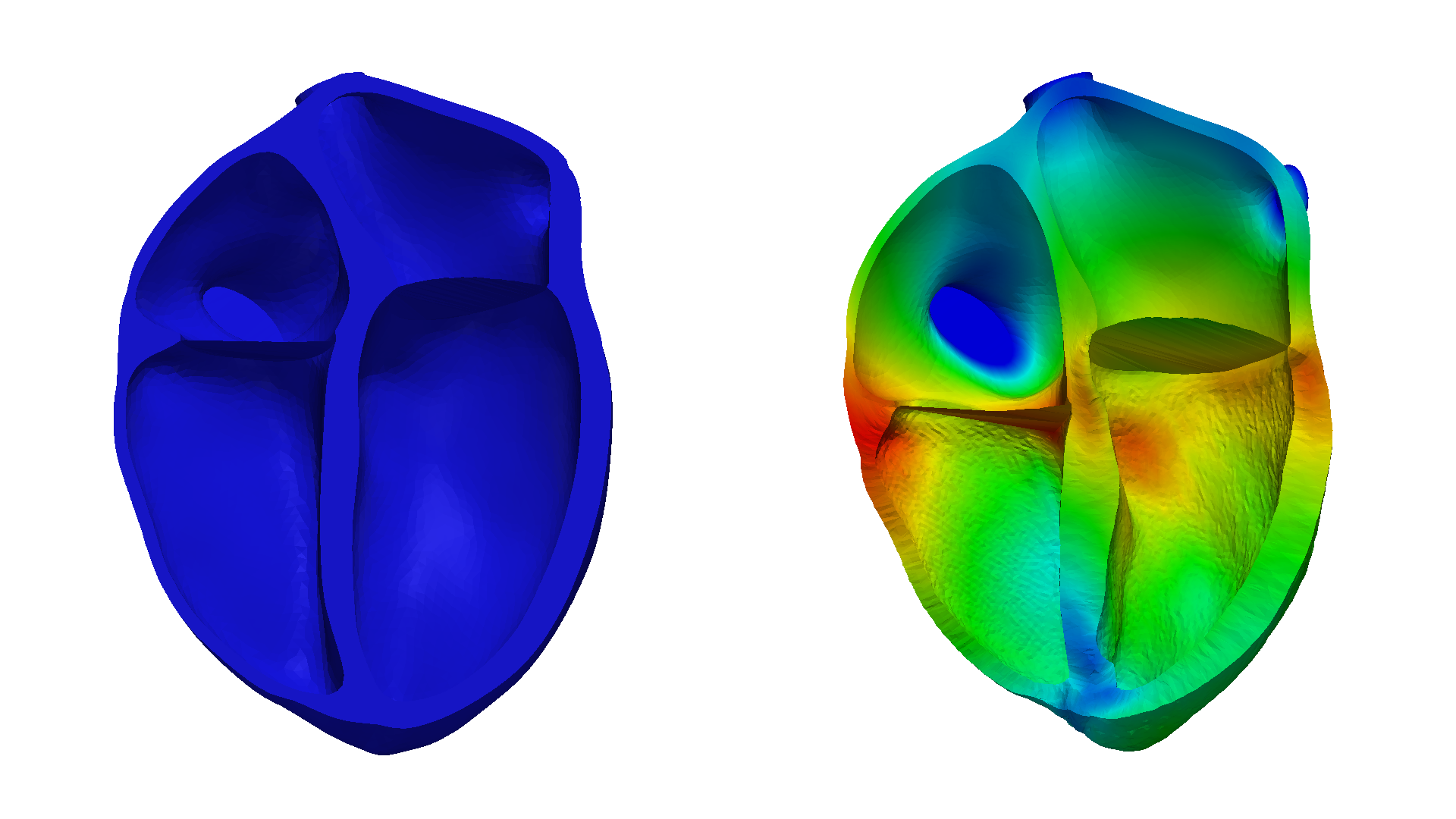}}~
\subfloat[Mode $i=2$.\label{fig_mode2}]{
\includegraphics[scale=\hscale,trim=39cm 2cm 5cm 3cm,clip]{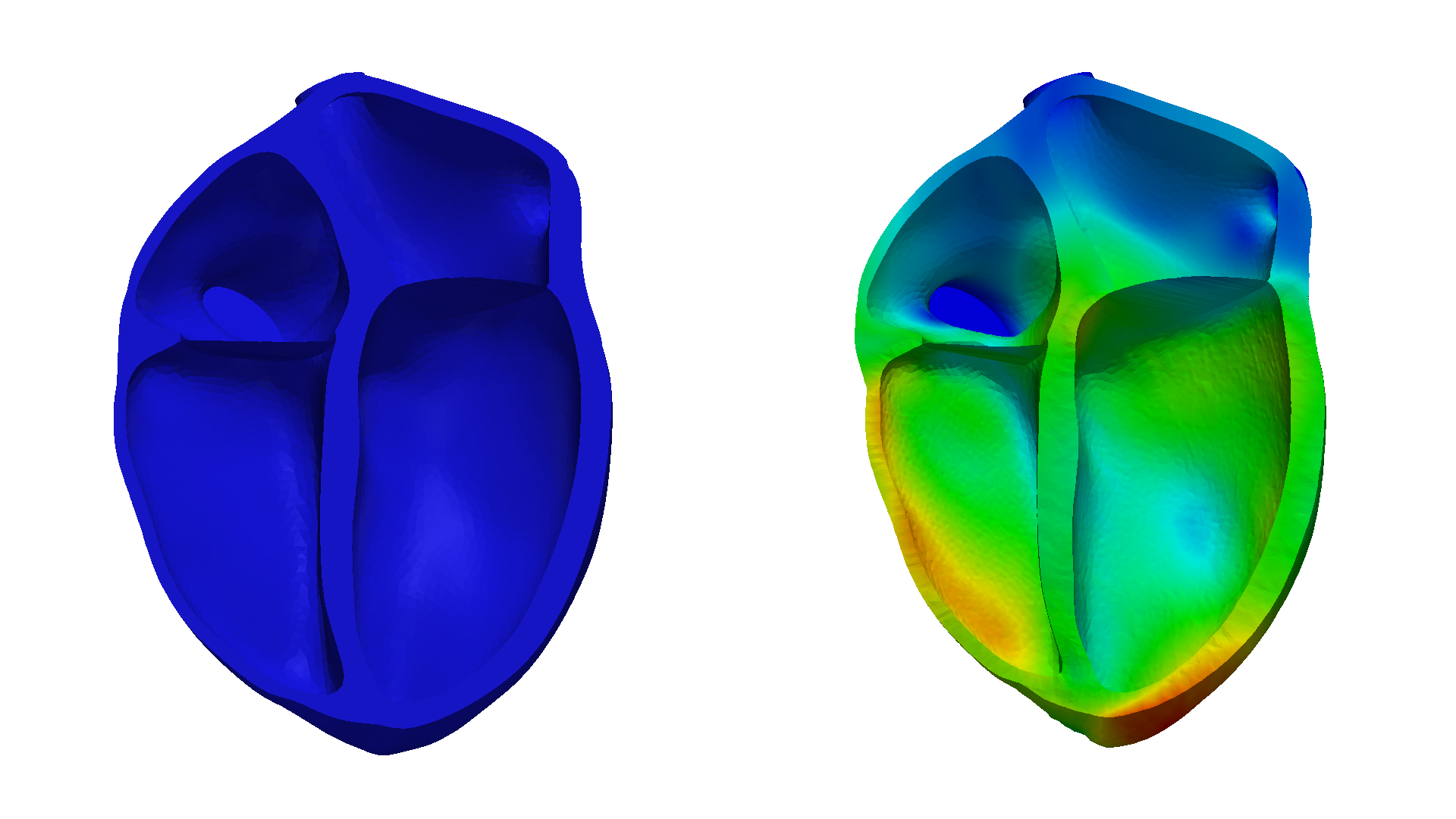}}~
\subfloat[Mode $i=3$.\label{fig_mode3}]{
\includegraphics[scale=\hscale,trim=39cm 2cm 5cm 3cm,clip]{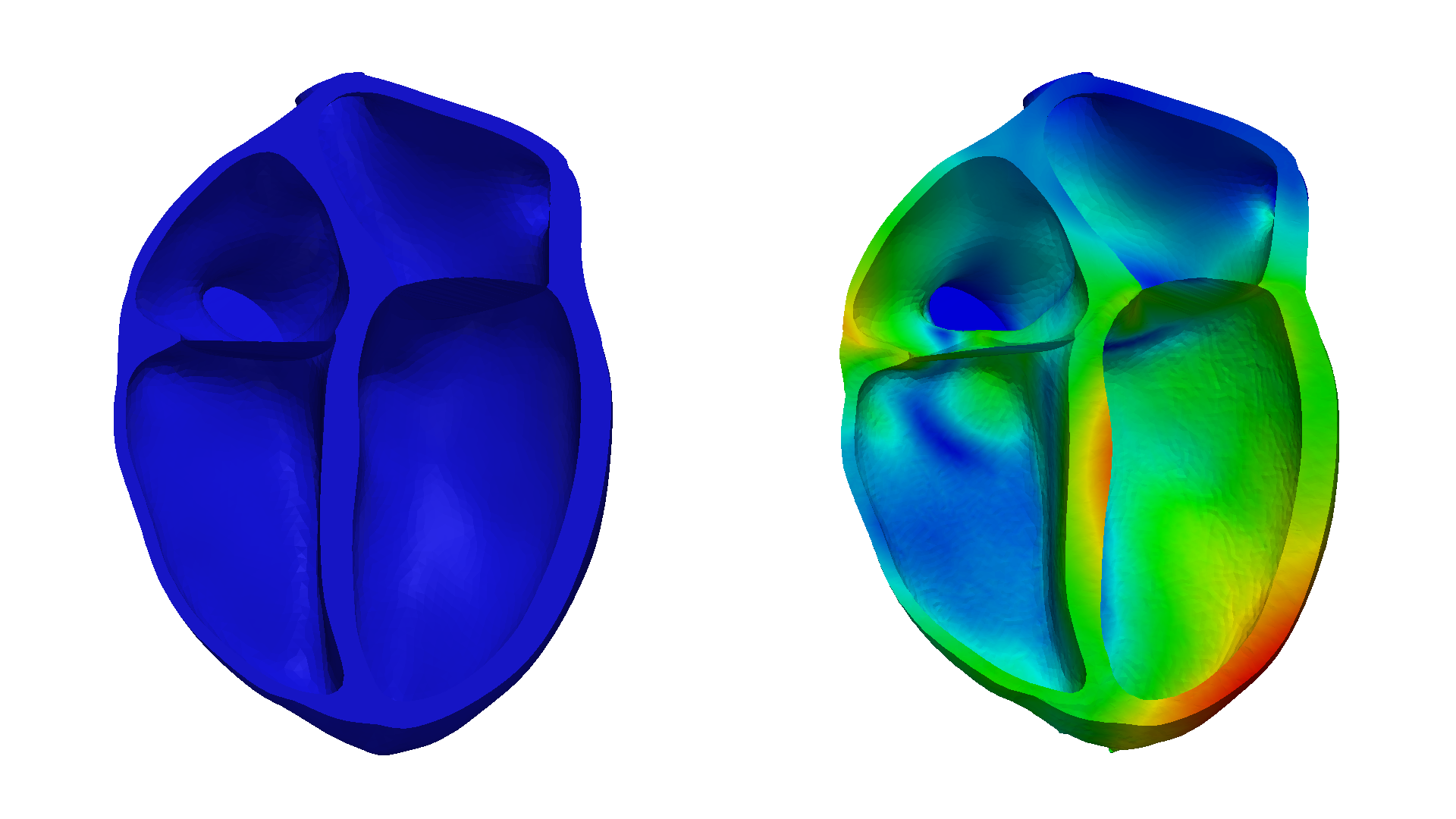}}
\caption{Visualized displacements in four-chamber-view. Displacements increase from blue to red regions.\label{fig_modes}}
\end{figure}

The first modes of the heart are visualized in figure~\ref{fig_modes}, where the heart is cut in four-chamber-view. The simulation in reference configuration and at end-systole are shown in figures~\ref{fig_modes_ref} and \ref{fig_modes_es}, respectively. Mode~$i=1$ in figure~\ref{fig_mode1} exhibits great similarity to the solution at end-systole and is characterized by a movement of the atrioventricular plane towards the apex with negligible change in outer shape of the heart. Mode~$i=2$ in figure~\ref{fig_mode2} consists of a more radial displacement of the outer walls of the ventricles and a pendulum motion of the intraventricular septum. Mode~$i=3$ in figure~\ref{fig_mode3} displays a rotating motion of the ventricles together with a large left-to-right movement of the intraventricular septum.

\subsubsection{Approximation quality}
\label{sec_mor_quality}
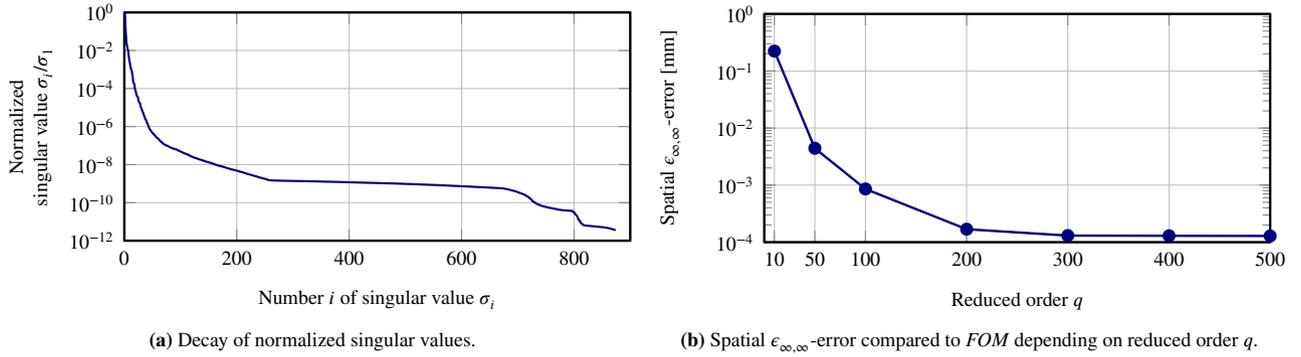
\begin{figure}
\setlength\figureheight{3cm}
\setlength\figurewidth{7cm}
\centering
\subfloat[Decay of normalized singular values.\label{fig_PODResult}]{
\input{tikz/PODResult.tex}}~
\subfloat[Spatial $\epsilon_{\infty,\infty}$-error compared to \fom{} depending on reduced order $q$.\label{fig_disp_err}]{
\input{tikz/err_infty_mor_all.tex}}
\caption{Accuracy of \rom{}.\label{fig_accuracy}}
\end{figure}

To quantify the overall approximation quality of \rom{} simulations, we calculate a spatial error compared to the \fom{} solution. We define here the spatial $\epsilon_{\infty,\infty}$-error
\begin{equation}
\epsilon_{\infty,\infty} = \max_{t_j} \left[ \max_k \norm{}{\ds_{\text{ROM}}^k(t_j) - \ds_{\text{FOM}}^k(t_j)} \right]
\end{equation}
with $\ds_{\text{ROM}}^k(t_j)$ and $\ds_{\text{FOM}}^k(t_j)$ as nodal displacements at node $k$ at time step $t_j$ of \rom{} and \fom{} respectively. The spatial $\epsilon_{\infty,\infty}$-error thus gives the highest displacement error at any node at any time step and is an upper bound for all spatial approximation errors. The $\epsilon_{\infty,\infty}$-error is shown in figure~\ref{fig_disp_err} depending on the number of reduced modes $q$. It is clearly evident that the approximation error strongly decreases, when more modes are used for the approximation. Remarkably, even for the very low number of 10 modes we obtain a solution whose largest approximation error at any node at any time step is below 1~mm, which is the order of magnitude of our MRI resolution from which the geometry was obtained. Furthermore, using \rom{} simulations with a reduced order of $q>300$ does not yield significant improvements in terms of accuracy. This is in agreement with the decay of the normalized singular values in figure~\ref{fig_PODResult}, where modes $q>300$ contain little more information than the preceding ones.

%The approximation error in figure~\ref{fig_disp_err} shows for $q\leq500$ a similar decay as the relative importance of the singular vectors in figure~\ref{fig_PODResult}. To display the correlation of both measures, we plot in figure~\ref{fig_correlation} the logarithmic approximation error against the logarithmic normalized singular values. The almost straight line indicates a monomial dependence of the measures on each other. The normalized singular values are thus in our case a good predictor of spatial approximation quality.
%%
%\begin{figure}
%\setlength\figureheight{3cm}
%\setlength\figurewidth{10cm}
%\centering
%\input{tikz/correlation.tex}
%\caption{Correlation of relative singular value $\sigma_q/\sigma_1$ with spatial $\epsilon_{\infty,\infty}$-error  compared to \fom{}.\label{fig_correlation}}
%\end{figure}

\begin{figure}
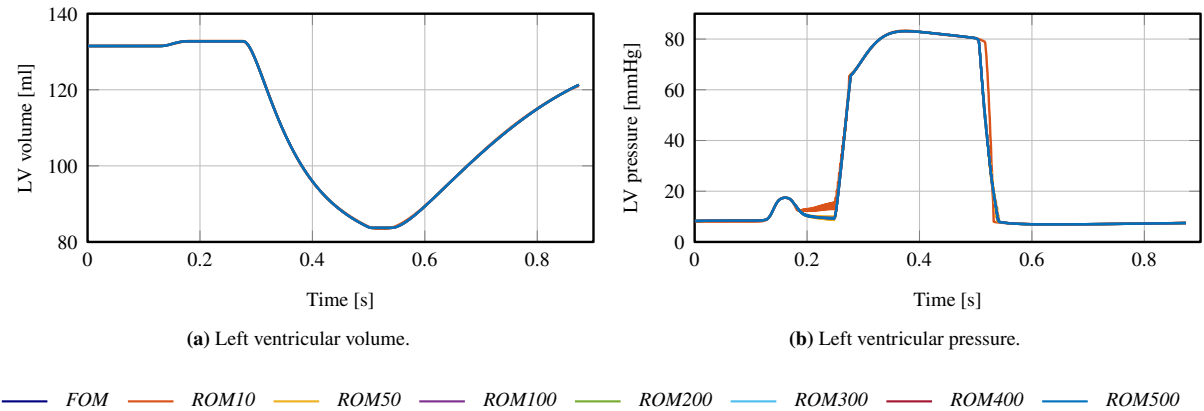

\setlength\figureheight{3cm}
\setlength\figurewidth{7cm}
\centering
\subfloat[Left ventricular volume.\label{fig_volume}]{
\input{tikz/volume_mor_all.tex}}~
\subfloat[Left ventricular pressure.\label{fig_pressure}]{
\input{tikz/lvp_mor_all.tex}}\\[2ex]
\input{./tikz/legend_q.tex}
\vspace{-2cm}
\caption{Scalar outputs of \fom{} and various \rom{}\textit{s} over time.\label{fig_scalar}}
\end{figure}

For many medical applications, it is not necessary to calculate an accurate spatial displacement field. There are rather a couple of scalar quantities which are used in clinical practice, e.g. as a cardiac performance indicator or for the prediction of disease progression. Such a quantity is the ejection fraction
\begin{equation}
\text{EF} = \frac{\max V - \min V}{\max V},
\label{eq_ef}
\end{equation}
which is calculated from left or right ventricular volume. To evaluate the approximation of the EF by a \rom{} simulation, we compare in figure~\ref{fig_volume} the left ventricular (LV) volume curves of the \fom{} simulation with \rom{} simulations of various reduced orders $q$. It shows that minimum and maximum volume are approximated well and the time curves are almost indistinguishable. We further compare left ventricular pressure over time for all simulations in figure~\ref{fig_pressure}. Again, key features such as maximum pressure are approximated well. Minor oscillations occur for \rom{10} and \rom{50} after the closure of the mitral valve at $t\approx0.2$. Furthermore, the closure of the aortic valve at $t\approx0.5$ is delayed slightly for simulation \rom{10}.

\subsubsection{Speedup}
\label{sec_speedup}
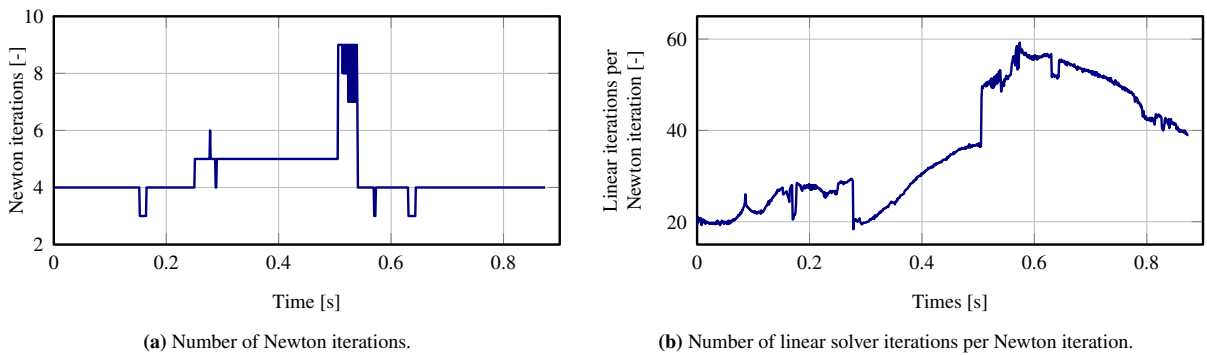
\begin{figure}
\setlength\figureheight{3cm}
\setlength\figurewidth{7cm}
\centering
\subfloat[Number of Newton iterations.\label{fig_newton}]{
\input{tikz/newton.tex}}\quad
\subfloat[Number of linear solver iterations per Newton iteration.\label{fig_linear}]{
\input{tikz/linear_solver.tex}}
\caption{Solver performance of \fom{} in each time step.\label{performance}}
\end{figure}

For performance measurements, we ran all \fom{} and \rom{} simulations on a single node of our Linux cluster. One node features 64 GB of RAM and two Intel Xeon E5-2680 "Haswell" processors, each equipped with 12 cores operating at a frequency of 2.5 GHz. In figure~\ref{performance} we give a brief overview of the numerical performance of the \fom{} simulation. We show in figure~\ref{fig_newton} the number of Newton iterations in each time step. The number of Newton iterations is between three and nine. It is elevated to five during ventricular systole and rises to nine at end-systole, where the aortic valve closes. The number of linear solver iterations of each Newton iteration at a given time step is shown in figure~\ref{fig_linear}. The number of linear iterations is between 20 and 60 and shows similar trends as the number of Newton iterations. This performance is reasonable and assures a good basis to which \rom{} simulations can be compared to. In the following, we compare exclusively simulation time and 
exclude time for creating the projection matrix~$\proj$. It is calculated once in a preliminary step using the same hardware and requires only about one minute.

\begin{figure}
\setlength\figureheight{3cm}
\setlength\figurewidth{4.8cm}
\centering
\subfloat[Speedup factor.\label{fig_speedup_factor}]{
\input{tikz/speedup_factor.tex}}~
\subfloat[Absolute components of computation time.\label{fig_speedup}]{
\input{tikz/speedup.tex}}~
\subfloat[Relative components of computation time.\label{fig_bar_time}]{
\input{tikz/bar_time.tex}}\\[2ex]
\input{./tikz/legend_area.tex}
\vspace{-2cm}
\caption{Simulation times of \fom{} and \rom{}.\label{fig_imes_mor_fom}}
\end{figure}
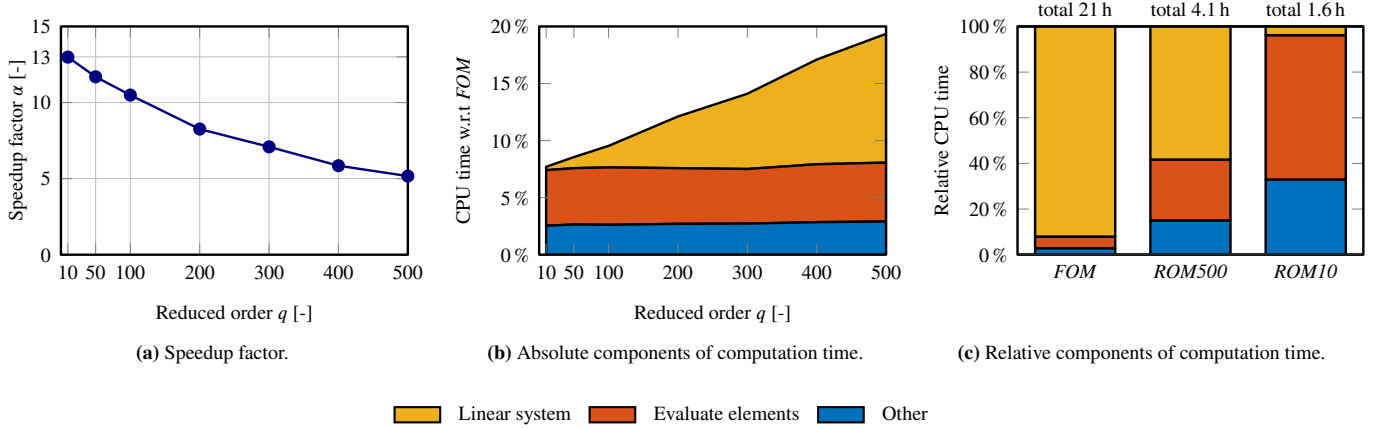

The computation time of \rom{} simulations with various reduced orders $q$ is compared to the total \fom{} simulation time in figure~\ref{fig_imes_mor_fom}. Firstly, we show in figure~\ref{fig_speedup_factor} the speedup factor $\alpha$ of \rom{} over \fom{} simulations. The effect of \ppod{} is evident for \rom{} simulations between $q=500$ and $q=10$, where we achieve a speedup of $\alpha\approx5$ and $\alpha\approx13$ over the \fom{} simulation, respectively. Note that while we achieve high speedups we did not lower hardware demands. The RAM consumption has actually increased slightly for \rom{} simulations, since we need to additionally store the projection matrix~$\proj$. This again motivates to pursue hyper-reduction in future studies.

We distinguish in figure~\ref{fig_speedup} between three components of total computation time. Component "Linear system" includes the time required for the multiplications of the projection matrix~$\proj$ with the blocks of the Jacobian matrix in \eqref{eq:RedLinSysV} as well as the time to solve the reduced linear system. This component strongly depends on the reduced order $q$ as it scales with the complexity of the matrix-matrix multiplications, which is the main time contributor. The solution time of the reduced linear system itself is negligible. Component "Evaluate elements" contains time spent during element evaluation to assemble the block Jacobian matrix and the right-hand side. As expected, this component is independent of $q$, since we still build the full system before projecting it to the $q$-dimensional subspace. Component "Other" sums up all other time spent during the simulation, e.g. file input and output or general overhead, and is also independent of $q$.

In figure~\ref{fig_bar_time} we show the relative distribution of simulation time for \fom{}, \rom{500}, and \rom{10}. For the 21 hours spent during a \fom{} simulation, 92$\%$ of simulation time are spent solving the linear system. This large proportion shows the potential for savings using \mor{} with \ppod{}. Reducing and solving the linear system in \rom{10} only makes up 4$\%$ of the simulation time. However, the new bottleneck is now the element evaluation at 63$\%$ of the simulation time. This, again, motivates the use of hyper-reduction methods, such as the discrete empirical interpolation method (DEIM) \cite{chaturantabut10}, for \rom{}\textit{s} with very few degrees of freedom.

\subsection{Parametric model order reduction}
\label{sec_res_pmor}
In this section we provide a quantitative comparison of several subspace interpolation methods introduced in section~\ref{sec_pmor} for parametric model order reduction (\pmor{}) to demonstrate the ability to evaluate the \rom{} simulations at parameter sets without prior \fom{} knowledge. We vary the contractility $\sigma$ in \eqref{eq_act_mat}, controlling the upper limit of the myocardium's active stress in fiber direction. It is a key parameter of cardiac contraction and has a major influence on elastodynamics as well as on several scalar cardiac measures. It is commonly calibrated to match the end-systolic volume of the left ventricle as measured in cine MRI. In this work, we vary the contractility $\sigma\in[280~\text{kPa},430~\text{kPa}]$, as this range produces \fom{} simulation results that are in agreement with cine MRI. We use $q=300$ for all \rom{} simulations in this section, since it was shown in figure~\ref{fig_disp_err} that no further improvements are made in approximation quality for 
$q>300$. In section~\ref{sec_pmor_disp} we demonstrate the approximation quality with respect to the spatial displacement field. However, in many clinical applications a full solution of the displacements is not needed. We therefore show the approximation quality of \pmor{} with respect to scalar cardiac quantities of clinical significance in section~\ref{sec_pmor_scalar}.

\subsubsection{Approximation of displacements}
\label{sec_pmor_disp}

\begin{figure}
\setlength\figureheight{3cm}
\setlength\figurewidth{10cm}
\centering
\subfloat[One sample point $\sigma_1=355$~kPa, increment $\Delta\sigma=15$~kPa.\label{fig_err_sample1}]{
\input{tikz/err_infty_mor300_sample1.tex}}\\
\subfloat[Two sample points $\sigma_k=\{280,430\}$~{[kPa]}, increment $\Delta\sigma=15$~kPa.\label{fig_err_sample2}]{
\input{tikz/err_infty_mor300_sample2.tex}}\\
\subfloat[Four sample points $\sigma_k=\{280,330,380,430\}$~{[kPa]}, increment $\Delta\sigma=5$~kPa.\label{fig_err_sample4}]{
\input{tikz/err_infty_mor300_sample4.tex}}\\[2ex]
\input{./tikz/legend_err.tex}
\vspace{-2cm}
\caption{Spatial $\epsilon_{\infty,\infty}$-error for direct \rom{300}, constant \rom{300}, and \prom{300} with different interpolation techniques for a varying number of sample points.\label{fig_err_sample124}}
\end{figure}

In figure~\ref{fig_err_sample124} we compare the spatial $\epsilon_{\infty,\infty}$-error for a varying contractility. We compare \pmor{} simulations using snapshots of \fom{} simulations from one, two, and four $\sigma$-sample points in figures \ref{fig_err_sample1}, \ref{fig_err_sample2}, and \ref{fig_err_sample4}, respectively. As a reference, we additionally show the error obtained from direct \rom{} simulations, i.e. \rom{} simulations where we use a projection matrix $\proj(\sigma)$ from the corresponding \fom{} evaluation for each $\sigma$. This information is however not used in the \pmor{} solutions displayed here and would not be available in a typical \mor{} scenario, as it would render \mor{} useless. The direct \rom{} approximation error is mostly independent of the choice for parameter $\sigma$.

We show in figure~\ref{fig_err_sample1} the \mor{} approximation with varying $\sigma$ and a constant projection matrix $\proj(\sigma_1)$ which was obtained from a single \fom{} simulation with sample point $\sigma_1 = 355$~kPa. Technically, this would not be considered in \pmor{}, since the projection matrix is not adapted to the parameter set. It can be observed that \mor{} simulations with $\sigma\ne\sigma_1$ provide reasonable results with a spatial error below 1~mm using $\proj(\sigma_1)$. However, with an increasing range of the parameter interval, the additional effort of subspace interpolation becomes advantageous. The approximation error of \mor{} simulations using two and four sample points are displayed in figures~\ref{fig_err_sample2} and \ref{fig_err_sample4}, respectively. Both studies show similar results. The error is highest between sample points and approaches the error of the direct \rom{} simulations close to the sample points. The \pmor{} approximation errors are reduced when using a 
finer resolution of sample points.   The subspace interpolation with the largest spatial error is obtained by the Grassmannian manifold and concatenation of bases (CoB) methods, coinciding in the middle between two sample points. An error of one order of magnitude smaller is achieved by using the concatenation of snapshots (CoS) and the adjusted direct interpolation methods, staying well below a spatial $\epsilon_{\infty,\infty}$-error of 1~mm.

The reason for the good performance of the adjusted direct interpolation method is due to the \ppod{} of both snapshot matrices of the left and right sample point yielding modes that allow for a distinctive pairing according to the Modal Assurance Criterion. However, the same subspace can be spanned by two sets of orthogonal basis vectors which are not necessarily linearly dependent on each other. If the information of a mode related to the left sample point is scattered among various modes corresponding to the right sample point, the direct interpolation method will most probably yield considerably worse results.

The CoS method is in our case far superior to the CoB method. As outlined in section~\ref{sec_pmor}, the CoS method allows for a direct usage of the snapshots at the sample points, while the CoB method uses the projection matrices computed at the sample points. The projection matrices, as compared to the snapshots, contain no information about the relative importance of the modes. The snapshots thus contain more information about the dynamics of the system, enabling the CoS method to select a more suitably interpolated subspace than the CoB method. Note, however, that the CoS method can only be applied when using it in combination with an observation-based reduction technique like \ppod{}.

\subsubsection{Approximation of scalar cardiac quantities}
\label{sec_pmor_scalar}

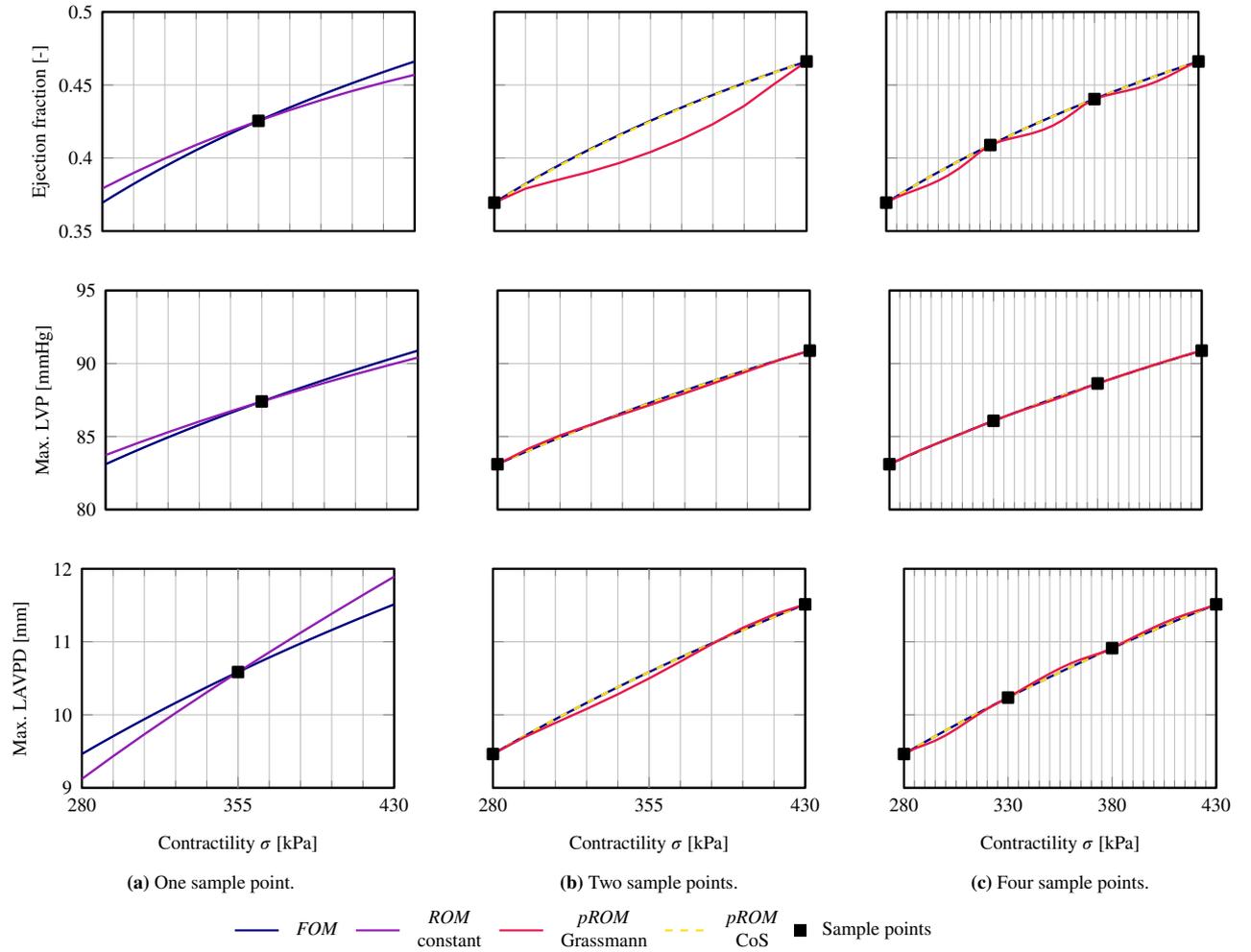
\begin{figure}
\centering
\setlength\figureheight{3cm}
\setlength\figurewidth{4.5cm}
\setlength\hsep{4ex}
\hspace{\hsep}\subfloat{\input{./tikz/sigma_vs_ef_sample1.tex}}\hspace{\hsep}
\subfloat{\input{./tikz/sigma_vs_ef_sample2.tex}}\hspace{\hsep}
\subfloat{\input{./tikz/sigma_vs_ef_sample4.tex}}\\\hspace{\hsep}
\subfloat{\input{./tikz/sigma_vs_pmax_sample1.tex}}\hspace{\hsep}
\subfloat{\input{./tikz/sigma_vs_pmax_sample2.tex}}\hspace{\hsep}
\subfloat{\input{./tikz/sigma_vs_pmax_sample4.tex}}\\\hspace{\hsep}
\setcounter{subfigure}{0}
\subfloat[One sample point.\label{fig_cardiac_sample1}]{\input{./tikz/sigma_vs_avpd_max_sample1.tex}}\hspace{\hsep}
\subfloat[Two sample points.\label{fig_cardiac_sample2}]{\input{./tikz/sigma_vs_avpd_max_sample2.tex}}\hspace{\hsep}
\subfloat[Four sample points.\label{fig_cardiac_sample4}]{\input{./tikz/sigma_vs_avpd_max_sample4.tex}}\\
%\hspace{\hsep}\footnotesize{Contractility $\sigma$ [kPa]}\\[2ex]
\input{./tikz/legend_scalar.tex}
\vspace{-2cm}
\caption{Scalar cardiac quantities ejection fraction (top), maximal left ventricular pressure (middle), and maximal left atrioventricular plane displacement (bottom) for varying contractility $\sigma$ in \fom{} and \rom{300}. \label{fig_cardiac_sample124}}
\end{figure}

In figure~\ref{fig_cardiac_sample124} we show scalar output quantities of our cardiac model, evaluated for the same \mor{} simulations as in figure~\ref{fig_err_sample124} using one, two, and four sample points in figures \ref{fig_cardiac_sample1}, \ref{fig_cardiac_sample2}, and \ref{fig_cardiac_sample4}, respectively. For each study we evaluate ejection fraction (EF), maximum left ventricular pressure (LVP), and maximum left atrioventricular plane displacement (LAVPD) and compare them to the the \fom{} results. All three output quantities are important determinants of cardiac viability. They are also chosen because they allow us to study the approximation of different outputs of our coupled 3D-0D elasto-hemodynamical model with \pmor{}. EF, as defined in \eqref{eq_ef}, is an integral value of the spatial displacement field, LVP is an output of the 0D windkessel model, and LAVPD is the average of a small subset of nodal directional displacements.

For one sample point we again compare the \fom{} solution in figure~\ref{fig_cardiac_sample1} to the solution of a \rom{} using a constant projection matrix. The parameter dependence of the three cardiac quantities on $\sigma$ is reproduced well by the constant \rom{} simulations. As expected, the deviations from the \fom{} solution are largest at evaluations furthest away from the sample point. However, the accuracy might still be sufficient for many applications. The cardiac quantities using two and four sample points are shown in figures~\ref{fig_cardiac_sample2} and \ref{fig_cardiac_sample4}, respectively. For clarity, we show here only the results of the subspace interpolation methods which performed best and worst in figure~\ref{fig_err_sample124}, i.e. the CoS and Grassmann method, respectively. The outputs oscillate visibly between sample points when using the Grassmann interpolation method, improving as the resolution of sample points is refined. As in figure~\ref{fig_err_sample124}, the CoS 
method performs well, leading to a good approximation of the scalar cardiac quantities between sample points.

\section{Application to inverse analysis}
\label{sec_inv_ana}

Models of cardiac elasto-hemodynamics commonly depend on a large set of parameters which need to be calibrated to patient-specific measurements using inverse analysis. This procedure is however computationally very expensive, as it requires many evaluations of the model \eqref{eq_fom}, which will be denoted forward model in the following. We propose in this section a novel approach for gradient-based inverse analysis utilizing the parametric reduced order model (\prom{}) developed in section~\ref{sec_res_pmor}. We replace the \fom{} forward evaluations typically required for the finite differences to obtain the gradient by \pmor{} forward evaluations \eqref{eq:RedLinSysV}. We illustrate this method using a simple Levenberg-Marquardt (LM) algorithm \cite{levenberg44,marquardt63} in section~\ref{sec_inv_ana_method}. The method however is applicable to any gradient-based optimization. In section~\ref{sec_inv_ana_res} we demonstrate its performance on a typical inverse analysis scenario using the cardiac forward 
problem introduced in section~\ref{sec_fom}.

\subsection{Parameter estimation based on reduced models}
\label{sec_inv_ana_method}

Given $m$ normalized model outputs $\vecrm{f}(\param)\in\mathbb{R}^m$ of a \fom{} depending on $n_p$ normalized parameters $\param\in\mathbb{R}^{n_p}$ and $m\geq n_p$ normalized measurements $\vecrm{y}$ we aim to minimize the squared sum $S$ of residuals $\vecrm{r}$
\begin{align}
\hat{\param} = \argmin_{\param}~S(\param), \quad\text{with~} S(\param) = \frac{1}{2} \norm{2}{\vecrm{r}(\param)}^2, \quad \vecrm{r}(\param) = \vecrm{y} - \fom{}~\vecrm{f}(\param), \quad\jac(\param) = \pfrac{\vecrm{r}(\param)}{\param}
\label{eq_optim}
\end{align}
to obtain the optimal set of parameters $\hat{\param}$. The Jacobian of the residual vector with respect to the parameter vector is $\jac\in\mathbb{R}^{m\times n_p}$. At the optimum $\hat{\param}$, the gradient $\nabla S = \T{\jac} \vecrm{r} = \vecrm{0}$ vanishes and the Hessian $\nabla^2 S > \vecrm{0}$ is positive definite. Using the LM algorithm, we obtain the iterative procedure
\begin{alignat}{3}
&\text{update~} \quad &&\param^{i+1} = \param^i + \Delta\param^{i+1}, \label{eq_lm_update}\\
&\text{with~} \quad &&\left[ \T{\jac}\jac + \lambda \text{\,diag} \left( \T{\jac}\jac \right) \right]^i \cdot \Delta \param^{i+1} = - \left[  \T{\jac} \vecrm{r} \right]^i, 
\quad \lambda^i = \lambda^{i-1} \cdot \norm{2}{\left[ \T{\jac} \vecrm{r} \right]^i} / \norm{2}{ \left[  \T{\jac} \vecrm{r} \right]^{i-1}} \label{eq_lm_iter}\\
&\text{until~} \quad &&\norm{2}{\left[ \T{\jac} \vecrm{r} \right]^i}<tol^{\param}_\text{grad}
\quad\text{and}\quad\norm{2}{\Delta\param^{i+1}}<tol^{\param}_\text{inc}, \label{eq_lm_tol}
\end{alignat}
at iteration $i+1$ with damping parameter $\lambda$. The LM algorithm approximates the Hessian as $\nabla^2 S \approx \T{\jac}\jac$. The damping parameter $\lambda$ should tend to zero as the parameter set $\param$ approaches the optimal solution $\hat{\param}$. For $\lambda\to\infty$ we approach the steepest descent method, for $\lambda=0$ we obtain the Gauss-Newton method. In general, for a nonlinear model the analytical derivatives of the model evaluations $\vecrm{f}$ with respect to the parameters $\param$ required for the Jacobian matrix are not easily available. The $n_p$ columns $\jac^i_p$ of the Jacobian matrix $\jac^i$ are thus typically approximated by finite differences
\begin{align}
\jac^i_p \approx \pm \frac{\text{\prom{}}~\vecrm{f}(\param^i_{\Delta p}) - \text{\prom{}}~\vecrm{f}(\param^i)}{\epsilon_p}, \quad\text{with~} \param^i_{\Delta p} = \param^i \pm \epsilon_p \vecrm{e}_p, \quad\forall p\in [1,\dots,n_p]
\label{eq_jac}
\end{align}
The gradient evaluation vector $\param^i_{\Delta p}$ is built from the $p$-th component $\epsilon_p$ of a finite distance vector $\vecrm{\epsilon} \in\mathbb{R}^p$ and the $p$-th unit vector $\vecrm{e}_p \in\mathbb{R}^p$ in the direction of each parameter. The sign in \eqref{eq_jac} is chosen for each parameter $p$ so that the evaluation with parameter set $\param^i_{\Delta p}$ is within the range of all previously evaluated parameter sets. 

Calculating the approximated Jacobian matrix requires $n_p+1$ evaluations of our forward model which is computationally expensive in case of a large number of parameters $n_p$. The \prom{} introduced in section~\ref{sec_res_pmor} is very accurate for parameters in the proximity of the sampled parameter sets, as was shown in section~\ref{sec_pmor_disp} for the cardiac contractility parameter $\sigma$. More importantly, the \pmor{} evaluations shown in section~\ref{sec_pmor_scalar} were not only able to accurately predict \fom{} scalar cardiac quantities but also their slope with respect to a changing contractility from only two \fom{} sample points. We therefore propose to use \prom{} evaluations of $\vecrm{f}$ in \eqref{eq_jac} instead of \fom{} evaluations. The iterative procedure for the inverse analysis is sketched in algorithm~\ref{algo_inv_ana}. Note that while using this approach, we still find a local minimum of the objective function $S$ in \eqref{eq_optim} with respect to the \fom{}.

\begin{algorithm}
\begin{spacing}{1.0}
\begin{algorithmic}[1]
%\Procedure{Inverse analysis}{$\param$}
\State initialize $\param^0, \lambda^0$
\State $i=0$
\While{convergence criterion from \eqref{eq_lm_tol} not fulfilled}
\vspace{.1cm}
\State evaluate \fom{} $\vecrm{f}(\param^i)$ and calculate residual $\vecrm{r}^i$ from \eqref{eq_optim}
\State store snapshots $\vecrm{D}(\param^i)$
\For{$p=0,\dots,n_p$}
\vspace{.1cm}
\State build reduced basis $\vecrm{V}(\param^i_{\Delta p})$  from \eqref{eq_interp_inv_ana}
\vspace{.1cm}
\State evaluate \prom{} $\vecrm{f}(\param^i_{\Delta p})$
\vspace{.1cm}
\EndFor
\State calculate Jacobian $\jac^i$ from \eqref{eq_jac}
\State update parameter vector $\param^{i+1}$ from \eqref{eq_lm_iter}
\State $i\gets i+1$
\vspace{.1cm}
\EndWhile\label{euclidendwhile}
\State \textbf{return} $\hat{\param}=\param^i$
%\EndProcedure
\end{algorithmic}
\end{spacing}
\caption{Inverse analysis with \pmor{}-gradient}\label{algo_inv_ana}
\end{algorithm}

Further note that with the strategy introduced in algorithm~\ref{algo_inv_ana} the $n_p+1$ gradient evaluations using the \prom{} simulations can only be computed after a complete evaluation of the \fom{} simulation. Considering a scenario of infinite available computing resources, our strategy would actually slightly increase computation time over the standard approach of using the \fom{} for all evaluations, as all $n_p+1$ model evaluations can here be run in parallel. However, considering the more likely scenario where computing resources are just sufficient to calculate one or few \fom{} simulations at a time, the strategy outline in algorithm~\ref{algo_inv_ana} leads to considerable time savings especially for a large number of parameters $n_p$.

Algorithm~\ref{algo_inv_ana} can be combined with any subspace interpolation method in step~7. We use here the weighted concatenation of snapshots method (CoS) introduced in section~\ref{sec_cos}, as it performed best in the experiments in section~\ref{sec_res_pmor} and is easily applicable to high dimensions of $n_p$. For the weights of the snapshot matrix for gradient evaluation $p$, we use a simple inverse distance weighting between two evaluation points
\begin{equation}
\begin{aligned}
\tilde{\matrm{D}}(\param^i_{\Delta p}) = \left[w_1 \, \matrm{D}(\param^i), \, w_2 \, \matrm{D}(\param^k) \right], \quad
w_1 = \frac{1/d_1}{1/d_1+1/d_2}, \quad w_2=1-w_1, \\
d_1 = \norm{2}{\param^i_{\Delta p} - \param^i}, \quad
d_2 = \norm{2}{\param^i_{\Delta p} - \param^k}, \quad
k = \argmin_{j\in[0,\dots,i-1[} \norm{2}{\param^i_{\Delta p} - \param^j },
\end{aligned}
\label{eq_interp_inv_ana}
\end{equation}
with normalized distances $d_1,d_2$ and weights $w_1,w_2$ for the current evaluation $i$ and the next closest evaluation $k$, respectively. Since at the beginning of the iteration $\epsilon_p\ll|\mu_p^i-\mu_p^k|$, the weight $w_1$ of the current snapshot matrix $\matrm{D}(\param^i)$ is always close to one, whereas the weight $w_2$ is close to zero. This can be interpreted that we "enrich" the snapshots of the current iteration with snapshots from a previous iteration to represent parametric dependence. As the optimization converges and the changes in parameters are close to the step size of the finite differences, the weights $w_1$ and $w_2$ equalize. For the first iteration of the optimization we rely on standard \mor{} evaluations using the constant projection matrix from the first \fom{} evaluation.

In the following, we give an equation for the speedup of gradient-based inverse analysis achieved by using \prom{} evaluations for the calculation of the Jacobian with respect to CPU time. Note that actual computation time depends on parallelization of model evaluations. We compare CPU time required to achieve convergence after $n_i$ iterations for a model with gradients calculated from \prom{} and \fom{} forward model evaluations, denoted by superscript \prom{} and \fom{}, respectively. We do not include the time spent during subspace generation, as it is negligibly small compared to \prom{} and \fom{} evaluation time. The total CPU times $T$ are
\begin{align}
T^{\text{\fom{}}} &= n_i^{\text{\fom{}}} (n_p+1)\, t^{\text{\fom{}}}, \\
T^{\text{\prom{}}} &= n_i^{\text{\prom{}}} \left[ t^{\text{\fom{}}} + (n_p+1) \, t^{\text{\prom{}}} \right],
\label{eq_timing1}
\end{align}
where $t$ is the time required for a single forward evaluation. It can be observed from \eqref{eq_timing1} that the number of parameters only scales the \prom{} evaluation time but not the \fom{} evaluation time. Using the speedup $\alpha$ of a single \prom{} evaluation over a \fom{} evaluation, we obtain the total speedup $\beta$ for the inverse problem with respect to CPU time as
\begin{align}
\beta = \frac{T^{\text{\fom{}}}}{T^{\text{\prom{}}}} = \underbrace{\frac{n_i^{\text{\fom{}}}}{n_i^{\text{\prom{}}}}}_{\approx 1} \cdot \underbrace{\frac{1}{\frac{1}{\alpha} + \frac{1}{1+n_p}}}_{\to\alpha \text{~for~} n_p\to\infty}, \quad\text{with~} \alpha = \frac{t^{\text{\fom{}}}}{t^{\text{\prom{}}}}.
\label{eq_timing2}
\end{align}
As will be shown in section~\ref{sec_inv_ana_res}, the first factor is close to one as the number of iterations is comparable for both approaches when using a reasonably large number of reduced modes~$q$. The second factor approaches $\alpha$ in the case of many parameters. Note that in practice there is a trade-off between the two factors. Choosing a very low-dimensional reduced model with few degrees of freedom $q$ results in a high single call speedup $\alpha$ but may increase the number of iterations for the inverse analysis, as the tangents are now approximated worse than with a higher $q$. Further note that, after their respective number of iterations, both approximations achieve the same convergence criterion, which is always evaluated using the \fom{}.

Other variants of algorithm~\ref{algo_inv_ana} are feasible, e.g. replacing all \fom{} evaluations by \prom{} approximations as the inverse analysis algorithm converges closer to the optimum. Such algorithms however require more advanced strategies to switch between both model evaluations. The algorithm presented here demonstrates the most simple and straightforward approach of including a \prom{} within a finite difference gradient-based inverse analysis.

\subsection{Numerical results}
\label{sec_inv_ana_res}

We demonstrate the ability of the inverse analysis method proposed in section~\ref{sec_inv_ana_method} to accurately and efficiently estimate parameters for a real-world cardiac estimation problem. We consider the case of a cardiac simulation which we want to calibrate to a given volume curve, i.e. measurements of left ventricular volume over time during one cardiac cycle. We have no prior solutions of our \fom{} and thus need to build our projection matrices from scratch starting at the first iteration of algorithm~\ref{algo_inv_ana}.

\begin{table}
\centering
\footnotesize
\setlength{\tabcolsep}{.3em}
\renewcommand{\arraystretch}{1.5}
\begin{tabular}{| l | c | c | c | c | c |}
\hline
&
$\sigma$ $\left[\text{kPa}\right]$ & 
$\alpha_{\text{max}}$ $\left[\frac{1}{\text{s}}\right]$ & 
$\alpha_{\text{min}}$ $\left[\frac{1}{\text{s}}\right]$ &
$t_{\text{sys}}$ $\left[\text{s}\right]$ & 
$t_{\text{dias}}$ $\left[\text{s}\right]$ \\
\hline
\hline
Initial & 200 & 15 & -15 & 0.35 & 0.60 \\
\hline
%Ground truth & 280 & 10 & -30 & 0.246 & 0.502 \\
Ground truth & 280 & 10 & -30 & 0.25 & 0.50 \\
\hline
\end{tabular}
\caption{Initial values and ground trouth of estimated parameters during inverse analysis ($n_p = 5$).\label{tab_inv_ana}}
\end{table}

We choose the solution displayed in figure~\ref{fig_volume} of a forward \fom{} simulation as our ground truth. As parameters we choose contractility $\sigma$ from \eqref{eq_act_mat} and myofiber activation rate $\alpha_{\text{max}}$, myofiber deactivation rate $\alpha_{\text{min}}$, onset of ventricular systole $t_{\text{sys}}$, and onset ventricular diastole $t_{\text{dias}}$ from \eqref{eq_activation}. We thus estimate all parameters necessary to determine the shape of the input function of our model, i.e. the active stress over time $\tau(t)$. The parameters $\sigma, \alpha_{\text{max}}$, and $\alpha_{\text{min}}$ control cardiac output. However, due to their large variation they are commonly calibrated to a given patient \cite{chabiniok12}. These parameters are interconnected with the timing parameters $t_{\text{sys}}$ and $t_{\text{dias}}$. The non-normalized parameters at the start of the inverse analysis and of the ground truth are listed in table~\ref{tab_inv_ana}. We initialize the damping 
parameter $\lambda^0=0.1$. We further choose the number of reduced modes $q=300$ as it offers a good trade-off between accuracy and speedup.

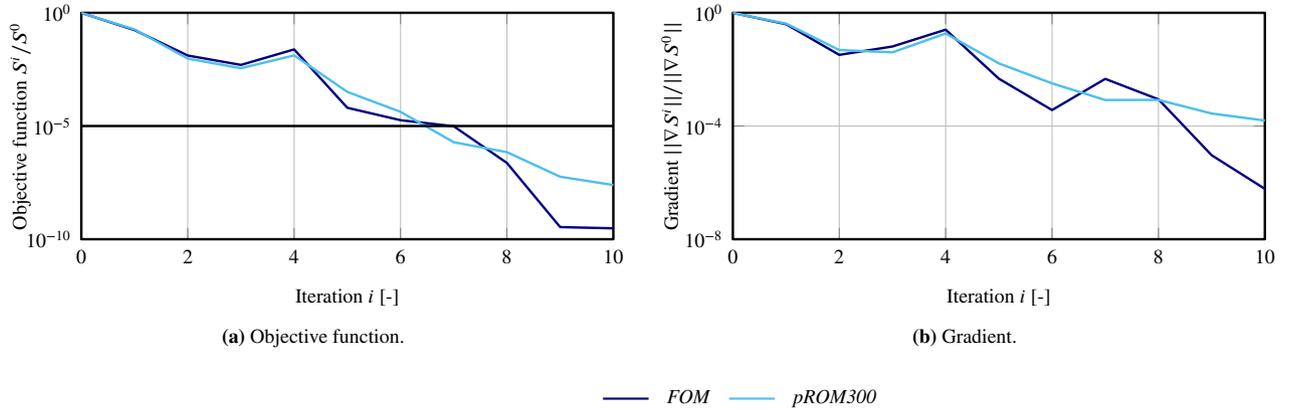
\begin{figure}
\setlength\figureheight{3cm}
\setlength\figurewidth{7cm}
\centering
\subfloat[Objective function.\label{fig_iter_cost}]{
\input{tikz/eva_cost.tex}}~
\subfloat[Gradient.\label{fig_iter_grad}]{
\input{tikz/eva_grad-norm.tex}}\\[2ex]
\input{./tikz/legend_inv_ana_performance.tex}
\vspace{-2.3cm}
\caption{Convergence behavior during gradient-based inverse analysis with finite differences for gradient calculation. Shown are objective function and gradient for each iteration, comparing the use of \fom{} and \prom{} for gradient calculation.\label{fig_iter}}
\end{figure}

In figure~\ref{fig_iter} we display the performance of the \prom{} inverse analysis using algorithm~\ref{algo_inv_ana} compared to the standard approach where the gradients are evaluated using the \fom{} only. Figure~\ref{fig_iter_cost} shows the decay of the objective function $S$ from \eqref{eq_optim}. We define a convergence criterion $S^i/S^0<10^{-5}$, which is achieved at $n_i^{\text{\fom{}}}=n_i^{\text{\prom{}}}=7$. In figure~\ref{fig_iter_grad} we compare the development of the gradient of the objective function with respect to the parameters. As we consider a synthetic case in the absence of noise, both objective function and gradient should approach zero as $i\to\infty$. As both measures are non-monotonically decreasing, this indicates a non-smooth optimization problem. However, the \prom{300}-gradient optimization is in excellent agreement with the \fom{}-gradient optimization.

\begin{figure}
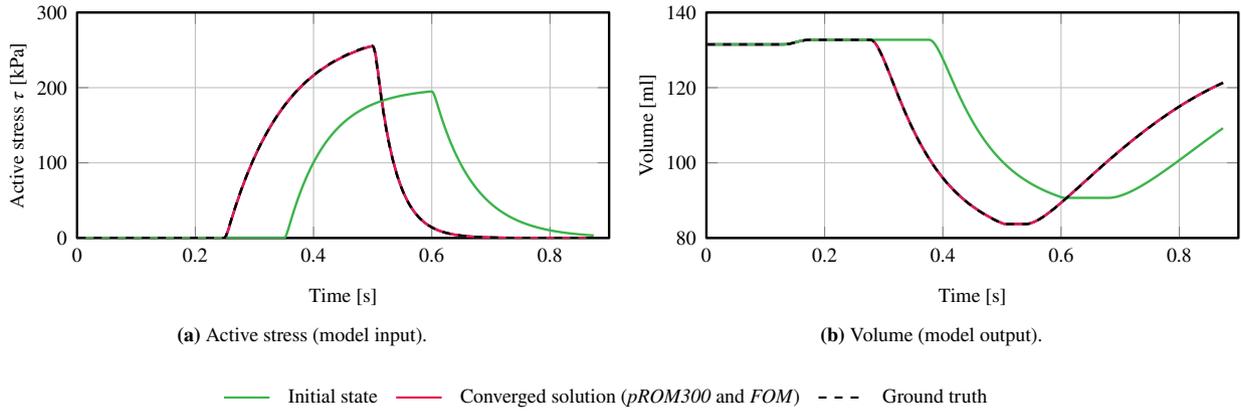

\setlength\figureheight{3cm}
\setlength\figurewidth{7cm}
\centering
\subfloat[Active stress (model input).\label{fig_iter_activation}]{
\input{tikz/eva_activation.tex}}~
\subfloat[Volume (model output).\label{fig_iter_solution}]{
\input{tikz/eva_solution.tex}}\\[2ex]
\input{./tikz/legend_inv_ana.tex}
\vspace{-2.3cm}
\caption{Initial state, converged solution, and ground truth of inverse analysis for model input (active stress) and model output (volume) over time.\label{fig_solution}}
\end{figure}

\begin{figure}
\setlength\figureheight{3cm}
\setlength\figurewidth{7cm}
\centering
\subfloat[\fom{} parameters.\label{fig_params_fom}]{
\input{tikz/eva_params_FOM.tex}}~
\subfloat[\prom{300} parameters.\label{fig_params_rom}]{
\input{tikz/eva_params_ROM300.tex}}\\[2ex]
\input{./tikz/legend_params.tex}
\vspace{-2.3cm}
\caption{Convergence of parameters in inverse analysis. Dotted lines indicate ground truth of parameter.\label{fig_params}}
\end{figure}
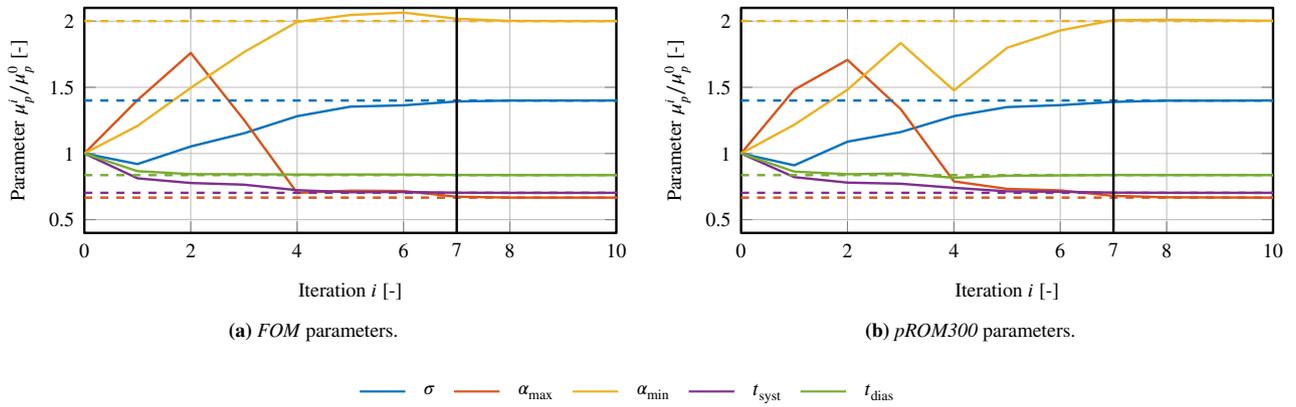

The start, ground truth, and the converged solutions of both methods after seven iterations are shown in figure~\ref{fig_solution}. Here, the activation function, i.e. the input of our model, and the volume, i.e. the output of our model, are shown in figures~\ref{fig_iter_activation} and \ref{fig_iter_solution}, respectively. It can be observed that both optimization methods match well with ground truth data for the given convergence criterion. The convergence of the five parameters relative to their initial values is shown in figure~\ref{fig_params} for both methods. Additionally, the iteration where the convergence criterion is achieved is indicated. Both methods show a similar trend towards the optimal parameters. With a single evaluation speedup of $\alpha\approx7.1$, we obtain an overall speedup in CPU time of the \prom{300} method over the \fom{} method of $\beta\approx3.3$, since $n_i^{\text{\fom{}}}/n_i^{\text{\rom{}}} = 1$ and $n_p=5$ in our case.% In this example, using \pmor{} simulations to 
evaluate the gradients reduces CPU time by 69\% while achieving the same accuracy.

\section{Conclusions}
\label{sec_conclusion}

% mor
In this work we proposed a new projection-based reduced order model for coupled structure-windkessel cardiac models, where we solely reduced the large structural dimension. Specifically, we used a nonlinear large deformation cardiac finite element model with pericardial boundary conditions. For subspace generation, we employed proper orthogonal decomposition (\ppod{}) applied to displacement snapshots of the full order model (\fom{}). We demonstrated the accuracy and speedup of the reduced order model (\rom{}) for a range of reduced dimensions $q\in\{10,\dots,500\}$. In that range, the approximation error was found to be between $2 \cdot 10^{-1}$~mm and $1 \cdot 10^{-4}$~mm, which is well below the resolution of state of the art cardiac imaging employed in current clinical practice. For these simulations, we achieved speedups between 13 and 5 over our \fom{}. For highly reduced models, it was shown that the new bottleneck in simulation time is element evaluation. This motivates the inclusion of hyper-
reduction methods, such as the discrete empirical interpolation method (DEIM) \cite{chaturantabut10} or the energy conserving mesh sampling and weighting method (ECSW) \cite{farhat15}, in future research. As the kinematics of a patient-specific heart can already be observed in motion MRI, it might be conceivable to incorporate this displacement information in the reduced space. 

% mor: applications
There exist many potential applications of \mor{} in cardiac many-query settings. One example is the task of obtaining a physiological periodic state, i.e. where left and right ventricular output per cardiac cycle match. In these scenarios, a cardiac simulation with constant parameters is run for multiple cycles, until the change from one cycle to the next is below a given tolerance. In \cite{hirschvogel16} it was reported that in some cases more than ten cycles were necessary until converge to a periodic state. After simulating one \fom{} cycle and calculating the projection matrix, all preceding cycles could be run using a \rom{} since the shape of the cardiac contraction will be similar to the first cycle. In this use case, \mor{} can lead to drastic time savings, especially as the individual cardiac cycles cannot be run in parallel.

% pmor
We further compared four different methods of parametric model order reduction (\pmor{}) to allow \rom{} evaluations at parameter sets without prior \fom{} knowledge. The \pmor{} methods were evaluated by varying cardiac contractility, an important determinant of cardiac performance. The weighted concatenation of snapshots method was found to approximate the displacements of the \fom{} best for this example. Additionally, we showed that the clinically important scalar cardiac quantities ejection fraction, maximum left ventricular pressure, and left atrioventricular plane displacement are also well approximated using \pmor{}. Next to model calibration and design exploration, a possible application of cardiac \pmor{} could be multifidelity uncertainty quantification \cite{biehler15}.

% inv ana
Finally, we introduced a novel method to include \pmor{} into a finite difference gradient-based inverse analysis. Using the Levenberg-Marquardt algorithm as an example, we proposed to use the \fom{} for all objective function evaluations and \pmor{} for all gradient evaluations, based on snapshots from the current and previous iterations. Using synthetic data in a real-world inverse analysis scenario, we demonstrated that \pmor{}-gradient-based optimization shows the same convergence properties as \fom{}-gradient-based optimization while achieving considerable CPU time savings. This method can be incorporated easily into existing optimization frameworks and could even be combined with commercial solvers for the structural problem. Using the inverse analysis approach proposed here has the advantage that we still calculate a full displacement field in each evaluation of the forward model. We can thus evaluate any spatial quantity, which is not possible when using 2D, 1D, or 0D surrogate models for the 3D 
structural model. In future research, this will allow us to compute a spatial approximation error with respect to cine or tagged MRI to estimate patient-specific parameters from clinical observations. To further improve \pmor{}-based gradient evaluation, we will include \pmor{}-gradients in a taylored optimization algorithm that benefits from multiple cheap gradient and second-derivative evaluations to speed up convergence.

% often adjoint problem, but bad for coupled problems (cite???)

%% das hier stimmt so nicht, z.b. wenn ich den gesunden zustand in der bildgebung sehe und den reduziere kann ich schwer vorhersagen fuer den kranken zustand treffen
%Biomechanical problems are especially suitable for projection-based model order reduction (\mor{}). In classical applications of \mor{} a common task is to find a system response subject to 
% classic MOR vs cardiac problems: no different system responses, heartbeat more or less constant
% POD is especially suitable to problems in cardiac mechanics, as the heartbeat can be observed from MRI
% why good for patient specific biomechanics: solution more or less constant (imaging), nearly incompressible -> high linear solver time
% suitable for bio applications: quasi-incompressible material, associated with bad performance of linear solver $\leadsto$ no problem in reduced system

\appendix{}
\section{Subspace interpolation methods}
\label{sec_app_interp}
In this appendix, we give a thorough mathematical description of the subspace interpolation methods utilized in this work. For a comprehensive review on these and other methods of parametric model order reduction, the reader is referred to \cite{benner2015survey}.

\subsection{Weighted concatenation of bases} 
\label{sec_cob}
A common and straightforward approach to obtain a global basis matrix $\proj$ from the precomputed local bases $\proj(\param_1), \ldots, \proj(\param_K)$ is given by the method ``\emph{concatenation of bases}''. With this technique, the local bases are at first simply concatenated side-by-side, followed by a SVD of the resulting matrix to compute the global basis $\proj$. This technique can be extended by introducing the weighting functions $w_k(\param^*)$ in the concatenation of bases, in order to compute a \emph{parameter-dependent} interpolated basis $\proj(\param^*)$ which takes the distance of the new query point~$\param^*$ with respect to the sample points~$\param_1, \ldots, \param_K$ into account. To this end, the matrices~$\proj(\param_1), \ldots, \proj(\param_K)$ are first weighted with weights $w_k(\param^*)$ and concatenated afterwards. Then, the SVD of the concatenated matrix~$\tilde{\proj}(\param^*)$
\begin{align} \label{eq:CoB-matrix}
\tilde{\proj}(\param^*)
= \left[ w_1(\param^*) \, \proj(\param_1), \dots, w_K(\param^*) \, \proj(\param_K) \right]
= \tilde{\matrm{U}}(\param^*) \tilde{\matrm{\Sigma}}(\param^*) \T{\tilde{\matrm{T}}(\param^*)} \in \mathbb{R}^{n \times K \cdot q}
\end{align}
is performed. The interpolated basis~$\proj(\param^*)$ is finally constructed by considering the first~$q$ left singular vectors~$\left\{\tilde{\vecrm{u}}_i(\param^*)\right\}_{i=1}^q$ that best represent the weighted and concatenated matrix~$\tilde{\proj}(\param^*)$:
\begin{align}
\proj(\param^*) = \left[\tilde{\vecrm{u}}_1(\param^*), \ldots, \tilde{\vecrm{u}}_q(\param^*)\right] \in \mathbb{R}^{n \times q}.
\end{align}
Please note that the described weighting procedure is purely optional. The advantage of the weighted approach is that subspaces near the interpolation point $\param^*$ are favored and stronger considered than subspaces describing the dynamics for far-distant sample points. However, this extended technique requires more computational effort than the classical concatenation approach, since a SVD has to be performed for every new $\param^*$ to compute the parameter-dependent interpolated basis $\proj(\param^*)$.

\subsection{Weighted concatenation of snapshots}
\label{sec_cos}
The concatenation of bases approach explained in the previous section provides a basis $\proj(\param^*)$ comprising the most important directions among the (weighted) basis vectors from all local bases. Note, however, that the bases~$\proj(\param_k) = \matrm{U}(\param_k)(:,1:q)$ for $k=1,\ldots,K$ are calculated in our case by means of the SVD-based technique of \ppod{} and they, therefore, essentially approximate the snapshot matrices~$\matrm{D}(\param_k)$
\begin{align}
\matrm{D}(\param_k) = \matrm{U}(\param_k) \matrm{\Sigma}(\param_k) \T{\matrm{T}(\param_k)}.
\end{align}
Since we are actually most interested in finding a basis that optimally approximates the system dynamics over a range of parameters, in our concrete \ppod{}-case it seems very reasonable to construct the interpolated basis~$\proj(\param^*)$ from a (weighted) concatenation of snapshots rather than from a (weighted) concatenation of bases. With the former technique, the matrix~$\proj(\param^*)$ is therefore constructed by considering the first~$q$ left singular vectors of the (weighted and) concatenated snapshot matrix~$\tilde{\matrm{D}}(\param^*)$:
\begin{align}
\tilde{\matrm{D}}(\param^*)
= \left[ w_1(\param^*) \, \matrm{D}(\param_1) , \dots, w_K(\param^*) \, \matrm{D}(\param_K) \right]
= \tilde{\matrm{U}}_{\tilde{\matrm{D}}}(\param^*) \tilde{\matrm{\Sigma}}_{\tilde{\matrm{D}}}(\param^*) \T{\tilde{\matrm{T}}_{\tilde{\matrm{D}}}(\param^*)} \in \mathbb{R}^{n \times K \cdot \ns}.
\label{eq_cos}
\end{align}
\paragraph*{Remark: Connection between the concatenation methods}
It can be shown that the just described (weighted) concatenation of snapshots approach corresponds to a \emph{modified} (weighted) concatenation of bases, where each vector~$\vecrm{v}_i(\param_k)$ is (further) weighted with the corresponding singular value~$\sigma_i(\param_k)$ for all non-zero singular values. Thus, the first~$q$ left singular vectors of the -- towards equation~\eqref{eq:CoB-matrix} -- modified matrix~$\tilde{\proj}(\param^*)$
\begin{align}
\tilde{\proj}(\param^*)
= \left[ w_1(\param^*) \, \proj(\param_1) \, \matrm{\Sigma}_q(\param_1) , \dots, w_K(\param^*) \, \proj(\param_K) \, \matrm{\Sigma}_q(\param_K)
\right]
= \tilde{\matrm{U}}(\param^*) \tilde{\matrm{\Sigma}}(\param^*) \T{\tilde{\matrm{T}}(\param^*)}, %\in \mathbb{R}^{n \times 2q}
\end{align}
where $\matrm{\Sigma}_q(\param_k) \!=\! \matrm{\Sigma}(\param_k)(1:q,1:q) \in \mathbb{R}^{q \times q}$, span the same interpolated subspace~$\mathcal{V}(\param^*)$ than the~$q$ leading vectors in~$\tilde{\matrm{U}}_{\tilde{\matrm{D}}}(\param^*)$. 

\subsection{Adjusted direct basis interpolation}
It is well-known that a straightforward interpolation of the basis vectors comprised in the local projection matrices $\proj(\param_1), \ldots, \proj(\param_K)$ does generally not yield a meaningful basis. This is due to the fact that the basis vectors $\left\{\vecrm{v}_i(\param_k)\right\}_{i=1}^q$ for different sample points span diverse subspaces, thus having a distinct physical interpretation and possibly pointing even in opposite directions in space. Therefore, the basis vectors should be first arranged to point in similar directions, thus spanning similar subspaces, before their entries are interpolated. This adjustment is performed using the Modal Assurance Criterion (MAC)\cite{allemang1982correlation,allemang2003modal}
\begin{align}
\text{MAC}(\vecrm{v}_i, \vecrm{v}_j) = \frac{|\T{\vecrm{v}_i} \cdot \vecrm{v}_j|^2}{\|\vecrm{v}_i\|^2_2 \cdot \|\vecrm{v}_j\|^2_2} \in [0,1],
\end{align}
which provides a measure for the similarity or linear dependence between the vectors $\vecrm{v}_i$ and $\vecrm{v}_j$. The maximal value of the MAC is 1, which corresponds to linear dependent vectors, whereas orthogonal vectors take the minimal value 0. Hence, the idea is to only interpolate vectors which are strongly correlated to each other and maximize the MAC. To do so, we first have to select a reference subspace with respect to which the adjustment of the bases should be performed. The reference subspace, spanned by the columns of $\matrm{R}_{\proj} \in \mathbb{R}^{n \times q}$, should ideally comprise the most important dynamics among all parameter sample points and be representative for all local bases. The simplest way to select $\matrm{R}_{\proj}$ is to take one particularly important local basis $\matrm{R}_{\proj} = \proj_{k_0}$ with $k_0 \in \{1,\ldots,K\}$. Another possibility is to construct the reference subspace similarly as described in section~\ref{sec_cob}, i.e. using the (
weighted) concatenation of bases approach, yielding $\matrm{R}_{\proj} = \tilde{\matrm{U}}(:,1:q)$ or $\matrm{R}_{\proj}(\param^*) = \tilde{\matrm{U}}(\param^*)(:,1:q)$. Once the reference subspace has been selected, the vectors~$\vecrm{v}_{i^*(j, k)}(\param_k)$ that fulfill 
\begin{align}
i^*(j, k) = \argmax_{i} \text{MAC}\big(\vecrm{v}_i(\param_k), \matrm{R}_{\proj}(:,j)\big) \quad \text{for } j=1,\ldots,q \ \text{ and } \ k=1,\ldots,K
\end{align}
are taken to be interpolated. Furthermore, the orientation of the vectors $\vecrm{v}_{i^*(j, k)}(\param_k)$ and $\matrm{R}_{\proj}(:,j)$ is equalized by adapting the sign, in order to avoid that an interpolation between vectors pointing in (almost) opposite directions results in a mutual cancellation. Finally, the interpolation of the vectors is given by
\begin{align}
\bar{\vecrm{v}}_j(\param^*) = \sum_{k=1}^{K} w_k(\param^*) \, \cdot \, \left[ \pm \, \vecrm{v}_{i^*(j, k)}(\param_k) \right] \quad \text{with} \ \ \sum_{k=1}^{K} w_k(\param^*) = 1.
\end{align}
The interpolation of orthonormal vectors does not necessarily yield a set of orthonormal vectors. Therefore, the interpolated vectors~$\left\{\bar{\vecrm{v}}_j(\param^*)\right\}_{j=1}^q$ are subsequently orthonormalized by employing the SVD of~$\bar{\proj}(\param^*)$
\begin{align}
\bar{\proj}(\param^*) = \left[\bar{\vecrm{v}}_1(\param^*), \ldots, \bar{\vecrm{v}}_q(\param^*) \right] = \bar{\matrm{U}}(\param^*) \bar{\matrm{\Sigma}}(\param^*) \T{\bar{\matrm{T}}(\param^*)}
\end{align}
and considering the first $q$ left singular vectors~$\left\{\bar{\vecrm{u}}_j(\param^*)\right\}_{j=1}^{q}$ for the interpolated basis~$\proj(\param^*) \in \mathbb{R}^{n \times q}$.
\paragraph*{Special case of two precomputed bases and one parameter}
In order to make the afore explained method more clear, we now briefly present the special case of two precomputed bases ($K\!=\!2$) and one single parameter ($n_p\!=\!1$). Let us assume that bases $\proj(\mu_1)$ and $\proj(\mu_2)$ have been computed at the parameter sample points $\mu_1$ and $\mu_2$, and that the new parameter value $\mu^*$ lies between these two samples. Suppose that we choose the reference basis e.g. as $\matrm{R}_{\proj} = \proj(\mu_2)$. Then, the vectors~$\vecrm{v}_{i^*(j)}(\mu_1)$ that fulfill 
\begin{align}
i^*(j) = \argmax_{i} \text{MAC}\big(\vecrm{v}_{i}(\mu_1), \vecrm{v}_j(\mu_2)\big) \quad \text{for } j=1,\ldots,q
\end{align}
are selected to be combined with the vectors~$\vecrm{v}_j(\mu_2)$. The interpolation reads
\begin{align}
\bar{\vecrm{v}}_j(\mu^*) = w(\mu^*) \, \cdot \, \left[ \pm \, \vecrm{v}_{i^*(j)}(\mu_1) \right] \ + \  \big(1-w(\mu^*)\big) \, \cdot \, \vecrm{v}_j(\mu_2)
\end{align}
with the weight
\begin{align} \label{eq:lin-weight}
w(\mu^*) = \frac{\mu^* - \mu_2}{\mu_1 - \mu_2} \in [0,1] \quad \text{for } \mu^* \in [\mu_1,\mu_2]\,,
\end{align}
providing that a linear interpolation is employed.

\subsection{Basis interpolation on a Grassmannian manifold}
%the underlying subspaces should be interpolated instead.
As discussed before, a direct interpolation of the local bases is not meaningful, since they span different subspaces. In addition to the afore explained adjustment of the bases before interpolation, one may also interpolate the underlying subspaces on a tangent space of a manifold. The method proposed by Amsallem and Farhat \cite{amsallem2008interpolation} constructs a basis matrix $\proj(\param^*)$ for a new parameter point $\param^*$ by interpolating the subspaces corresponding to the bases $\{\proj(\param_k)\}_{k=1}^K$ on the tangent space to the Grassmannian manifold $\mathcal{G}_q(\mathbb{R}^n)$.

The first step of the approach consists in choosing a local basis matrix $\proj_{k_0}$ for the reference point $\mathcal{V}_{k_0} \in \mathcal{G}_{q}(\mathbb{R}^n)$, at which the tangent space $\mathcal{T}_{\mathcal{V}_{k_0}}$ to the manifold $\mathcal{G}_q(\mathbb{R}^n)$ is constructed. Afterwards, all subspaces $\mathcal{V}(\param_k)$ spanned by the local bases $\proj(\param_k)$ are mapped onto this tangent space by the so-called logarithmic mapping: $\text{span}(\matrm{\Gamma}_k) = \textrm{Log}_{\mathcal{V}_{k_0}}(\mathcal{V}_k) \in \mathcal{T}_{\mathcal{V}_{k_0}}$. This is done basically by computing $K$ thin SVDs
\begin{align}
\big(\matrm{I} - \proj_{k_0} \T{\proj_{k_0}}\big) \, \proj(\param_k) \, \big(\T{\proj_{k_0}} \proj(\param_k)\big)^{-1} = \matrm{U}(\param_k) \, \matrm{\Sigma}(\param_k) \, \T{\matrm{T}(\param_k)} \qquad \text{for } k=1,\ldots,K
\end{align}    
and then calculating
\begin{align}
\matrm{\Gamma}(\param_k) = \matrm{U}(\param_k) \, \arctan\big(\matrm{\Sigma}(\param_k)\big) \, \T{\matrm{T}(\param_k)}.
\end{align}
In order to compute the orthonormal basis $\proj(\param^*)$ for a new parameter point $\param^*$, the matrices $\left\{\matrm{\Gamma}(\param_k)\right\}^K_{k=1}$ are first interpolated using the weights $w_k(\param^*)$ to obtain
\begin{align}
\matrm{\Gamma}^* = \matrm{\Gamma}(\param^*) = \sum_{k=1}^K w_k(\param^*) \, \matrm{\Gamma}(\param_k).
\end{align}
The interpolated subspace $\text{span}(\matrm{\Gamma}^*) \in \mathcal{T}_{\mathcal{V}_{k_0}}$ is then mapped back to the original manifold $\mathcal{G}_{q}(\mathbb{R}^n)$ by the so-called exponential mapping: $\mathcal{V}(\param^*) = \textrm{Exp}_{\mathcal{V}_{k_0}}\big(\text{span}(\matrm{\Gamma}^*)\big) \in \mathcal{G}_{q}(\mathbb{R}^n)$. The back-mapping step is numerically achieved by computing a thin SVD
\begin{align}
\matrm{\Gamma}(\param^*) = \matrm{U}(\param^*) \, \matrm{\Sigma}(\param^*) \, \T{\matrm{T}(\param^*)},
\end{align}
followed by
\begin{align}
\proj(\param^*) = \proj_{k_0} \matrm{T}(\param^*) \cos\big(\matrm{\Sigma}(\param^*)\big) + \matrm{U}(\param^*) \sin\big(\matrm{\Sigma}(\param^*)\big).
\end{align}
The special case of two precomputed bases ($K\!=\!2$) and one single parameter ($n_p\!=\!1$) is extensively described in \cite{amsallem2008interpolation}.

\bibliographystyle{WileyNJD-AMA}
\bibliography{references}
\end{document}

%% file: tikz/coupled_mor.tex
\definecolor{st}{RGB}{254,196,79}
\definecolor{wk}{RGB}{49,163,84}

\begin{tikzpicture}

\shade[left color=st,right color=st, draw=black] (5,1) rectangle (0,6) node[pos=.5] {$\pfrac{\rs}{\ds}$};
\shade[bottom color=wk,top color=st, draw=black] (5,1) rectangle (6,6) node[pos=.5, rotate=90] {$\pfrac{\rs}{\p}$};
\shade[left color=st,right color=wk, draw=black] (0,0) rectangle (5,1) node[pos=.5] {$\pfrac{\rd}{\ds}$};
\shade[left color=wk,right color=wk, draw=black] (5,0) rectangle (6,1) node[pos=.5] {$\pfrac{\rd}{\p}$};

\draw[thick,->] (6.5,1) -- (9.5,1) node[above, pos=.5] {Structural \mor{}};

\shade[left color=st,right color=st, draw=black] (12,1) rectangle (10,3) node[pos=.5] {$\projt \pfrac{\rs}{\ds} \proj$};
\shade[bottom color=wk,top color=st, draw=black] (12,1) rectangle (13,3) node[pos=.5, rotate=90] {$\projt \pfrac{\rs}{\p}$};
\shade[left color=st,right color=wk, draw=black] (10,0) rectangle (12,1) node[pos=.5] {$\pfrac{\rd}{\ds} \proj$};
\shade[left color=wk,right color=wk, draw=black] (12,0) rectangle (13,1) node[pos=.5] {$\pfrac{\rd}{\p}$};

\end{tikzpicture}

%% file: tikz/PODResult.tex
\begin{tikzpicture}

\begin{axis}[%
width=0.951\figurewidth,
height=\figureheight,
at={(0\figurewidth,0\figureheight)},
scale only axis,
xmin=0,
xmax=900,
xlabel={Number $i$ of singular value $\sigma_i$},
xmajorgrids,
ymode=log,
ymin=1e-12,
ymax=1,
yminorticks=true,
ytick={1e-12,1e-10,1e-8,1e-6,1e-4,1e-2,1},
ylabel style={align=center},
ylabel style={align=center},
ylabel={Normalized\\singular value $\sigma_i$/$\sigma_1$},
ymajorgrids,
yminorgrids,
axis background/.style={fill=white},
legend style={legend cell align=left,align=left,draw=white!15!black,font=\footnotesize},
title style={font=\footnotesize},xlabel style={font=\footnotesize},ylabel style={font=\footnotesize},ticklabel style={font=\footnotesize},line join=round
]
\addplot [color=std,solid,thick]
  table[row sep=crcr]{%
1	1\\
2	0.17860387879548\\
3	0.0621680744098263\\
4	0.0223691571499461\\
5	0.0176160729436508\\
6	0.0114094936443827\\
7	0.00966028301361063\\
8	0.00465628077721463\\
9	0.00303769906605705\\
10	0.00212573950822899\\
11	0.00134435473233158\\
12	0.00108558740706047\\
13	0.000881039324269823\\
14	0.000625500132849112\\
15	0.000396769125428014\\
16	0.000202993581194093\\
17	0.000189735999981947\\
18	0.000162783918408444\\
19	9.18432537574996e-05\\
20	7.80958939240895e-05\\
21	5.63093515347215e-05\\
22	4.31041674710365e-05\\
23	3.91605284623985e-05\\
24	3.69534359236266e-05\\
25	2.46322052851177e-05\\
26	2.00599148410698e-05\\
27	1.78696880713065e-05\\
28	1.64693115041654e-05\\
29	1.11678241473187e-05\\
30	9.81524271010001e-06\\
31	8.29332544779236e-06\\
32	6.49188806457926e-06\\
33	6.00529754081434e-06\\
34	4.74109742783709e-06\\
35	3.85924096190519e-06\\
36	3.37036662242233e-06\\
37	2.51658789937391e-06\\
38	2.35854583356151e-06\\
39	2.1672557806018e-06\\
40	1.64357498194024e-06\\
41	1.49198001044942e-06\\
42	1.27168324294855e-06\\
43	1.01240377801755e-06\\
44	8.46495764955732e-07\\
45	7.70271414972295e-07\\
46	6.79329061335892e-07\\
47	6.15019446046977e-07\\
48	5.6959633948059e-07\\
49	5.12272144177018e-07\\
50	4.59198462890682e-07\\
51	4.33802592030304e-07\\
52	4.17002830861522e-07\\
53	3.70785602746941e-07\\
54	3.47020526599488e-07\\
55	3.27526792683828e-07\\
56	2.94753695781717e-07\\
57	2.69342799916124e-07\\
58	2.56725991382908e-07\\
59	2.5223953014208e-07\\
60	2.39261424426885e-07\\
61	2.1416160545849e-07\\
62	2.02816540996522e-07\\
63	1.8495195891062e-07\\
64	1.714924138957e-07\\
65	1.64007367309852e-07\\
66	1.52794794910318e-07\\
67	1.49275116471164e-07\\
68	1.34073453102911e-07\\
69	1.29037552765199e-07\\
70	1.20017658478607e-07\\
71	1.1534975041566e-07\\
72	1.12081798952168e-07\\
73	1.09182512187232e-07\\
74	1.03857879588174e-07\\
75	1.01202428501449e-07\\
76	9.95593598075046e-08\\
77	9.62757758436127e-08\\
78	9.27166645392802e-08\\
79	8.9135426184596e-08\\
80	8.68827366465449e-08\\
81	8.17303952799896e-08\\
82	7.9639366201522e-08\\
83	7.7550133305442e-08\\
84	7.5357852000141e-08\\
85	7.48383730921394e-08\\
86	7.03723136723571e-08\\
87	7.01016580775694e-08\\
88	6.92179659551301e-08\\
89	6.64962239616318e-08\\
90	6.45979570360514e-08\\
91	6.41675529149416e-08\\
92	6.34801425476078e-08\\
93	6.2745617122215e-08\\
94	5.9579602481844e-08\\
95	5.68030057173119e-08\\
96	5.55904668973092e-08\\
97	5.35529723687834e-08\\
98	5.217116686922e-08\\
99	5.07938754542702e-08\\
100	4.88238934009583e-08\\
101	4.74517862327569e-08\\
102	4.55151122864609e-08\\
103	4.45214594582654e-08\\
104	4.27282095766064e-08\\
105	4.19173029422807e-08\\
106	4.14470663082619e-08\\
107	3.86304232291859e-08\\
108	3.836973200188e-08\\
109	3.70611560844725e-08\\
110	3.63782377339387e-08\\
111	3.53438460776314e-08\\
112	3.45350779408623e-08\\
113	3.434289887582e-08\\
114	3.20602266688518e-08\\
115	3.10776597003431e-08\\
116	2.96836181760788e-08\\
117	2.92137558138041e-08\\
118	2.80514428063737e-08\\
119	2.76169885459745e-08\\
120	2.59753387134798e-08\\
121	2.5635863731437e-08\\
122	2.51039969199249e-08\\
123	2.45125201842608e-08\\
124	2.35349758635204e-08\\
125	2.27915013155262e-08\\
126	2.25502042556611e-08\\
127	2.20960910824495e-08\\
128	2.18096030167931e-08\\
129	2.0994714567232e-08\\
130	2.04911847824247e-08\\
131	2.01974268674239e-08\\
132	1.93705849026482e-08\\
133	1.90302168832367e-08\\
134	1.87709096994483e-08\\
135	1.84936479374638e-08\\
136	1.80599382343429e-08\\
137	1.75360241844346e-08\\
138	1.74938223507148e-08\\
139	1.65438310332319e-08\\
140	1.63260531091808e-08\\
141	1.60900600462774e-08\\
142	1.58362204231037e-08\\
143	1.51297655844172e-08\\
144	1.49401643103213e-08\\
145	1.4389846332258e-08\\
146	1.42753286676777e-08\\
147	1.40902276426941e-08\\
148	1.38919348831247e-08\\
149	1.3633254447291e-08\\
150	1.31460388850104e-08\\
151	1.28539555253059e-08\\
152	1.27442314405359e-08\\
153	1.25193803824682e-08\\
154	1.23304122923674e-08\\
155	1.18573656988032e-08\\
156	1.17171455701035e-08\\
157	1.12301147224431e-08\\
158	1.11880755369954e-08\\
159	1.07716871173075e-08\\
160	1.04533833427349e-08\\
161	1.01652928605076e-08\\
162	9.98333312231146e-09\\
163	9.77323301089545e-09\\
164	9.59572477831846e-09\\
165	9.32495825216449e-09\\
166	9.15712911813796e-09\\
167	8.94405141294352e-09\\
168	8.77795929411238e-09\\
169	8.59902510400623e-09\\
170	8.54471057625752e-09\\
171	8.23013712199982e-09\\
172	8.20987074375736e-09\\
173	8.0681017182441e-09\\
174	8.00346944220596e-09\\
175	7.84581739317812e-09\\
176	7.57643717486043e-09\\
177	7.53149615671697e-09\\
178	7.38129485612345e-09\\
179	7.25859564939744e-09\\
180	7.08668828187624e-09\\
181	6.91382737717307e-09\\
182	6.68542961818728e-09\\
183	6.62680315702504e-09\\
184	6.50482645719879e-09\\
185	6.26930183323574e-09\\
186	6.13660986857482e-09\\
187	6.00597253933997e-09\\
188	5.92045353743008e-09\\
189	5.8481060504366e-09\\
190	5.71999745150399e-09\\
191	5.63847661301483e-09\\
192	5.57141090393354e-09\\
193	5.40401620910163e-09\\
194	5.33191719466827e-09\\
195	5.27512454444842e-09\\
196	5.17077606176161e-09\\
197	5.10664883053061e-09\\
198	5.01502284052268e-09\\
199	4.90542079935012e-09\\
200	4.85501937429777e-09\\
201	4.75128132236036e-09\\
202	4.66504069094879e-09\\
203	4.57368009441259e-09\\
204	4.50979830146899e-09\\
205	4.279877672822e-09\\
206	4.22346801226726e-09\\
207	4.21600217393962e-09\\
208	4.14874885046203e-09\\
209	4.05419537983484e-09\\
210	4.02547363619002e-09\\
211	3.90683892388376e-09\\
212	3.83510012336913e-09\\
213	3.76382590781037e-09\\
214	3.66477472200352e-09\\
215	3.52886073527245e-09\\
216	3.47758474964542e-09\\
217	3.41011632755026e-09\\
218	3.27276701529924e-09\\
219	3.24994950003134e-09\\
220	3.19442299151419e-09\\
221	3.1531131597095e-09\\
222	3.06403273272824e-09\\
223	3.03944477980732e-09\\
224	2.94148858580606e-09\\
225	2.90696739158511e-09\\
226	2.84587865299733e-09\\
227	2.78002620202551e-09\\
228	2.74375172221337e-09\\
229	2.67438578434641e-09\\
230	2.60230139252085e-09\\
231	2.57879846796769e-09\\
232	2.55341815472972e-09\\
233	2.47638019114682e-09\\
234	2.43943080949112e-09\\
235	2.4322559809207e-09\\
236	2.33837964844833e-09\\
237	2.28752132514746e-09\\
238	2.23038211838894e-09\\
239	2.17372996578932e-09\\
240	2.1185138234724e-09\\
241	2.09246377485216e-09\\
242	2.07005718878704e-09\\
243	2.03994206751229e-09\\
244	1.9933725867608e-09\\
245	1.96249177788685e-09\\
246	1.92369226710901e-09\\
247	1.8807999223082e-09\\
248	1.85080773585943e-09\\
249	1.82865393647574e-09\\
250	1.78729810453907e-09\\
251	1.75487315384666e-09\\
252	1.73184019203153e-09\\
253	1.72861826101714e-09\\
254	1.63582020183052e-09\\
255	1.57299885106826e-09\\
256	1.56332339084359e-09\\
257	1.55301479759201e-09\\
258	1.53485622359579e-09\\
259	1.51384977580497e-09\\
260	1.50257690909096e-09\\
261	1.4836033094287e-09\\
262	1.46950847035009e-09\\
263	1.46541980798066e-09\\
264	1.46115438324993e-09\\
265	1.4605180569247e-09\\
266	1.45994023826128e-09\\
267	1.45949682981188e-09\\
268	1.45669740048229e-09\\
269	1.45551196629358e-09\\
270	1.45247845195949e-09\\
271	1.44921783152602e-09\\
272	1.44791741754359e-09\\
273	1.44643113568392e-09\\
274	1.44447891647013e-09\\
275	1.43576099987983e-09\\
276	1.43304660038954e-09\\
277	1.43234362713142e-09\\
278	1.43079063754433e-09\\
279	1.42797943152261e-09\\
280	1.42449305974278e-09\\
281	1.41992087787358e-09\\
282	1.41911823314924e-09\\
283	1.41787321360977e-09\\
284	1.41651480609609e-09\\
285	1.41485255259391e-09\\
286	1.4123495411556e-09\\
287	1.41204725733064e-09\\
288	1.41068452848943e-09\\
289	1.4097799100501e-09\\
290	1.40576800154446e-09\\
291	1.40463259715336e-09\\
292	1.40161266036371e-09\\
293	1.40106768961109e-09\\
294	1.39876591508286e-09\\
295	1.39808125100735e-09\\
296	1.39566946625335e-09\\
297	1.3932780393079e-09\\
298	1.39202524383474e-09\\
299	1.39078393062194e-09\\
300	1.38758313711265e-09\\
301	1.38696080176019e-09\\
302	1.38435215748567e-09\\
303	1.38318023207571e-09\\
304	1.3800980067688e-09\\
305	1.3787635237569e-09\\
306	1.37771778706696e-09\\
307	1.37623513556781e-09\\
308	1.37143954290799e-09\\
309	1.37065114770828e-09\\
310	1.36980258906965e-09\\
311	1.36883103853573e-09\\
312	1.36719193989445e-09\\
313	1.36398898785639e-09\\
314	1.36335956053103e-09\\
315	1.36118661049629e-09\\
316	1.35967316302304e-09\\
317	1.35797612742037e-09\\
318	1.35473144125223e-09\\
319	1.35369301477355e-09\\
320	1.35152496016983e-09\\
321	1.34981152720308e-09\\
322	1.34870305533406e-09\\
323	1.34677304881302e-09\\
324	1.34569297482212e-09\\
325	1.34317234019956e-09\\
326	1.34157391403816e-09\\
327	1.33925906255555e-09\\
328	1.33690951188378e-09\\
329	1.3363810092379e-09\\
330	1.33316520859188e-09\\
331	1.33157208616197e-09\\
332	1.33010312168574e-09\\
333	1.32532129959872e-09\\
334	1.32473202974279e-09\\
335	1.32214423025602e-09\\
336	1.3209414701106e-09\\
337	1.31841815710758e-09\\
338	1.31670019669177e-09\\
339	1.31526737311685e-09\\
340	1.31175408221659e-09\\
341	1.31098773439963e-09\\
342	1.30951681126622e-09\\
343	1.30724223085916e-09\\
344	1.30475616839613e-09\\
345	1.30203699400136e-09\\
346	1.30193274002714e-09\\
347	1.30042788674472e-09\\
348	1.29880368650636e-09\\
349	1.29445427376542e-09\\
350	1.29238764412232e-09\\
351	1.28948052915128e-09\\
352	1.28825377628701e-09\\
353	1.28686468630773e-09\\
354	1.28411352581391e-09\\
355	1.28259928445744e-09\\
356	1.28030971322425e-09\\
357	1.27730603287166e-09\\
358	1.27489410338736e-09\\
359	1.27233113556104e-09\\
360	1.26938672534784e-09\\
361	1.26838902038469e-09\\
362	1.26677039988963e-09\\
363	1.26569656689168e-09\\
364	1.26192433812525e-09\\
365	1.26057629098722e-09\\
366	1.25803743143329e-09\\
367	1.2544350010444e-09\\
368	1.25119703941179e-09\\
369	1.25052941926261e-09\\
370	1.25008311814383e-09\\
371	1.24779884902209e-09\\
372	1.24642643833752e-09\\
373	1.2445436635515e-09\\
374	1.24363784211709e-09\\
375	1.23999727395977e-09\\
376	1.23860423771072e-09\\
377	1.23610243448171e-09\\
378	1.23402531250254e-09\\
379	1.2321016805357e-09\\
380	1.22860720305156e-09\\
381	1.2284830998495e-09\\
382	1.22635551310954e-09\\
383	1.22474708013157e-09\\
384	1.22048620004939e-09\\
385	1.21824114416903e-09\\
386	1.21791193512006e-09\\
387	1.21511509838754e-09\\
388	1.21216291118103e-09\\
389	1.20951127091921e-09\\
390	1.20823325371151e-09\\
391	1.20633371663553e-09\\
392	1.20118190913127e-09\\
393	1.19841011648565e-09\\
394	1.19588718485027e-09\\
395	1.19462227648537e-09\\
396	1.19295804044249e-09\\
397	1.19015093767865e-09\\
398	1.18772092167071e-09\\
399	1.18579025793121e-09\\
400	1.1844078547334e-09\\
401	1.18043785670935e-09\\
402	1.17989215337522e-09\\
403	1.17740562917422e-09\\
404	1.17425020842217e-09\\
405	1.1711485291004e-09\\
406	1.16809210425548e-09\\
407	1.16562756931945e-09\\
408	1.16469853321839e-09\\
409	1.16245373485178e-09\\
410	1.16173722586107e-09\\
411	1.15718807768419e-09\\
412	1.15456490035007e-09\\
413	1.15340367187738e-09\\
414	1.15209321484256e-09\\
415	1.1501428011079e-09\\
416	1.14622906810347e-09\\
417	1.14407137922164e-09\\
418	1.14122620145416e-09\\
419	1.13915263645181e-09\\
420	1.13775942525753e-09\\
421	1.13751901888276e-09\\
422	1.13242366289449e-09\\
423	1.13177662286065e-09\\
424	1.12951116124493e-09\\
425	1.12661685124603e-09\\
426	1.12432644062219e-09\\
427	1.12356301994809e-09\\
428	1.12203048986926e-09\\
429	1.11861851521314e-09\\
430	1.11557143668504e-09\\
431	1.11399412255307e-09\\
432	1.11223063128412e-09\\
433	1.10974522876745e-09\\
434	1.10767673519718e-09\\
435	1.10614814026567e-09\\
436	1.10421589985834e-09\\
437	1.10222068177989e-09\\
438	1.10044236831047e-09\\
439	1.09928691176052e-09\\
440	1.09715062724031e-09\\
441	1.09558000766503e-09\\
442	1.09353510838493e-09\\
443	1.09143245951819e-09\\
444	1.09003270976783e-09\\
445	1.08567315829455e-09\\
446	1.08462370972987e-09\\
447	1.08380220994886e-09\\
448	1.07797010421545e-09\\
449	1.07689300281693e-09\\
450	1.07471837047412e-09\\
451	1.07305741059771e-09\\
452	1.07106465852431e-09\\
453	1.06687405007581e-09\\
454	1.06672001311486e-09\\
455	1.06376624997699e-09\\
456	1.06266518130614e-09\\
457	1.06112434394851e-09\\
458	1.05969543454421e-09\\
459	1.05770123815791e-09\\
460	1.05668639152158e-09\\
461	1.05373666532856e-09\\
462	1.05046226887009e-09\\
463	1.05021957712143e-09\\
464	1.04691031406796e-09\\
465	1.04572550312272e-09\\
466	1.04179880413215e-09\\
467	1.04034448343171e-09\\
468	1.0384185151716e-09\\
469	1.03697190680145e-09\\
470	1.03558850454658e-09\\
471	1.03500378306936e-09\\
472	1.03250016069791e-09\\
473	1.03021414306608e-09\\
474	1.02812707497474e-09\\
475	1.02605980517387e-09\\
476	1.02440412224666e-09\\
477	1.02230185870042e-09\\
478	1.02086782640132e-09\\
479	1.01989808124308e-09\\
480	1.01828815641003e-09\\
481	1.01373695304545e-09\\
482	1.01191369240896e-09\\
483	1.01164116399991e-09\\
484	1.00864021950152e-09\\
485	1.0076757671462e-09\\
486	1.00311009977912e-09\\
487	1.00170354078475e-09\\
488	9.99237852449406e-10\\
489	9.98110980460557e-10\\
490	9.96125181022363e-10\\
491	9.92432825371168e-10\\
492	9.8995284320691e-10\\
493	9.87192849640766e-10\\
494	9.85586303129283e-10\\
495	9.82960643708975e-10\\
496	9.80828126913832e-10\\
497	9.79659732794459e-10\\
498	9.73498551503603e-10\\
499	9.72850121785732e-10\\
500	9.7076386425815e-10\\
501	9.66862795111408e-10\\
502	9.65226333817061e-10\\
503	9.61670301944575e-10\\
504	9.59693072865113e-10\\
505	9.58502949657004e-10\\
506	9.56252048271421e-10\\
507	9.54828918828504e-10\\
508	9.54761966078731e-10\\
509	9.50131724561436e-10\\
510	9.48046220187479e-10\\
511	9.44773284200765e-10\\
512	9.42746515712149e-10\\
513	9.40324058377333e-10\\
514	9.3624321814343e-10\\
515	9.32314442372798e-10\\
516	9.30140311333082e-10\\
517	9.29158114838472e-10\\
518	9.26279835476112e-10\\
519	9.2544745589999e-10\\
520	9.22555631838185e-10\\
521	9.215246415569e-10\\
522	9.17535207398794e-10\\
523	9.1579319555761e-10\\
524	9.14942787322972e-10\\
525	9.12606506293539e-10\\
526	9.11900165100201e-10\\
527	9.06994415735996e-10\\
528	9.05336741558486e-10\\
529	9.02263505087848e-10\\
530	9.00877185673765e-10\\
531	8.9802529262205e-10\\
532	8.9719209871192e-10\\
533	8.93040856723858e-10\\
534	8.91953694238681e-10\\
535	8.89810000592575e-10\\
536	8.88198226438401e-10\\
537	8.84786698335538e-10\\
538	8.81612530012344e-10\\
539	8.79564233296772e-10\\
540	8.77355496751715e-10\\
541	8.75740896925929e-10\\
542	8.73905995848209e-10\\
543	8.70594588407477e-10\\
544	8.6723035208426e-10\\
545	8.65406049018539e-10\\
546	8.63960012671612e-10\\
547	8.59855943312793e-10\\
548	8.5885965520908e-10\\
549	8.56749636292174e-10\\
550	8.54483106170015e-10\\
551	8.51188961532792e-10\\
552	8.497593934653e-10\\
553	8.48553106758937e-10\\
554	8.46802577343371e-10\\
555	8.42037372730142e-10\\
556	8.41492319539441e-10\\
557	8.36183883891902e-10\\
558	8.3480994711173e-10\\
559	8.33385633046662e-10\\
560	8.28816783812367e-10\\
561	8.27465108119296e-10\\
562	8.24431723577632e-10\\
563	8.23097350868274e-10\\
564	8.21690320295277e-10\\
565	8.17504094614778e-10\\
566	8.15867270435744e-10\\
567	8.15000111300452e-10\\
568	8.09401559503311e-10\\
569	8.09307808790038e-10\\
570	8.0475757245016e-10\\
571	8.02524940749921e-10\\
572	8.00369674195808e-10\\
573	7.98300946376139e-10\\
574	7.93487323981428e-10\\
575	7.92220070874314e-10\\
576	7.90779256457759e-10\\
577	7.88439284127311e-10\\
578	7.86020310203533e-10\\
579	7.81578981414831e-10\\
580	7.7989913137653e-10\\
581	7.74967735684731e-10\\
582	7.72467487524347e-10\\
583	7.71323277384687e-10\\
584	7.70143087106764e-10\\
585	7.66977935229133e-10\\
586	7.64797258297113e-10\\
587	7.61310119501898e-10\\
588	7.59540669638124e-10\\
589	7.58331169809433e-10\\
590	7.55891932056695e-10\\
591	7.5254520554422e-10\\
592	7.50301916898614e-10\\
593	7.49425265415784e-10\\
594	7.46370066686802e-10\\
595	7.4329009982063e-10\\
596	7.40708295500046e-10\\
597	7.37277274411771e-10\\
598	7.36143964406295e-10\\
599	7.31995506245492e-10\\
600	7.31899867918819e-10\\
601	7.29594902599984e-10\\
602	7.26543561860447e-10\\
603	7.24826391694597e-10\\
604	7.23848869359632e-10\\
605	7.20892575024478e-10\\
606	7.20741300328445e-10\\
607	7.17531016719076e-10\\
608	7.13460393947666e-10\\
609	7.08506767051022e-10\\
610	7.07331784342549e-10\\
611	7.07133608827475e-10\\
612	7.03940594636984e-10\\
613	7.0257068666171e-10\\
614	7.0037760995155e-10\\
615	6.96599380311843e-10\\
616	6.93029325440283e-10\\
617	6.91267104305233e-10\\
618	6.89099366028595e-10\\
619	6.85706469602286e-10\\
620	6.83309048360196e-10\\
621	6.81724126321937e-10\\
622	6.77688111416251e-10\\
623	6.75936679037237e-10\\
624	6.74167189402338e-10\\
625	6.715706301214e-10\\
626	6.6976037296737e-10\\
627	6.6778560858931e-10\\
628	6.66116383642806e-10\\
629	6.64100772911452e-10\\
630	6.62056349643267e-10\\
631	6.58900522570545e-10\\
632	6.5604593853786e-10\\
633	6.53498071306058e-10\\
634	6.52403197737516e-10\\
635	6.51984936685341e-10\\
636	6.50454561354137e-10\\
637	6.47565592892584e-10\\
638	6.45320227434177e-10\\
639	6.44606023530007e-10\\
640	6.44081273875699e-10\\
641	6.40105559339495e-10\\
642	6.37175980075744e-10\\
643	6.34432843674246e-10\\
644	6.32378721599507e-10\\
645	6.30451470794668e-10\\
646	6.29110955027622e-10\\
647	6.25526759647996e-10\\
648	6.24638155844892e-10\\
649	6.18875208657741e-10\\
650	6.15714153716913e-10\\
651	6.14553845215093e-10\\
652	6.12646258847831e-10\\
653	6.11594815705632e-10\\
654	6.06402499812349e-10\\
655	6.05980539364911e-10\\
656	6.04177984056927e-10\\
657	6.03773465209931e-10\\
658	6.00275081964984e-10\\
659	5.95965525076134e-10\\
660	5.95493177567444e-10\\
661	5.9269255487763e-10\\
662	5.9065096349077e-10\\
663	5.88559797160813e-10\\
664	5.85214236238639e-10\\
665	5.84155526689325e-10\\
666	5.8388889314364e-10\\
667	5.8155597603403e-10\\
668	5.78815887048322e-10\\
669	5.75126853501827e-10\\
670	5.72327991825034e-10\\
671	5.69040751617221e-10\\
672	5.68138438277531e-10\\
673	5.65557370178658e-10\\
674	5.59634746489299e-10\\
675	5.55784093958997e-10\\
676	5.41843943577757e-10\\
677	5.38525826368284e-10\\
678	5.33051573836595e-10\\
679	5.29213420632992e-10\\
680	5.19678248903221e-10\\
681	5.11031649926567e-10\\
682	5.0322929795254e-10\\
683	4.92179592950841e-10\\
684	4.86298740155512e-10\\
685	4.79334312396805e-10\\
686	4.70157850675215e-10\\
687	4.6507468689212e-10\\
688	4.62928141314067e-10\\
689	4.46650367938907e-10\\
690	4.35659309298183e-10\\
691	4.32260323850023e-10\\
692	4.26433920102338e-10\\
693	4.11034223121933e-10\\
694	4.08192751389936e-10\\
695	3.96094437715208e-10\\
696	3.92313399785973e-10\\
697	3.8966629473197e-10\\
698	3.77944809708012e-10\\
699	3.67739446350804e-10\\
700	3.5390868596866e-10\\
701	3.41271266433844e-10\\
702	3.39847504059534e-10\\
703	3.31572951414223e-10\\
704	3.17415361006188e-10\\
705	3.1707208990591e-10\\
706	3.07074670517553e-10\\
707	2.98499374533226e-10\\
708	2.91681789767941e-10\\
709	2.78555855797059e-10\\
710	2.75446872631532e-10\\
711	2.6716203902971e-10\\
712	2.61961219965076e-10\\
713	2.52860688515305e-10\\
714	2.45445005115639e-10\\
715	2.25136936762174e-10\\
716	2.17822879357858e-10\\
717	2.15562886997558e-10\\
718	2.05978596582418e-10\\
719	1.94729144420287e-10\\
720	1.89703149104693e-10\\
721	1.70390455841941e-10\\
722	1.65963446959692e-10\\
723	1.56694011363313e-10\\
724	1.41412152236177e-10\\
725	1.35771431988515e-10\\
726	1.20327228768395e-10\\
727	1.18726643696693e-10\\
728	1.08446229560929e-10\\
729	1.06744732140528e-10\\
730	1.03196945234074e-10\\
731	9.89239564976872e-11\\
732	9.36912161880119e-11\\
733	8.80703312513341e-11\\
734	8.71785796931318e-11\\
735	8.34188825440316e-11\\
736	8.05983953203059e-11\\
737	7.8909234744745e-11\\
738	7.87447773408738e-11\\
739	7.42729574251358e-11\\
740	7.21109529575432e-11\\
741	7.06760662686872e-11\\
742	6.89115180325753e-11\\
743	6.72773841924004e-11\\
744	6.64301240965469e-11\\
745	6.50593765925478e-11\\
746	6.44468260154408e-11\\
747	6.26569241180168e-11\\
748	6.16373229084523e-11\\
749	6.04883851958615e-11\\
750	5.89845334523273e-11\\
751	5.84886952916536e-11\\
752	5.69445787511872e-11\\
753	5.64689801679566e-11\\
754	5.55675415491437e-11\\
755	5.48051443855146e-11\\
756	5.43462707184783e-11\\
757	5.26504235417698e-11\\
758	5.25199774776112e-11\\
759	5.22924404162582e-11\\
760	5.16692516454281e-11\\
761	5.08408589561465e-11\\
762	4.98852277472206e-11\\
763	4.91680630821833e-11\\
764	4.88352988039502e-11\\
765	4.77710474124998e-11\\
766	4.73455245002821e-11\\
767	4.71377505777436e-11\\
768	4.56946694176959e-11\\
769	4.53153602469662e-11\\
770	4.42444510903123e-11\\
771	4.37041117617677e-11\\
772	4.34186278308801e-11\\
773	4.29344097260824e-11\\
774	4.23670476439864e-11\\
775	4.20677457774986e-11\\
776	4.15262861746126e-11\\
777	4.10226955604253e-11\\
778	4.07001982020641e-11\\
779	4.04467630361535e-11\\
780	4.02342166067977e-11\\
781	3.99432972129881e-11\\
782	3.97129621403683e-11\\
783	3.96489743865288e-11\\
784	3.94518911060924e-11\\
785	3.9189893106107e-11\\
786	3.89565807365843e-11\\
787	3.86231303129631e-11\\
788	3.82612813569262e-11\\
789	3.81492477787918e-11\\
790	3.80388589068113e-11\\
791	3.79218912351891e-11\\
792	3.77777533141138e-11\\
793	3.7632815840953e-11\\
794	3.75066231256098e-11\\
795	3.74896373632132e-11\\
796	3.7333858253384e-11\\
797	3.69203349986139e-11\\
798	3.25536529306051e-11\\
799	3.16026179509499e-11\\
800	3.14874708728257e-11\\
801	2.9179607572383e-11\\
802	2.4444838097669e-11\\
803	2.35549765100981e-11\\
804	2.21142646436595e-11\\
805	2.07038662319543e-11\\
806	1.79988832550229e-11\\
807	1.56767206650528e-11\\
808	1.24279725742003e-11\\
809	1.21353408686431e-11\\
810	1.18948884321988e-11\\
811	9.49806430773077e-12\\
812	8.9450040242722e-12\\
813	8.11061123792354e-12\\
814	7.71603918615274e-12\\
815	7.44798242763228e-12\\
816	6.76847542652598e-12\\
817	6.70204262063749e-12\\
818	6.38592970985464e-12\\
819	6.35546138220704e-12\\
820	6.31674371287993e-12\\
821	6.24400205607616e-12\\
822	6.20724501132777e-12\\
823	6.15006072656894e-12\\
824	6.13241346514239e-12\\
825	6.07421877200743e-12\\
826	6.04362308264349e-12\\
827	6.01240602646542e-12\\
828	5.972379204987e-12\\
829	5.9355831130431e-12\\
830	5.88535487437611e-12\\
831	5.83468911378566e-12\\
832	5.81611880512134e-12\\
833	5.75619080360628e-12\\
834	5.72366286241658e-12\\
835	5.70458222188103e-12\\
836	5.66774883966566e-12\\
837	5.63070129565454e-12\\
838	5.57595895454905e-12\\
839	5.55165615132228e-12\\
840	5.52780806818253e-12\\
841	5.46705506763417e-12\\
842	5.43066374195641e-12\\
843	5.41454553715043e-12\\
844	5.37557172276875e-12\\
845	5.32446897805154e-12\\
846	5.26925784544021e-12\\
847	5.24113455616933e-12\\
848	5.21408185693741e-12\\
849	5.18318967252033e-12\\
850	5.17422130189283e-12\\
851	5.13800432739934e-12\\
852	5.06800949563672e-12\\
853	5.0065197470901e-12\\
854	4.94148401564744e-12\\
855	4.91901975368169e-12\\
856	4.8495452336154e-12\\
857	4.82460166762381e-12\\
858	4.7531384131693e-12\\
859	4.71068374051637e-12\\
860	4.66257266622013e-12\\
861	4.59732350586202e-12\\
862	4.4850279962629e-12\\
863	4.4283499058859e-12\\
864	4.38448334594605e-12\\
865	4.2885335374603e-12\\
866	4.25629855321311e-12\\
867	4.08360037489019e-12\\
868	4.05304315911681e-12\\
869	3.91074246736573e-12\\
870	3.83830110842261e-12\\
871	3.81965198035987e-12\\
872	3.77423768750698e-12\\
873	3.63351718532285e-12\\
874	3.48498874008525e-12\\
};

\end{axis}
\end{tikzpicture}%

%% file: tikz/err_infty_mor_all.tex
%\definecolor{mycolor1}{rgb}{0.00000,0.44700,0.74100}%
%\definecolor{mycolor2}{rgb}{0.85000,0.32500,0.09800}%
%
\begin{tikzpicture}

\begin{axis}[%
width=0.951\figurewidth,
height=\figureheight,
at={(0\figurewidth,0\figureheight)},
scale only axis,
xmin=0,
xmax=500,
xtick={10,50,100,200,300,400,500},
xlabel style={font=\color{white!15!black}},
xlabel={Reduced order $q$},
ymode=log,
ymin=1e-04,
ymax=1,
ytick={1e-4,1e-3,1e-2,1e-1,1},
yminorticks=true,
ylabel style={font=\color{white!15!black}},
ylabel={Spatial $\epsilon_{\infty,\infty}$-error [mm]},
axis background/.style={fill=white},
xmajorgrids,
ymajorgrids,
%yminorgrids,
reverse legend,
legend style={legend cell align=left, align=left, draw=white!15!black},
title style={font=\footnotesize},xlabel style={font=\footnotesize},ylabel style={font=\footnotesize},ticklabel style={font=\footnotesize},line join=round,legend style={at={(1.02,0.5)},anchor=west,legend cell align=left,align=left,font=\footnotesize},every axis legend/.code={\let\addlegendentry\relax}
]
%\addplot [color=mycolor2]
%  table[row sep=crcr]{%
%10	0.076148832755416\\
%50	0.000424468652315124\\
%100	0.000132963717166797\\
%200	5.68103401294507e-05\\
%300	4.06535386171894e-05\\
%400	4.04801234180263e-05\\
%500	4.02587025879358e-05\\
%};
%\addlegendentry{Mean $L_\infty$-error}

\addplot [color=std, mark=*, mark options={solid, fill=std}]
  table[row sep=crcr]{%
10	0.222997552690448\\
50	0.00441760406001255\\
100	0.000855848845799557\\
200	0.000168709748084224\\
300	0.000130498675336395\\
400	0.000129433292203173\\
500	0.000128524504046671\\
};
%\addlegendentry{$L_{\infty{},\infty}$-error}

\end{axis}
\end{tikzpicture}%

%% file: tikz/legend_q.tex
\definecolor{mycolor1}{rgb}{0.00000,0.44700,0.74100}%
\definecolor{mycolor2}{rgb}{0.85000,0.32500,0.09800}%
\definecolor{mycolor3}{rgb}{0.92900,0.69400,0.12500}%
\definecolor{mycolor4}{rgb}{0.49400,0.18400,0.55600}%
\definecolor{mycolor5}{rgb}{0.46600,0.67400,0.18800}%
\definecolor{mycolor6}{rgb}{0.30100,0.74500,0.93300}%
\definecolor{mycolor7}{rgb}{0.63500,0.07800,0.18400}%
\begin{tikzpicture}

\begin{axis}[%
width=\figurewidth,
height=0.876\figureheight,
at={(0\figurewidth,0\figureheight)},
hide axis,
scale only axis,
xmin=10,
xmax=100,
ymin=0,
ymax=0.4,
legend columns=-1,
legend style={draw=none,column sep=1ex, font=\footnotesize}
]
\addlegendimage{color=fom,thick}
\addlegendentry{\fom{}}
\addlegendimage{color=rom10,thick}
\addlegendentry{\rom{10}}
\addlegendimage{color=rom50,thick}
\addlegendentry{\rom{50}}
\addlegendimage{color=rom100,thick}
\addlegendentry{\rom{100}}
\addlegendimage{color=rom200,thick}
\addlegendentry{\rom{200}}
\addlegendimage{color=rom300,thick}
\addlegendentry{\rom{300}}
\addlegendimage{color=rom400,thick}
\addlegendentry{\rom{400}}
\addlegendimage{color=rom500,thick}
\addlegendentry{\rom{500}}

\end{axis}
\end{tikzpicture}%

%% file: tikz/newton.tex
\begin{tikzpicture}

\begin{axis}[%
width=0.951\figurewidth,
height=\figureheight,
at={(0\figurewidth,0\figureheight)},
scale only axis,
xmin=0,
xmax=0.9,
xlabel style={font=\color{white!15!black}},
xlabel={Time [s]},
ymin=2,
ymax=10,
xmajorgrids,
ymajorgrids,
ylabel style={font=\color{white!15!black}},
ylabel={Newton iterations [-]},
ylabel style={at={(axis description cs:0.15,.5)},anchor=south},
axis background/.style={fill=white},
legend style={legend cell align=left, align=left, draw=white!15!black},
title style={font=\footnotesize},xlabel style={font=\footnotesize},ylabel style={font=\footnotesize},ticklabel style={font=\footnotesize},line join=round,every axis legend/.code={\let\addlegendentry\relax}
]
\addplot [color=std]
  table[row sep=crcr]{%
0.001	4\\
0.002	4\\
0.003	4\\
0.004	4\\
0.005	4\\
0.006	4\\
0.007	4\\
0.008	4\\
0.009	4\\
0.01	4\\
0.011	4\\
0.012	4\\
0.013	4\\
0.014	4\\
0.015	4\\
0.016	4\\
0.017	4\\
0.018	4\\
0.019	4\\
0.02	4\\
0.021	4\\
0.022	4\\
0.023	4\\
0.024	4\\
0.025	4\\
0.026	4\\
0.027	4\\
0.028	4\\
0.029	4\\
0.03	4\\
0.031	4\\
0.032	4\\
0.033	4\\
0.034	4\\
0.035	4\\
0.036	4\\
0.037	4\\
0.038	4\\
0.039	4\\
0.04	4\\
0.041	4\\
0.042	4\\
0.043	4\\
0.044	4\\
0.045	4\\
0.046	4\\
0.047	4\\
0.048	4\\
0.049	4\\
0.05	4\\
0.051	4\\
0.052	4\\
0.053	4\\
0.054	4\\
0.055	4\\
0.056	4\\
0.057	4\\
0.058	4\\
0.059	4\\
0.06	4\\
0.061	4\\
0.062	4\\
0.063	4\\
0.064	4\\
0.065	4\\
0.066	4\\
0.067	4\\
0.068	4\\
0.069	4\\
0.07	4\\
0.071	4\\
0.072	4\\
0.073	4\\
0.074	4\\
0.075	4\\
0.076	4\\
0.077	4\\
0.078	4\\
0.079	4\\
0.08	4\\
0.081	4\\
0.082	4\\
0.083	4\\
0.084	4\\
0.085	4\\
0.086	4\\
0.087	4\\
0.088	4\\
0.089	4\\
0.09	4\\
0.091	4\\
0.092	4\\
0.093	4\\
0.094	4\\
0.095	4\\
0.096	4\\
0.097	4\\
0.098	4\\
0.099	4\\
0.1	4\\
0.101	4\\
0.102	4\\
0.103	4\\
0.104	4\\
0.105	4\\
0.106	4\\
0.107	4\\
0.108	4\\
0.109	4\\
0.11	4\\
0.111	4\\
0.112	4\\
0.113	4\\
0.114	4\\
0.115	4\\
0.116	4\\
0.117	4\\
0.118	4\\
0.119	4\\
0.12	4\\
0.121	4\\
0.122	4\\
0.123	4\\
0.124	4\\
0.125	4\\
0.126	4\\
0.127	4\\
0.128	4\\
0.129	4\\
0.13	4\\
0.131	4\\
0.132	4\\
0.133	4\\
0.134	4\\
0.135	4\\
0.136	4\\
0.137	4\\
0.138	4\\
0.139	4\\
0.14	4\\
0.141	4\\
0.142	4\\
0.143	4\\
0.144	4\\
0.145	4\\
0.146	4\\
0.147	4\\
0.148	4\\
0.149	4\\
0.15	4\\
0.151	4\\
0.152	4\\
0.153	3\\
0.154	3\\
0.155	3\\
0.156	3\\
0.157	3\\
0.158	3\\
0.159	3\\
0.16	3\\
0.161	3\\
0.162	3\\
0.163	3\\
0.164	3\\
0.165	4\\
0.166	4\\
0.167	4\\
0.168	4\\
0.169	4\\
0.17	4\\
0.171	4\\
0.172	4\\
0.173	4\\
0.174	4\\
0.175	4\\
0.176	4\\
0.177	4\\
0.178	4\\
0.179	4\\
0.18	4\\
0.181	4\\
0.182	4\\
0.183	4\\
0.184	4\\
0.185	4\\
0.186	4\\
0.187	4\\
0.188	4\\
0.189	4\\
0.19	4\\
0.191	4\\
0.192	4\\
0.193	4\\
0.194	4\\
0.195	4\\
0.196	4\\
0.197	4\\
0.198	4\\
0.199	4\\
0.2	4\\
0.201	4\\
0.202	4\\
0.203	4\\
0.204	4\\
0.205	4\\
0.206	4\\
0.207	4\\
0.208	4\\
0.209	4\\
0.21	4\\
0.211	4\\
0.212	4\\
0.213	4\\
0.214	4\\
0.215	4\\
0.216	4\\
0.217	4\\
0.218	4\\
0.219	4\\
0.22	4\\
0.221	4\\
0.222	4\\
0.223	4\\
0.224	4\\
0.225	4\\
0.226	4\\
0.227	4\\
0.228	4\\
0.229	4\\
0.23	4\\
0.231	4\\
0.232	4\\
0.233	4\\
0.234	4\\
0.235	4\\
0.236	4\\
0.237	4\\
0.238	4\\
0.239	4\\
0.24	4\\
0.241	4\\
0.242	4\\
0.243	4\\
0.244	4\\
0.245	4\\
0.246	4\\
0.247	4\\
0.248	4\\
0.249	4\\
0.25	4\\
0.251	5\\
0.252	5\\
0.253	5\\
0.254	5\\
0.255	5\\
0.256	5\\
0.257	5\\
0.258	5\\
0.259	5\\
0.26	5\\
0.261	5\\
0.262	5\\
0.263	5\\
0.264	5\\
0.265	5\\
0.266	5\\
0.267	5\\
0.268	5\\
0.269	5\\
0.27	5\\
0.271	5\\
0.272	5\\
0.273	5\\
0.274	5\\
0.275	5\\
0.276	5\\
0.277	5\\
0.278	6\\
0.279	5\\
0.28	5\\
0.281	5\\
0.282	5\\
0.283	5\\
0.284	5\\
0.285	5\\
0.286	5\\
0.287	5\\
0.288	4\\
0.289	4\\
0.29	5\\
0.291	5\\
0.292	5\\
0.293	5\\
0.294	5\\
0.295	5\\
0.296	5\\
0.297	5\\
0.298	5\\
0.299	5\\
0.3	5\\
0.301	5\\
0.302	5\\
0.303	5\\
0.304	5\\
0.305	5\\
0.306	5\\
0.307	5\\
0.308	5\\
0.309	5\\
0.31	5\\
0.311	5\\
0.312	5\\
0.313	5\\
0.314	5\\
0.315	5\\
0.316	5\\
0.317	5\\
0.318	5\\
0.319	5\\
0.32	5\\
0.321	5\\
0.322	5\\
0.323	5\\
0.324	5\\
0.325	5\\
0.326	5\\
0.327	5\\
0.328	5\\
0.329	5\\
0.33	5\\
0.331	5\\
0.332	5\\
0.333	5\\
0.334	5\\
0.335	5\\
0.336	5\\
0.337	5\\
0.338	5\\
0.339	5\\
0.34	5\\
0.341	5\\
0.342	5\\
0.343	5\\
0.344	5\\
0.345	5\\
0.346	5\\
0.347	5\\
0.348	5\\
0.349	5\\
0.35	5\\
0.351	5\\
0.352	5\\
0.353	5\\
0.354	5\\
0.355	5\\
0.356	5\\
0.357	5\\
0.358	5\\
0.359	5\\
0.36	5\\
0.361	5\\
0.362	5\\
0.363	5\\
0.364	5\\
0.365	5\\
0.366	5\\
0.367	5\\
0.368	5\\
0.369	5\\
0.37	5\\
0.371	5\\
0.372	5\\
0.373	5\\
0.374	5\\
0.375	5\\
0.376	5\\
0.377	5\\
0.378	5\\
0.379	5\\
0.38	5\\
0.381	5\\
0.382	5\\
0.383	5\\
0.384	5\\
0.385	5\\
0.386	5\\
0.387	5\\
0.388	5\\
0.389	5\\
0.39	5\\
0.391	5\\
0.392	5\\
0.393	5\\
0.394	5\\
0.395	5\\
0.396	5\\
0.397	5\\
0.398	5\\
0.399	5\\
0.4	5\\
0.401	5\\
0.402	5\\
0.403	5\\
0.404	5\\
0.405	5\\
0.406	5\\
0.407	5\\
0.408	5\\
0.409	5\\
0.41	5\\
0.411	5\\
0.412	5\\
0.413	5\\
0.414	5\\
0.415	5\\
0.416	5\\
0.417	5\\
0.418	5\\
0.419	5\\
0.42	5\\
0.421	5\\
0.422	5\\
0.423	5\\
0.424	5\\
0.425	5\\
0.426	5\\
0.427	5\\
0.428	5\\
0.429	5\\
0.43	5\\
0.431	5\\
0.432	5\\
0.433	5\\
0.434	5\\
0.435	5\\
0.436	5\\
0.437	5\\
0.438	5\\
0.439	5\\
0.44	5\\
0.441	5\\
0.442	5\\
0.443	5\\
0.444	5\\
0.445	5\\
0.446	5\\
0.447	5\\
0.448	5\\
0.449	5\\
0.45	5\\
0.451	5\\
0.452	5\\
0.453	5\\
0.454	5\\
0.455	5\\
0.456	5\\
0.457	5\\
0.458	5\\
0.459	5\\
0.46	5\\
0.461	5\\
0.462	5\\
0.463	5\\
0.464	5\\
0.465	5\\
0.466	5\\
0.467	5\\
0.468	5\\
0.469	5\\
0.47	5\\
0.471	5\\
0.472	5\\
0.473	5\\
0.474	5\\
0.475	5\\
0.476	5\\
0.477	5\\
0.478	5\\
0.479	5\\
0.48	5\\
0.481	5\\
0.482	5\\
0.483	5\\
0.484	5\\
0.485	5\\
0.486	5\\
0.487	5\\
0.488	5\\
0.489	5\\
0.49	5\\
0.491	5\\
0.492	5\\
0.493	5\\
0.494	5\\
0.495	5\\
0.496	5\\
0.497	5\\
0.498	5\\
0.499	5\\
0.5	5\\
0.501	5\\
0.502	5\\
0.503	5\\
0.504	5\\
0.505	5\\
0.506	9\\
0.507	9\\
0.508	9\\
0.509	9\\
0.51	9\\
0.511	9\\
0.512	9\\
0.513	9\\
0.514	8\\
0.515	9\\
0.516	8\\
0.517	9\\
0.518	8\\
0.519	9\\
0.52	8\\
0.521	9\\
0.522	8\\
0.523	9\\
0.524	7\\
0.525	9\\
0.526	7\\
0.527	9\\
0.528	7\\
0.529	9\\
0.53	7\\
0.531	9\\
0.532	7\\
0.533	9\\
0.534	7\\
0.535	9\\
0.536	7\\
0.537	9\\
0.538	7\\
0.539	9\\
0.54	7\\
0.541	4\\
0.542	4\\
0.543	4\\
0.544	4\\
0.545	4\\
0.546	4\\
0.547	4\\
0.548	4\\
0.549	4\\
0.55	4\\
0.551	4\\
0.552	4\\
0.553	4\\
0.554	4\\
0.555	4\\
0.556	4\\
0.557	4\\
0.558	4\\
0.559	4\\
0.56	4\\
0.561	4\\
0.562	4\\
0.563	4\\
0.564	4\\
0.565	4\\
0.566	4\\
0.567	4\\
0.568	4\\
0.569	4\\
0.57	3\\
0.571	3\\
0.572	3\\
0.573	4\\
0.574	4\\
0.575	4\\
0.576	4\\
0.577	4\\
0.578	4\\
0.579	4\\
0.58	4\\
0.581	4\\
0.582	4\\
0.583	4\\
0.584	4\\
0.585	4\\
0.586	4\\
0.587	4\\
0.588	4\\
0.589	4\\
0.59	4\\
0.591	4\\
0.592	4\\
0.593	4\\
0.594	4\\
0.595	4\\
0.596	4\\
0.597	4\\
0.598	4\\
0.599	4\\
0.6	4\\
0.601	4\\
0.602	4\\
0.603	4\\
0.604	4\\
0.605	4\\
0.606	4\\
0.607	4\\
0.608	4\\
0.609	4\\
0.61	4\\
0.611	4\\
0.612	4\\
0.613	4\\
0.614	4\\
0.615	4\\
0.616	4\\
0.617	4\\
0.618	4\\
0.619	4\\
0.62	4\\
0.621	4\\
0.622	4\\
0.623	4\\
0.624	4\\
0.625	4\\
0.626	4\\
0.627	4\\
0.628	4\\
0.629	4\\
0.63	4\\
0.631	3\\
0.632	3\\
0.633	3\\
0.634	3\\
0.635	3\\
0.636	3\\
0.637	3\\
0.638	3\\
0.639	3\\
0.64	3\\
0.641	3\\
0.642	3\\
0.643	3\\
0.644	4\\
0.645	4\\
0.646	4\\
0.647	4\\
0.648	4\\
0.649	4\\
0.65	4\\
0.651	4\\
0.652	4\\
0.653	4\\
0.654	4\\
0.655	4\\
0.656	4\\
0.657	4\\
0.658	4\\
0.659	4\\
0.66	4\\
0.661	4\\
0.662	4\\
0.663	4\\
0.664	4\\
0.665	4\\
0.666	4\\
0.667	4\\
0.668	4\\
0.669	4\\
0.67	4\\
0.671	4\\
0.672	4\\
0.673	4\\
0.674	4\\
0.675	4\\
0.676	4\\
0.677	4\\
0.678	4\\
0.679	4\\
0.68	4\\
0.681	4\\
0.682	4\\
0.683	4\\
0.684	4\\
0.685	4\\
0.686	4\\
0.687	4\\
0.688	4\\
0.689	4\\
0.69	4\\
0.691	4\\
0.692	4\\
0.693	4\\
0.694	4\\
0.695	4\\
0.696	4\\
0.697	4\\
0.698	4\\
0.699	4\\
0.7	4\\
0.701	4\\
0.702	4\\
0.703	4\\
0.704	4\\
0.705	4\\
0.706	4\\
0.707	4\\
0.708	4\\
0.709	4\\
0.71	4\\
0.711	4\\
0.712	4\\
0.713	4\\
0.714	4\\
0.715	4\\
0.716	4\\
0.717	4\\
0.718	4\\
0.719	4\\
0.72	4\\
0.721	4\\
0.722	4\\
0.723	4\\
0.724	4\\
0.725	4\\
0.726	4\\
0.727	4\\
0.728	4\\
0.729	4\\
0.73	4\\
0.731	4\\
0.732	4\\
0.733	4\\
0.734	4\\
0.735	4\\
0.736	4\\
0.737	4\\
0.738	4\\
0.739	4\\
0.74	4\\
0.741	4\\
0.742	4\\
0.743	4\\
0.744	4\\
0.745	4\\
0.746	4\\
0.747	4\\
0.748	4\\
0.749	4\\
0.75	4\\
0.751	4\\
0.752	4\\
0.753	4\\
0.754	4\\
0.755	4\\
0.756	4\\
0.757	4\\
0.758	4\\
0.759	4\\
0.76	4\\
0.761	4\\
0.762	4\\
0.763	4\\
0.764	4\\
0.765	4\\
0.766	4\\
0.767	4\\
0.768	4\\
0.769	4\\
0.77	4\\
0.771	4\\
0.772	4\\
0.773	4\\
0.774	4\\
0.775	4\\
0.776	4\\
0.777	4\\
0.778	4\\
0.779	4\\
0.78	4\\
0.781	4\\
0.782	4\\
0.783	4\\
0.784	4\\
0.785	4\\
0.786	4\\
0.787	4\\
0.788	4\\
0.789	4\\
0.79	4\\
0.791	4\\
0.792	4\\
0.793	4\\
0.794	4\\
0.795	4\\
0.796	4\\
0.797	4\\
0.798	4\\
0.799	4\\
0.8	4\\
0.801	4\\
0.802	4\\
0.803	4\\
0.804	4\\
0.805	4\\
0.806	4\\
0.807	4\\
0.808	4\\
0.809	4\\
0.81	4\\
0.811	4\\
0.812	4\\
0.813	4\\
0.814	4\\
0.815	4\\
0.816	4\\
0.817	4\\
0.818	4\\
0.819	4\\
0.82	4\\
0.821	4\\
0.822	4\\
0.823	4\\
0.824	4\\
0.825	4\\
0.826	4\\
0.827	4\\
0.828	4\\
0.829	4\\
0.83	4\\
0.831	4\\
0.832	4\\
0.833	4\\
0.834	4\\
0.835	4\\
0.836	4\\
0.837	4\\
0.838	4\\
0.839	4\\
0.84	4\\
0.841	4\\
0.842	4\\
0.843	4\\
0.844	4\\
0.845	4\\
0.846	4\\
0.847	4\\
0.848	4\\
0.849	4\\
0.85	4\\
0.851	4\\
0.852	4\\
0.853	4\\
0.854	4\\
0.855	4\\
0.856	4\\
0.857	4\\
0.858	4\\
0.859	4\\
0.86	4\\
0.861	4\\
0.862	4\\
0.863	4\\
0.864	4\\
0.865	4\\
0.866	4\\
0.867	4\\
0.868	4\\
0.869	4\\
0.87	4\\
0.871	4\\
0.872	4\\
0.873	4\\
0.874	4\\
};
\addlegendentry{data1}

\end{axis}
\end{tikzpicture}%

%% file: tikz/linear_solver.tex
\definecolor{mycolor1}{rgb}{0.00000,0.44700,0.74100}%
\begin{tikzpicture}

\begin{axis}[%
width=0.951\figurewidth,
height=\figureheight,
at={(0\figurewidth,0\figureheight)},
scale only axis,
xmin=0,
xmax=0.9,
xlabel style={font=\color{white!15!black}},
xlabel={Times [s]},
ymin=15,
ymax=65,
xmajorgrids,
ymajorgrids,
ylabel style={font=\color{white!15!black}},
ylabel style={align=center},
ylabel style={at={(axis description cs:0.1,.5)},anchor=south},
ylabel={Linear iterations per\\Newton iteration [-]},
axis background/.style={fill=white},
legend style={legend cell align=left, align=left, draw=white!15!black},
title style={font=\footnotesize},xlabel style={font=\footnotesize},ylabel style={font=\footnotesize},ticklabel style={font=\footnotesize},line join=round,every axis legend/.code={\let\addlegendentry\relax}
]
\addplot [color=std]
  table[row sep=crcr]{%
0.001	19.5\\
0.002	21\\
0.003	20\\
0.004	20.25\\
0.005	20.75\\
0.006	20.75\\
0.007	20.75\\
0.008	19.25\\
0.009	20\\
0.01	20.25\\
0.011	20.5\\
0.012	20\\
0.013	19.75\\
0.014	20.25\\
0.015	19.5\\
0.016	20\\
0.017	19.75\\
0.018	20\\
0.019	19.75\\
0.02	19.75\\
0.021	19.75\\
0.022	19.5\\
0.023	19.75\\
0.024	19.5\\
0.025	19.75\\
0.026	19.75\\
0.027	19.75\\
0.028	19.5\\
0.029	19.75\\
0.03	19.75\\
0.031	19.5\\
0.032	19.5\\
0.033	19.5\\
0.034	20\\
0.035	19.5\\
0.036	20.25\\
0.037	20\\
0.038	20\\
0.039	19.5\\
0.04	19.75\\
0.041	19.25\\
0.042	19.75\\
0.043	19.75\\
0.044	20.25\\
0.045	20\\
0.046	20\\
0.047	19.75\\
0.048	19.5\\
0.049	20\\
0.05	19.75\\
0.051	20\\
0.052	19.75\\
0.053	19.5\\
0.054	20\\
0.055	19.5\\
0.056	20.25\\
0.057	19.5\\
0.058	19.75\\
0.059	20\\
0.06	19.75\\
0.061	20.25\\
0.062	20\\
0.063	20.25\\
0.064	20.5\\
0.065	20.5\\
0.066	20.25\\
0.067	20.5\\
0.068	20.5\\
0.069	20.75\\
0.07	20.75\\
0.071	21\\
0.072	21\\
0.073	21.5\\
0.074	21.25\\
0.075	21.5\\
0.076	21.5\\
0.077	22\\
0.078	21.75\\
0.079	22\\
0.08	22.5\\
0.081	22.5\\
0.082	23\\
0.083	23.25\\
0.084	23.5\\
0.085	24.25\\
0.086	26\\
0.087	24\\
0.088	23.5\\
0.089	23.25\\
0.09	23.25\\
0.091	23\\
0.092	23\\
0.093	22.75\\
0.094	22.75\\
0.095	22.5\\
0.096	22.5\\
0.097	22.25\\
0.098	22.25\\
0.099	22.25\\
0.1	22.25\\
0.101	22\\
0.102	22.25\\
0.103	22\\
0.104	22.25\\
0.105	22\\
0.106	21.75\\
0.107	22\\
0.108	22\\
0.109	22\\
0.11	22.25\\
0.111	22\\
0.112	22\\
0.113	21.75\\
0.114	22.5\\
0.115	22\\
0.116	22\\
0.117	22.25\\
0.118	22.25\\
0.119	22.5\\
0.12	22.75\\
0.121	23\\
0.122	23.25\\
0.123	23.5\\
0.124	23.5\\
0.125	23.75\\
0.126	24.25\\
0.127	24.5\\
0.128	23.75\\
0.129	24.25\\
0.13	24.75\\
0.131	24.75\\
0.132	24.5\\
0.133	25\\
0.134	25.25\\
0.135	25.25\\
0.136	25.5\\
0.137	25.75\\
0.138	25.75\\
0.139	26\\
0.14	26\\
0.141	26.5\\
0.142	26.25\\
0.143	27\\
0.144	26.75\\
0.145	27\\
0.146	26.75\\
0.147	27\\
0.148	27\\
0.149	27.25\\
0.15	27.25\\
0.151	27.25\\
0.152	27.5\\
0.153	26\\
0.154	26\\
0.155	26.3333333333333\\
0.156	26.3333333333333\\
0.157	26.3333333333333\\
0.158	26.3333333333333\\
0.159	26.6666666666667\\
0.16	26\\
0.161	26\\
0.162	25.3333333333333\\
0.163	25\\
0.164	24.3333333333333\\
0.165	27\\
0.166	27.5\\
0.167	27.75\\
0.168	28\\
0.169	28\\
0.17	20.5\\
0.171	20.5\\
0.172	21\\
0.173	21.5\\
0.174	21.5\\
0.175	22\\
0.176	24\\
0.177	28.25\\
0.178	28.5\\
0.179	28.25\\
0.18	28.25\\
0.181	28.25\\
0.182	28\\
0.183	28\\
0.184	28\\
0.185	27.25\\
0.186	27\\
0.187	26.75\\
0.188	27.25\\
0.189	27.75\\
0.19	27.5\\
0.191	28\\
0.192	27.25\\
0.193	27.75\\
0.194	27.5\\
0.195	28\\
0.196	27.5\\
0.197	28\\
0.198	27.5\\
0.199	28.25\\
0.2	27.25\\
0.201	28\\
0.202	27\\
0.203	28\\
0.204	27\\
0.205	28\\
0.206	27.25\\
0.207	28\\
0.208	27.25\\
0.209	28\\
0.21	27.25\\
0.211	28\\
0.212	27.25\\
0.213	27.75\\
0.214	27.5\\
0.215	27.5\\
0.216	27.25\\
0.217	27.25\\
0.218	27.25\\
0.219	27.25\\
0.22	26.75\\
0.221	27\\
0.222	26.75\\
0.223	26.5\\
0.224	26.5\\
0.225	26.25\\
0.226	25.5\\
0.227	26.5\\
0.228	25.25\\
0.229	26.25\\
0.23	25.5\\
0.231	26.25\\
0.232	25.75\\
0.233	26.25\\
0.234	26.25\\
0.235	26.5\\
0.236	26.5\\
0.237	26.75\\
0.238	26.5\\
0.239	26.25\\
0.24	26.5\\
0.241	26.75\\
0.242	26.5\\
0.243	26.5\\
0.244	26.5\\
0.245	26.75\\
0.246	26\\
0.247	26.75\\
0.248	26\\
0.249	26.75\\
0.25	27\\
0.251	28.4\\
0.252	28.4\\
0.253	28.4\\
0.254	28.6\\
0.255	28.8\\
0.256	28.6\\
0.257	28.8\\
0.258	28.6\\
0.259	28.6\\
0.26	28.2\\
0.261	28.6\\
0.262	28.4\\
0.263	28.6\\
0.264	28.6\\
0.265	28.6\\
0.266	28.6\\
0.267	28.6\\
0.268	28.8\\
0.269	29\\
0.27	29\\
0.271	29\\
0.272	29.2\\
0.273	29.4\\
0.274	29.2\\
0.275	29.4\\
0.276	29.2\\
0.277	29\\
0.278	18.3333333333333\\
0.279	20.2\\
0.28	20.2\\
0.281	20.4\\
0.282	20.6\\
0.283	20.2\\
0.284	20.2\\
0.285	20\\
0.286	20\\
0.287	20.4\\
0.288	20.5\\
0.289	20.75\\
0.29	20\\
0.291	19.8\\
0.292	19.6\\
0.293	19.4\\
0.294	19.6\\
0.295	19.6\\
0.296	19.8\\
0.297	19.8\\
0.298	19.8\\
0.299	19.8\\
0.3	19.8\\
0.301	19.8\\
0.302	20\\
0.303	20\\
0.304	20\\
0.305	20\\
0.306	20.2\\
0.307	20.4\\
0.308	20.6\\
0.309	20.4\\
0.31	20.6\\
0.311	20.6\\
0.312	20.8\\
0.313	21\\
0.314	20.8\\
0.315	20.6\\
0.316	21.4\\
0.317	21\\
0.318	21.2\\
0.319	21.4\\
0.32	21.2\\
0.321	21.6\\
0.322	21.6\\
0.323	21.8\\
0.324	21.8\\
0.325	22.2\\
0.326	22\\
0.327	22\\
0.328	22.4\\
0.329	22.4\\
0.33	22.4\\
0.331	22.8\\
0.332	22.8\\
0.333	22.8\\
0.334	23\\
0.335	23\\
0.336	23\\
0.337	23.2\\
0.338	23.2\\
0.339	23.4\\
0.34	23.2\\
0.341	23.6\\
0.342	23.8\\
0.343	24\\
0.344	24.2\\
0.345	24.2\\
0.346	24.4\\
0.347	24.2\\
0.348	24.2\\
0.349	24.2\\
0.35	24.2\\
0.351	24\\
0.352	23.8\\
0.353	24.4\\
0.354	24.4\\
0.355	24.4\\
0.356	25\\
0.357	24.8\\
0.358	25.2\\
0.359	25.2\\
0.36	25.2\\
0.361	25.6\\
0.362	25.8\\
0.363	25.8\\
0.364	25.8\\
0.365	26\\
0.366	26\\
0.367	26\\
0.368	26.2\\
0.369	26.6\\
0.37	26.8\\
0.371	27\\
0.372	27\\
0.373	27.2\\
0.374	27.2\\
0.375	27.4\\
0.376	27.6\\
0.377	27.8\\
0.378	27.8\\
0.379	28.2\\
0.38	28.4\\
0.381	28.4\\
0.382	28.4\\
0.383	28.6\\
0.384	28.6\\
0.385	29\\
0.386	28.8\\
0.387	29\\
0.388	29.2\\
0.389	29.4\\
0.39	29.2\\
0.391	29.8\\
0.392	29.8\\
0.393	29.8\\
0.394	30\\
0.395	30\\
0.396	30\\
0.397	30.2\\
0.398	30\\
0.399	30.6\\
0.4	30.6\\
0.401	30.4\\
0.402	30.6\\
0.403	30.8\\
0.404	31\\
0.405	30.8\\
0.406	31.2\\
0.407	31.2\\
0.408	31.2\\
0.409	31.4\\
0.41	31.4\\
0.411	31.8\\
0.412	31.6\\
0.413	31.6\\
0.414	31.8\\
0.415	32\\
0.416	32\\
0.417	32\\
0.418	31.8\\
0.419	32.2\\
0.42	32\\
0.421	32.4\\
0.422	32.6\\
0.423	32.4\\
0.424	32.4\\
0.425	32.8\\
0.426	32.6\\
0.427	32.8\\
0.428	32.8\\
0.429	32.8\\
0.43	33\\
0.431	33\\
0.432	33\\
0.433	33\\
0.434	33.2\\
0.435	33\\
0.436	33.4\\
0.437	33.4\\
0.438	33.6\\
0.439	33.6\\
0.44	33.6\\
0.441	33.6\\
0.442	33.6\\
0.443	33.8\\
0.444	33.6\\
0.445	34.2\\
0.446	34\\
0.447	34\\
0.448	34.4\\
0.449	34.4\\
0.45	34.4\\
0.451	34.8\\
0.452	34.8\\
0.453	34.8\\
0.454	34.8\\
0.455	34.8\\
0.456	35.2\\
0.457	34.8\\
0.458	35.2\\
0.459	35.2\\
0.46	35.2\\
0.461	35.4\\
0.462	35.4\\
0.463	35.6\\
0.464	35.6\\
0.465	35.8\\
0.466	35.8\\
0.467	36\\
0.468	35.8\\
0.469	35.8\\
0.47	36\\
0.471	36.2\\
0.472	36\\
0.473	36\\
0.474	36\\
0.475	36\\
0.476	36\\
0.477	36.4\\
0.478	36\\
0.479	36.4\\
0.48	36.4\\
0.481	36.4\\
0.482	36.2\\
0.483	36.6\\
0.484	37\\
0.485	36.8\\
0.486	36.2\\
0.487	36.6\\
0.488	36.6\\
0.489	36.4\\
0.49	36.8\\
0.491	36.6\\
0.492	36.6\\
0.493	36.8\\
0.494	36.6\\
0.495	36.8\\
0.496	36.8\\
0.497	36.8\\
0.498	37\\
0.499	36.6\\
0.5	37\\
0.501	37.2\\
0.502	36.8\\
0.503	36.4\\
0.504	36.4\\
0.505	36.6\\
0.506	47.2222222222222\\
0.507	49.6666666666667\\
0.508	49.2222222222222\\
0.509	49.6666666666667\\
0.51	49.7777777777778\\
0.511	49.8888888888889\\
0.512	50.2222222222222\\
0.513	50.2222222222222\\
0.514	49.625\\
0.515	50.5555555555556\\
0.516	49.75\\
0.517	50.6666666666667\\
0.518	50.25\\
0.519	50.6666666666667\\
0.52	50.125\\
0.521	50.8888888888889\\
0.522	50.125\\
0.523	51.3333333333333\\
0.524	49.4285714285714\\
0.525	51.3333333333333\\
0.526	49.2857142857143\\
0.527	51.3333333333333\\
0.528	49.4285714285714\\
0.529	51.7777777777778\\
0.53	50\\
0.531	51.7777777777778\\
0.532	50.2857142857143\\
0.533	52.2222222222222\\
0.534	50.2857142857143\\
0.535	52.3333333333333\\
0.536	50.7142857142857\\
0.537	52.8888888888889\\
0.538	51\\
0.539	53.2222222222222\\
0.54	51.5714285714286\\
0.541	48.5\\
0.542	50\\
0.543	49.5\\
0.544	50\\
0.545	50.75\\
0.546	49.75\\
0.547	51.25\\
0.548	51\\
0.549	51.5\\
0.55	52\\
0.551	51.25\\
0.552	51.75\\
0.553	51.25\\
0.554	51.75\\
0.555	52\\
0.556	52\\
0.557	52\\
0.558	52.25\\
0.559	53.25\\
0.56	54.75\\
0.561	55\\
0.562	56\\
0.563	56.25\\
0.564	55.75\\
0.565	54.75\\
0.566	54.25\\
0.567	57.5\\
0.568	58\\
0.569	58.5\\
0.57	53.3333333333333\\
0.571	54\\
0.572	53.6666666666667\\
0.573	58.75\\
0.574	59.25\\
0.575	57.75\\
0.576	57.75\\
0.577	57\\
0.578	57.75\\
0.579	57.25\\
0.58	57\\
0.581	56.75\\
0.582	56.5\\
0.583	57\\
0.584	56.5\\
0.585	57\\
0.586	56.5\\
0.587	56.25\\
0.588	56\\
0.589	56\\
0.59	56.5\\
0.591	56.5\\
0.592	55.75\\
0.593	55.75\\
0.594	56\\
0.595	56.25\\
0.596	55.5\\
0.597	55.5\\
0.598	56.25\\
0.599	55.5\\
0.6	56\\
0.601	56.5\\
0.602	55.75\\
0.603	56\\
0.604	56.25\\
0.605	55.5\\
0.606	56.25\\
0.607	55.75\\
0.608	56\\
0.609	56\\
0.61	56.25\\
0.611	56\\
0.612	56\\
0.613	56\\
0.614	56.25\\
0.615	56.5\\
0.616	56.25\\
0.617	56.25\\
0.618	56.75\\
0.619	56.5\\
0.62	56.5\\
0.621	56.25\\
0.622	56.75\\
0.623	56.75\\
0.624	56.5\\
0.625	56.5\\
0.626	56\\
0.627	56.25\\
0.628	56.25\\
0.629	56.5\\
0.63	56.75\\
0.631	51.6666666666667\\
0.632	52.3333333333333\\
0.633	52.3333333333333\\
0.634	52.3333333333333\\
0.635	52\\
0.636	51.6666666666667\\
0.637	52\\
0.638	52\\
0.639	52\\
0.64	52\\
0.641	51.3333333333333\\
0.642	52\\
0.643	51.6666666666667\\
0.644	55.25\\
0.645	55.5\\
0.646	55\\
0.647	55\\
0.648	55.5\\
0.649	55\\
0.65	55.25\\
0.651	55\\
0.652	55\\
0.653	55.25\\
0.654	54.75\\
0.655	54.5\\
0.656	55\\
0.657	54.5\\
0.658	54.25\\
0.659	55\\
0.66	54.75\\
0.661	54.75\\
0.662	54.75\\
0.663	54.5\\
0.664	54.25\\
0.665	54.5\\
0.666	54.5\\
0.667	54.25\\
0.668	54.25\\
0.669	54.25\\
0.67	54\\
0.671	54.5\\
0.672	54\\
0.673	54\\
0.674	53.75\\
0.675	53.75\\
0.676	53.25\\
0.677	54\\
0.678	53.5\\
0.679	53.75\\
0.68	53.75\\
0.681	53.75\\
0.682	53.5\\
0.683	53.5\\
0.684	53.5\\
0.685	53.5\\
0.686	52.75\\
0.687	53.25\\
0.688	53.25\\
0.689	53.5\\
0.69	53\\
0.691	53\\
0.692	53.25\\
0.693	53.25\\
0.694	53\\
0.695	53.25\\
0.696	53.25\\
0.697	52.75\\
0.698	53\\
0.699	53\\
0.7	53\\
0.701	52.5\\
0.702	53\\
0.703	52.25\\
0.704	53\\
0.705	52.25\\
0.706	52.5\\
0.707	52.75\\
0.708	52\\
0.709	52.75\\
0.71	52.25\\
0.711	52.75\\
0.712	52.25\\
0.713	51.75\\
0.714	52.25\\
0.715	51.75\\
0.716	52.25\\
0.717	52\\
0.718	51.75\\
0.719	51.75\\
0.72	51.75\\
0.721	51.75\\
0.722	51.5\\
0.723	52\\
0.724	51.5\\
0.725	51.5\\
0.726	51.5\\
0.727	51.75\\
0.728	51.5\\
0.729	51.25\\
0.73	51.5\\
0.731	51\\
0.732	51.25\\
0.733	51.5\\
0.734	50.75\\
0.735	51\\
0.736	51.5\\
0.737	51\\
0.738	50.5\\
0.739	50.75\\
0.74	50.5\\
0.741	50.5\\
0.742	50\\
0.743	50.5\\
0.744	51\\
0.745	50.75\\
0.746	50.25\\
0.747	50\\
0.748	50\\
0.749	50.5\\
0.75	49.75\\
0.751	50\\
0.752	49.75\\
0.753	49.5\\
0.754	49.75\\
0.755	49.5\\
0.756	49.5\\
0.757	49\\
0.758	49.25\\
0.759	49.25\\
0.76	49.5\\
0.761	48.75\\
0.762	48.75\\
0.763	49.25\\
0.764	49\\
0.765	49\\
0.766	48.5\\
0.767	48.5\\
0.768	48.75\\
0.769	48.25\\
0.77	48.25\\
0.771	48.25\\
0.772	48.5\\
0.773	48.25\\
0.774	48.25\\
0.775	47.75\\
0.776	48\\
0.777	47.75\\
0.778	47.25\\
0.779	46.25\\
0.78	46.75\\
0.781	47.25\\
0.782	46.5\\
0.783	47\\
0.784	46\\
0.785	46\\
0.786	46.25\\
0.787	46\\
0.788	45.5\\
0.789	45.75\\
0.79	45\\
0.791	45.25\\
0.792	44\\
0.793	43\\
0.794	44.25\\
0.795	43.5\\
0.796	43\\
0.797	42.5\\
0.798	43\\
0.799	43\\
0.8	42.25\\
0.801	42.5\\
0.802	42.5\\
0.803	42.5\\
0.804	42.25\\
0.805	42.75\\
0.806	42.25\\
0.807	42\\
0.808	42.75\\
0.809	42.5\\
0.81	42.75\\
0.811	42\\
0.812	42.25\\
0.813	42.5\\
0.814	41.25\\
0.815	42.75\\
0.816	43.5\\
0.817	43.5\\
0.818	42.75\\
0.819	43.25\\
0.82	43.25\\
0.821	43.5\\
0.822	43.25\\
0.823	43\\
0.824	43.25\\
0.825	43\\
0.826	42.75\\
0.827	40.5\\
0.828	40.25\\
0.829	40\\
0.83	40.25\\
0.831	41\\
0.832	42.25\\
0.833	42.75\\
0.834	41.75\\
0.835	42.25\\
0.836	42\\
0.837	42.25\\
0.838	41.25\\
0.839	42.5\\
0.84	40.5\\
0.841	42.25\\
0.842	41.75\\
0.843	42.5\\
0.844	41.5\\
0.845	41.5\\
0.846	41.5\\
0.847	41.75\\
0.848	41.25\\
0.849	41.25\\
0.85	41.5\\
0.851	41.5\\
0.852	41.25\\
0.853	40.25\\
0.854	40.75\\
0.855	40.75\\
0.856	40.25\\
0.857	39.75\\
0.858	40.25\\
0.859	39.75\\
0.86	40\\
0.861	39.75\\
0.862	40\\
0.863	39.75\\
0.864	39.75\\
0.865	40\\
0.866	39.5\\
0.867	39.5\\
0.868	40\\
0.869	39.25\\
0.87	39.5\\
0.871	39.25\\
0.872	39\\
0.873	39\\
0.874	39\\
};
\addlegendentry{data1}

\end{axis}
\end{tikzpicture}%

%% file: tikz/speedup_factor.tex
\begin{tikzpicture}

\begin{axis}[%
width=0.951\figurewidth,
height=\figureheight,
at={(0\figurewidth,0\figureheight)},
scale only axis,
xmin=0,
xmax=500,
xtick={ 10,  50, 100, 200, 300, 400, 500},
xlabel={Reduced order $q$ [-]},
xmajorgrids,
ymin=0,
ymax=15,
ytick={ 0,  5, 10, 13, 15},
ylabel={Speedup factor $\alpha$ [-]},
ylabel style={at={(axis description cs:0.2,.5)},anchor=south},
ymajorgrids,
axis background/.style={fill=white},
legend style={legend cell align=left,align=left,draw=white!15!black},
title style={font=\footnotesize},xlabel style={font=\footnotesize},ylabel style={font=\footnotesize},ticklabel style={font=\footnotesize},line join=round,every axis legend/.code={\let\addlegendentry\relax}
]
\addplot [color=std, mark=*, mark options={solid, fill=std}]
  table[row sep=crcr]{%
500	5.16677966101695\\
400	5.84881043745203\\
300	7.0893023255814\\
200	8.25587693640992\\
100	10.4885769336636\\
50	11.6904433195275\\
10	12.9851763503152\\
};
\end{axis}
\end{tikzpicture}%

%% file: tikz/speedup.tex
\begin{tikzpicture}

\begin{axis}[%
width=0.951\figurewidth,
height=\figureheight,
at={(0\figurewidth,0\figureheight)},
scale only axis,
area style,
%reverse legend,
stack plots=y,
xmin=0,
xmax=500,
xtick={ 10,  50, 100, 200, 300, 400, 500},
xlabel style={font=\color{white!15!black}},
xlabel={Reduced order $q$ [-]},
ymin=0,
ymax=20,
ytick={0,5,10,15,20},
yticklabels={{0\,\%},{5\,\%},{10\,\%},{15\,\%},{20\,\%}},
ylabel style={at={(axis description cs:0.1,.5)},anchor=south},
ylabel style={font=\color{white!15!black}},
ylabel style={align=center},
ylabel={CPU time w.r.t \fom{}},
axis background/.style={fill=white},
title style={font=\footnotesize},xlabel style={font=\footnotesize},ylabel style={font=\footnotesize},ticklabel style={font=\footnotesize},line join=round,
every axis legend/.code={\let\addlegendentry\relax}
]
\addplot[fill=t_other, draw=black] table[row sep=crcr]{%
500	2.91049731006434\\
400	2.8501115339194\\
300	2.72552158509381\\
200	2.70353365700037\\
100	2.61746883611075\\
50	2.64507020076105\\
10	2.53780015745965\\
}
\closedcycle;
%\addlegendentry{Other}

\addplot[fill=t_eval, draw=black] table[row sep=crcr]{%
500	5.15680356908542\\
400	5.07676157984517\\
300	4.78414906180291\\
200	4.87862485238158\\
100	5.03214801207191\\
50	4.93504789397717\\
10	4.86812754231728\\
}
\closedcycle;
%\addlegendentry{Evaluate elements}

\addplot[fill=t_solve, draw=black] table[row sep=crcr]{%
500	11.287114551896\\
400	9.17062065345759\\
300	6.59608975200106\\
200	4.53042514105763\\
100	1.88456501771421\\
50	0.973877443905005\\
10	0.29516139614224\\
}
\closedcycle;
%\addlegendentry{Reduce and solve\\linear system}

\end{axis}
\end{tikzpicture}%

%% file: tikz/bar_time.tex
\begin{tikzpicture}

\begin{axis}[%
width=0.948\figurewidth,
height=\figureheight,
at={(0\figurewidth,0\figureheight)},
scale only axis,
reverse legend,
bar width=30,
xmin=0.5,
xmax=3.5,
xtick={1,2,3},
xticklabels={{\fom{}},{\rom{500}},{\rom{10}}},
xlabel={\vphantom{Reduced order $q$}},
ymin=0,
ymax=100,
ytick={0,20,40,60,80,100},
yticklabels={{0\,\%},{20\,\%},{40\,\%},{60\,\%},{80\,\%},{100\,\%}},
ylabel style={at={(axis description cs:0.1,.5)},anchor=south},
ylabel style={font=\color{white!15!black}},
ylabel={Relative CPU time},
ylabel style={align=center},
axis background/.style={fill=white},
legend style={at={(1.02,0.5)},anchor=west,legend cell align=left,align=left,draw=white!15!black,font=\footnotesize,cells={align=center}},
title style={font=\footnotesize},xlabel style={font=\footnotesize},ylabel style={font=\footnotesize},ticklabel style={font=\footnotesize},line join=round,
every axis legend/.code={\let\addlegendentry\relax}
]
\addplot[ybar stacked, fill=t_other, draw=black, area legend] table[row sep=crcr] {%
1	2.87925469098545\\
2	15.037898305085\\
3	32.9537825864713\\
};
\addplot[forget plot, color=white!15!black] table[row sep=crcr] {%
0.5	0\\
3.5	0\\
};
\addlegendentry{Other}

\addplot[ybar stacked, fill=t_eval, draw=black, area legend] table[row sep=crcr] {%
1	5.07151292481302\\
2	26.6440677966102\\
3	63.2134946328165\\
};
\addplot[forget plot, color=white!15!black] table[row sep=crcr] {%
0.5	0\\
3.5	0\\
};
\addlegendentry{Evaluate elements}

\addplot[ybar stacked, fill=t_solve, draw=black, area legend] table[row sep=crcr] {%
1	92.0492323842015\\
2	58.3180338983049\\
3	3.83272278071223\\
};
\addplot[forget plot, color=white!15!black] table[row sep=crcr] {%
0.5	0\\
3.5	0\\
};
\addlegendentry{Linear system}

\end{axis}

\begin{axis}[%
width=0.948\figurewidth,
height=\figureheight,
at={(0\figurewidth,0\figureheight)},
scale only axis,
hide y axis,
xmin=0,
xmax=1,
xtick={0.166666666666667,0.5,0.833333333333333},
xticklabels={{total 21\,h},{total 4.1\,h},{total 1.6\,h}},
xtick style={draw=none},
ymin=0,
ymax=1,
axis x line*=top,
axis y line*=left,
legend style={legend cell align=left, align=left, draw=white!15!black, fill=white!94!black},
title style={font=\footnotesize},xlabel style={font=\footnotesize},ylabel style={font=\footnotesize},ticklabel style={font=\footnotesize},line join=round,every axis legend/.code={\let\addlegendentry\relax}
]
\end{axis}
\end{tikzpicture}%

%% file: tikz/legend_area.tex
\definecolor{mycolor1}{rgb}{0.00000,0.44700,0.74100}%
\definecolor{mycolor2}{rgb}{0.85000,0.32500,0.09800}%
\definecolor{mycolor3}{rgb}{0.92900,0.69400,0.12500}%
\begin{tikzpicture}

\begin{axis}[%
width=\figurewidth,
height=0.876\figureheight,
at={(0\figurewidth,0\figureheight)},
hide axis,
scale only axis,
xmin=10,
xmax=100,
ymin=0,
ymax=0.4,
legend columns=-1,
legend style={draw=none,column sep=1ex, font=\footnotesize}
]
\addlegendimage{color=black, fill=mycolor3, thick, area legend}
\addlegendentry{Linear system}
\addlegendimage{color=black, fill=mycolor2,thick, area legend}
\addlegendentry{Evaluate elements}
\addlegendimage{color=black, fill=mycolor1,thick, area legend}
\addlegendentry{Other}

\end{axis}
\end{tikzpicture}%

%% file: tikz/err_infty_mor300_sample1.tex
\begin{tikzpicture}

\begin{axis}[%
width=0.951\figurewidth,
height=\figureheight,
at={(0\figurewidth,0\figureheight)},
scale only axis,
xmin=280,
xmax=430,
xlabel style={font=\color{white!15!black}},
xlabel={$\text{Contractility }\sigma\text{ [kPa]}$},
xtick={280,310,...,430},
minor tick num=1,
extra x ticks={355},
ymode=log,
ymin=0.0001,
ymax=10,
yminorticks=true,
ytick={1e-4,1e-3,1e-2,1e-1,1e0,1e1},
ylabel style={font=\color{white!15!black}},
ylabel={Spatial $\epsilon_{\infty,\infty}$-error [mm]},
axis background/.style={fill=white},
xminorgrids,
xmajorgrids,
ymajorgrids,
%yminorgrids,
legend style={legend cell align=left, align=left, draw=white!15!black},
title style={font=\footnotesize},xlabel style={font=\footnotesize},ylabel style={font=\footnotesize},ticklabel style={font=\footnotesize},line join=round,legend style={at={(1.02,0.5)},anchor=west,legend cell align=left,align=left,font=\footnotesize},cells={align=center},
every axis legend/.code={\let\addlegendentry\relax}
]
\addplot [color=rom_direct]
  table[row sep=crcr]{%
280	0.000130500464726172\\
295	0.000130199312694098\\
310	0.000130764637859939\\
325	0.00013086410716768\\
340	0.000131074678249239\\
355	0.000131119845913769\\
370	0.000131328276601752\\
385	0.000131438437574175\\
400	0.000130416637691223\\
415	0.000130811562420321\\
430	0.000131831199436232\\
};
\addlegendentry{Direct ROM}

\addplot [color=rom_const]
  table[row sep=crcr]{%
280	1.15773906944159\\
295	0.92414167378404\\
310	0.689630991372062\\
325	0.456476743541271\\
340	0.226373097876085\\
355	0.000131119845913769\\
370	0.221983103241051\\
385	0.439070896935746\\
400	0.650946282267181\\
415	0.857429089332031\\
430	1.05844823865893\\
};
\addlegendentry{const}

%\addplot [color=black] coordinates {(0.0001,355) (10,355)};

\end{axis}
\end{tikzpicture}%

%% file: tikz/err_infty_mor300_sample2.tex
\definecolor{mycolor1}{rgb}{0.00000,0.44700,0.74100}%
\definecolor{mycolor2}{rgb}{0.85000,0.32500,0.09800}%
\definecolor{mycolor3}{rgb}{0.92900,0.69400,0.12500}%
\definecolor{mycolor4}{rgb}{0.49400,0.18400,0.55600}%
\begin{tikzpicture}

\begin{axis}[%
width=0.951\figurewidth,
height=\figureheight,
at={(0\figurewidth,0\figureheight)},
scale only axis,
xmin=280,
xmax=430,
xlabel style={font=\color{white!15!black}},
xlabel={$\text{Contractility }\sigma\text{ [kPa]}$},
xtick={280,310,...,430},
minor tick num=1,
ymode=log,
ymin=0.0001,
ymax=10,
yminorticks=true,
ytick={1e-4,1e-3,1e-2,1e-1,1e0,1e1},
ylabel style={font=\color{white!15!black}},
ylabel={Spatial $\epsilon_{\infty,\infty}$-error [mm]},,
axis background/.style={fill=white},
xmajorgrids,
xminorgrids,
ymajorgrids,
%yminorgrids,
legend style={legend cell align=left, align=left, draw=white!15!black},
title style={font=\footnotesize},xlabel style={font=\footnotesize},ylabel style={font=\footnotesize},ticklabel style={font=\footnotesize},line join=round,legend style={at={(1.02,0.5)},anchor=west,legend cell align=left,align=left,font=\footnotesize},cells={align=center},
every axis legend/.code={\let\addlegendentry\relax}
]
\addplot [color=rom_direct]
  table[row sep=crcr]{%
280	0.000130500464726172\\
295	0.000130199312694098\\
310	0.000130764637859939\\
325	0.00013086410716768\\
340	0.000131074678249239\\
355	0.000131119845913769\\
370	0.000131328276601752\\
385	0.000131438437574175\\
400	0.000130416637691223\\
415	0.000130811562420321\\
430	0.000131831199436232\\
};
\addlegendentry{Direct ROM}

\addplot [color=rom_direct_interp]
  table[row sep=crcr]{%
280	0.000130500464726172\\
295	0.322343856318994\\
310	0.301136707336174\\
325	0.324123335985299\\
340	0.300648795939091\\
355	0.269672599294487\\
370	0.234186728330038\\
385	0.181584466409565\\
400	0.141149062061766\\
415	0.153061306894406\\
430	0.000131831199436232\\
};
\addlegendentry{Direct Interp}

\addplot [color=rom_cos, dashed]
  table[row sep=crcr]{%
280	0.000130500464726172\\
295	0.315161295076808\\
310	0.294189961688769\\
325	0.300087705090901\\
340	0.270779110489433\\
355	0.225665886432787\\
370	0.159044017301455\\
385	0.146180160229274\\
400	0.150711273539943\\
415	0.135208786865232\\
430	0.000131831199436232\\
};
\addlegendentry{CoB SV}

\addplot [color=rom_cob]
  table[row sep=crcr]{%
280	0.000130500464726172\\
295	0.328826956685932\\
310	0.428914441126067\\
325	0.843499690001742\\
340	1.65815734072021\\
355	2.28951758425897\\
370	1.94119226028605\\
385	1.02276575807942\\
400	0.255873281287621\\
415	0.175945989122563\\
430	0.000131831199436232\\
};
\addlegendentry{CoB}

\addplot [color=rom_grassmann]
  table[row sep=crcr]{%
280	0.000130500464726172\\
295	0.444563149194952\\
310	1.21211565354429\\
325	1.77458609448118\\
340	2.11652129269866\\
355	2.28951680637448\\
370	2.31141937610762\\
385	2.16845671487282\\
400	1.809792331161\\
415	1.06940168324016\\
430	0.000131831199436232\\
};
\addlegendentry{Grassmann}

\end{axis}
\end{tikzpicture}%

%% file: tikz/err_infty_mor300_sample4.tex
\begin{tikzpicture}

\begin{axis}[%
width=0.951\figurewidth,
height=\figureheight,
at={(0\figurewidth,0\figureheight)},
scale only axis,
xmin=280,
xmax=430,
xlabel style={font=\color{white!15!black}},
xlabel={Contractility $\sigma$ [kPa]},
xtick={280,330,...,430},
%extra x ticks={330,380},
minor tick num=9,
ymode=log,
ymin=0.0001,
ymax=10,
yminorticks=true,
ytick={1e-4,1e-3,1e-2,1e-1,1e0,1e1},
ylabel style={font=\color{white!15!black}},
ylabel={Spatial $\epsilon_{\infty,\infty}$-error [mm]},
axis background/.style={fill=white},
xmajorgrids,
xminorgrids,
ymajorgrids,
%yminorgrids,
legend style={legend cell align=left, align=left, draw=white!15!black},
title style={font=\footnotesize},xlabel style={font=\footnotesize},ylabel style={font=\footnotesize},ticklabel style={font=\footnotesize},line join=round,legend style={at={(1.02,0.5)},anchor=west,legend cell align=left,align=left,font=\footnotesize},cells={align=center},
every axis legend/.code={\let\addlegendentry\relax}
]

\addplot [color=rom_direct_interp]
  table[row sep=crcr]{%
280	0.000130500464726172\\
285	0.0255773560413124\\
290	0.0501764679344977\\
295	0.288014071860316\\
300	0.268487937526989\\
305	0.236633290294382\\
310	0.170812204386445\\
315	0.120889603753406\\
320	0.0745139073129785\\
325	0.0337416792189508\\
330	0.000130759716665636\\
335	0.0223915389913002\\
340	0.0443428647678083\\
345	0.0617471072025998\\
350	0.0728041923936214\\
355	0.0765069420793907\\
360	0.0723598069859799\\
365	0.0606261926580977\\
370	0.0402088829012704\\
375	0.0164289249591225\\
380	0.000130712256603018\\
385	0.0148477896198651\\
390	0.0294636676872745\\
395	0.0430045131596132\\
400	0.0540784938097762\\
405	0.0598067315158001\\
410	0.0598618115017943\\
415	0.0798277432562356\\
420	0.0973007881046853\\
425	0.0675759306612084\\
430	0.000131831199436232\\
};
\addlegendentry{Direct Interp}

\addplot [color=rom_cos, dashed]
  table[row sep=crcr]{%
280	0.000130500464726172\\
285	0.0253373092085515\\
290	0.0463059046705023\\
295	0.298889351314195\\
300	0.284247002343815\\
305	0.257287106252674\\
310	0.178731860978602\\
315	0.0818645242814839\\
320	0.044327060404369\\
325	0.0227183446198322\\
330	0.000130759716665636\\
335	0.0200055595372555\\
340	0.0403136652748748\\
345	0.057289317710276\\
350	0.0686823197967597\\
355	0.0740796476739169\\
360	0.0734565955004905\\
365	0.0664381553821249\\
370	0.0514816190615596\\
375	0.0266507501316837\\
380	0.000130712256603018\\
385	0.0150100078031385\\
390	0.0297764364316751\\
395	0.0431135702823942\\
400	0.053740244224557\\
405	0.0593689140186287\\
410	0.0568665099663346\\
415	0.0539764359771366\\
420	0.0685049723133968\\
425	0.0541880610786396\\
430	0.000131831199436232\\
};
\addlegendentry{CoB SV}

\addplot [color=rom_cob]
  table[row sep=crcr]{%
280	0.000130500464726172\\
285	0.0746531819106323\\
290	0.130237193784788\\
295	0.334236472073014\\
300	0.550091585328265\\
305	0.914790373589211\\
310	0.620719842196158\\
315	0.3446705755845\\
320	0.109800878119131\\
325	0.0695290467124228\\
330	0.000130759716665636\\
335	0.0703055746960598\\
340	0.121813008518741\\
345	0.210803879998888\\
350	0.57152815685826\\
355	0.857743289713889\\
360	0.401783713564038\\
365	0.189172468786997\\
370	0.117683019478015\\
375	0.0630045960829086\\
380	0.000130712256603018\\
385	0.06381133617293\\
390	0.108735413975404\\
395	0.145129297693481\\
400	0.333571518552479\\
405	0.549082497124369\\
410	0.372873691111435\\
415	0.111800582378679\\
420	0.109596235383906\\
425	0.0879526038646822\\
430	0.000131831199436232\\
};
\addlegendentry{CoB}

\addplot [color=rom_grassmann]
  table[row sep=crcr]{%
280	0.000130500464726172\\
285	0.103293274905427\\
290	0.353998122662911\\
295	0.613317374266026\\
300	0.808890543958191\\
305	0.91478977540727\\
310	0.916696073094121\\
315	0.800740829548466\\
320	0.557432913666328\\
325	0.240034572022788\\
330	0.000130759716665636\\
335	0.100096107608721\\
340	0.304118905475306\\
345	0.529376434131304\\
350	0.723744442535739\\
355	0.857744173161642\\
360	0.903007026915081\\
365	0.829693533218062\\
370	0.61045801537603\\
375	0.257491925334892\\
380	0.000130712256603018\\
385	0.0873723076357341\\
390	0.189700713512399\\
395	0.341659380899199\\
400	0.471310934926892\\
405	0.549083055103868\\
410	0.558566144633311\\
415	0.48893571002481\\
420	0.336696169047744\\
425	0.181632462747187\\
430	0.000131831199436232\\
};
\addlegendentry{Grassmann}

\addplot [color=rom_direct]
  table[row sep=crcr]{%
280	0.000130500464726172\\
285	0.000130665888013983\\
290	0.000130725903048871\\
295	0.000130199312694098\\
300	0.000130818449808602\\
305	0.000131124424122691\\
310	0.000130764637859939\\
};
\addlegendentry{Direct ROM}

\addplot [color=rom_direct, forget plot]
  table[row sep=crcr]{%
310	0.000130764637859939\\
320	0.000130426340592994\\
};
\addplot [color=rom_direct, forget plot]
  table[row sep=crcr]{%
320	0.000130426340592994\\
325	0.00013086410716768\\
330	0.000130759716665636\\
335	0.000130805284770161\\
340	0.000131074678249239\\
345	0.000130545404855991\\
350	0.000130485325096401\\
355	0.000131119845913769\\
360	0.000131105753977983\\
365	0.000130964912291265\\
370	0.000131328276601752\\
375	0.000130280538569951\\
380	0.000130712256603018\\
385	0.000131438437574175\\
390	0.0001314024608364\\
395	0.000130858138476318\\
400	0.000130416637691223\\
405	0.000131065629348215\\
410	0.000131676748530013\\
415	0.000130811562420321\\
420	0.000131127721479264\\
425	0.000131522938747539\\
430	0.000131831199436232\\
};
\end{axis}
\end{tikzpicture}%

%% file: tikz/legend_err.tex
\definecolor{mycolor1}{rgb}{0.00000,0.44700,0.74100}%
\definecolor{mycolor2}{rgb}{0.85000,0.32500,0.09800}%
\definecolor{mycolor3}{rgb}{0.92900,0.69400,0.12500}%
\definecolor{mycolor4}{rgb}{0.49400,0.18400,0.55600}%
\begin{tikzpicture}

\begin{axis}[%
width=\figurewidth,
height=0.876\figureheight,
at={(0\figurewidth,0\figureheight)},
hide axis,
scale only axis,
xmin=10,
xmax=100,
ymin=0,
ymax=0.4,
legend columns=-1,
legend style={draw=none,column sep=1ex, font=\footnotesize,cells={align=center}}
]
\addlegendimage{color=rom_direct,thick}
\addlegendentry{\rom{}\\direct}
\addlegendimage{color=rom_const,thick}
\addlegendentry{\rom{}\\constant}
\addlegendimage{color=rom_grassmann,thick}
\addlegendentry{\prom{}\\Grassmann}
\addlegendimage{color=rom_cob,thick}
\addlegendentry{\prom{}\\CoB}
\addlegendimage{color=rom_direct_interp,thick}
\addlegendentry{\prom{}\\direct interpolation}
\addlegendimage{color=rom_cos,thick, dashed}
\addlegendentry{\prom{}\\CoS}
\end{axis}
\end{tikzpicture}%

%\addplot [color=blue]
%\addlegendentry{Direct ROM}
%
%\addplot [color=mycolor1]
%\addlegendentry{CoB SV}
%
%\addplot [color=mycolor2]
%\addlegendentry{Direct Interp}
%
%\addplot [color=mycolor3]
%\addlegendentry{CoB}
%
%\addplot [color=mycolor4]
%\addlegendentry{Grassmann}

%% file: tikz/sigma_vs_ef_sample1.tex
\begin{tikzpicture}[trim axis left,trim axis right]

\begin{axis}[%
width=0.951\figurewidth,
height=\figureheight,
at={(0\figurewidth,0\figureheight)},
scale only axis,
xmin=280,
xmax=430,
xtick={280,430},
minor x tick num=9,
xminorgrids,
xlabel style={font=\color{white}},
ymin=0.35,
ymax=0.5,
ylabel style={font=\color{white!15!black}},
ylabel={Ejection fraction [-]},
ylabel style={at={(axis description cs:0.15,.5)},anchor=south},
xticklabels={,,},
axis background/.style={fill=white},
xmajorgrids,
ymajorgrids,
legend style={at={(0.03,0.97)}, anchor=north west, legend cell align=left, align=left, draw=white!15!black},
title style={font=\footnotesize},xlabel style={font=\footnotesize},ylabel style={font=\footnotesize},ticklabel style={font=\footnotesize},line join=round,every axis legend/.code={\let\addlegendentry\relax}
]
\addplot [color=fom]
  table[row sep=crcr]{%
280	0.369410253043035\\
285	0.37380314968201\\
290	0.378088579183474\\
295	0.382268291744098\\
300	0.386347749771799\\
305	0.39032955527101\\
310	0.394217934514138\\
315	0.398015455865787\\
320	0.401724651303498\\
325	0.405350374119022\\
330	0.408893842761627\\
335	0.412358641671888\\
340	0.415747450201582\\
345	0.419061947586122\\
350	0.422304977680014\\
355	0.425479613302391\\
360	0.428586953773377\\
365	0.431629481707372\\
370	0.434609997129592\\
375	0.437529391684804\\
380	0.440389566667662\\
385	0.443192714851152\\
390	0.44594109003728\\
395	0.448635476309595\\
400	0.451277492555556\\
405	0.453869293256192\\
410	0.456412294458721\\
415	0.458907397441533\\
420	0.46135673134845\\
425	0.463762694613125\\
430	0.466125353397793\\
};
\addlegendentry{FOM}

\addplot [color=rom_const]
  table[row sep=crcr]{%
280	0.379152313403052\\
295	0.389815925760527\\
310	0.399730201786435\\
325	0.408952104612917\\
340	0.417524293254796\\
355	0.425479597083264\\
370	0.432845918957734\\
385	0.439649301672492\\
400	0.445932585228778\\
415	0.45171311659559\\
430	0.457020631173168\\
};
\addlegendentry{MOR no Interpolation}

\addplot [color=black, draw=none, mark=square*, mark options={solid, fill=black, black}]
  table[row sep=crcr]{%
355	0.425479597083264\\
};
\addlegendentry{Sample Point}

\end{axis}
\end{tikzpicture}%

%% file: tikz/sigma_vs_ef_sample2.tex
\begin{tikzpicture}[trim axis left,trim axis right]

\begin{axis}[%
width=0.951\figurewidth,
height=\figureheight,
at={(0\figurewidth,0\figureheight)},
scale only axis,
xmin=280,
xmax=430,
xtick={280,430},
minor x tick num=9,
xminorgrids,
xlabel style={font=\color{white!15!black}},
ymin=0.35,
ymax=0.5,
ylabel style={font=\color{white!15!black}},
xticklabels={,,},
yticklabels={,,},
axis background/.style={fill=white},
xmajorgrids,
ymajorgrids,
legend style={at={(0.03,0.97)}, anchor=north west, legend cell align=left, align=left, draw=white!15!black},
title style={font=\footnotesize},xlabel style={font=\footnotesize},ylabel style={font=\footnotesize},ticklabel style={font=\footnotesize},line join=round,every axis legend/.code={\let\addlegendentry\relax}
]
\addplot [color=fom]
  table[row sep=crcr]{%
280	0.369410253043035\\
285	0.37380314968201\\
290	0.378088579183474\\
295	0.382268291744098\\
300	0.386347749771799\\
305	0.39032955527101\\
310	0.394217934514138\\
315	0.398015455865787\\
320	0.401724651303498\\
325	0.405350374119022\\
330	0.408893842761627\\
335	0.412358641671888\\
340	0.415747450201582\\
345	0.419061947586122\\
350	0.422304977680014\\
355	0.425479613302391\\
360	0.428586953773377\\
365	0.431629481707372\\
370	0.434609997129592\\
375	0.437529391684804\\
380	0.440389566667662\\
385	0.443192714851152\\
390	0.44594109003728\\
395	0.448635476309595\\
400	0.451277492555556\\
405	0.453869293256192\\
410	0.456412294458721\\
415	0.458907397441533\\
420	0.46135673134845\\
425	0.463762694613125\\
430	0.466125353397793\\
};
\addlegendentry{FOM}

\addplot [color=rom_cos, dashed]
  table[row sep=crcr]{%
280	0.369410199794652\\
295	0.382063706360695\\
310	0.393883428095398\\
325	0.404953799493135\\
340	0.415341937657589\\
355	0.425103442779707\\
370	0.434288664729585\\
385	0.442941271045658\\
400	0.451113202562362\\
415	0.458830932449834\\
430	0.46612525034337\\
};
\addlegendentry{pMOR Concatenation of snapshots}

\addplot [color=rom_grassmann]
  table[row sep=crcr]{%
280	0.369410199794652\\
295	0.379068980374042\\
310	0.384848364944442\\
325	0.390253285516934\\
340	0.396571144732069\\
355	0.404081777261908\\
370	0.412889975357806\\
385	0.423220688939276\\
400	0.435645168783099\\
415	0.451258343827102\\
430	0.46612525034337\\
};
\addlegendentry{pMOR Grassmann}

\addplot [color=black, draw=none, mark=square*, mark options={solid, fill=black, black}]
  table[row sep=crcr]{%
280	0.369410199794652\\
430	0.46612525034337\\
};
\addlegendentry{Sample Points}

\end{axis}
\end{tikzpicture}%

%% file: tikz/sigma_vs_ef_sample4.tex
\begin{tikzpicture}[trim axis left, trim axis right]

\begin{axis}[%
width=0.951\figurewidth,
height=\figureheight,
at={(0\figurewidth,0\figureheight)},
scale only axis,
xmin=280,
xmax=430,
xtick={280, 330, 380, 430},
minor x tick num=9,
xminorgrids,
xlabel style={font=\color{white!15!black}},
ymin=0.35,
ymax=0.5,
ylabel style={font=\color{white!15!black}},
xticklabels={,,},
yticklabels={,,},
axis background/.style={fill=white},
xmajorgrids,
ymajorgrids,
legend style={at={(0.03,0.97)}, anchor=north west, legend cell align=left, align=left, draw=white!15!black},
title style={font=\footnotesize},xlabel style={font=\footnotesize},ylabel style={font=\footnotesize},ticklabel style={font=\footnotesize},line join=round,every axis legend/.code={\let\addlegendentry\relax}
]
\addplot [color=fom]
  table[row sep=crcr]{%
280	0.369410253043035\\
285	0.37380314968201\\
290	0.378088579183474\\
295	0.382268291744098\\
300	0.386347749771799\\
305	0.39032955527101\\
310	0.394217934514138\\
315	0.398015455865787\\
320	0.401724651303498\\
325	0.405350374119022\\
330	0.408893842761627\\
335	0.412358641671888\\
340	0.415747450201582\\
345	0.419061947586122\\
350	0.422304977680014\\
355	0.425479613302391\\
360	0.428586953773377\\
365	0.431629481707372\\
370	0.434609997129592\\
375	0.437529391684804\\
380	0.440389566667662\\
385	0.443192714851152\\
390	0.44594109003728\\
395	0.448635476309595\\
400	0.451277492555556\\
405	0.453869293256192\\
410	0.456412294458721\\
415	0.458907397441533\\
420	0.46135673134845\\
425	0.463762694613125\\
430	0.466125353397793\\
};
\addlegendentry{FOM}

\addplot [color=rom_cos, dashed]
  table[row sep=crcr]{%
280	0.369410199794652\\
285	0.373780553203253\\
290	0.378049455910205\\
295	0.382219619566596\\
300	0.386294606500689\\
305	0.390276640467943\\
310	0.394168994497107\\
315	0.397974180647079\\
320	0.401695227260645\\
325	0.405334731421631\\
330	0.408893837391706\\
335	0.412341296944888\\
340	0.415717481832851\\
345	0.41902437698448\\
350	0.422264195693213\\
355	0.42543882419453\\
360	0.428549363025147\\
365	0.431597879507612\\
370	0.434586579295445\\
375	0.437517110678322\\
380	0.440389580856546\\
385	0.443179679315328\\
390	0.445918191503365\\
395	0.448606359013536\\
400	0.45124557726558\\
405	0.45383706612753\\
410	0.456382005390922\\
415	0.458881674809266\\
420	0.461337903483661\\
425	0.463752605008054\\
430	0.46612525034337\\
};
\addlegendentry{pMOR Concatenation of snapshots}

\addplot [color=rom_grassmann]
  table[row sep=crcr]{%
280	0.369410199794652\\
285	0.373104166928613\\
290	0.375962074894542\\
295	0.378556392910843\\
300	0.381295671971853\\
305	0.384474531520919\\
310	0.38830021099082\\
315	0.39290383884126\\
320	0.398299521942215\\
325	0.404106029827798\\
330	0.408893837391706\\
335	0.41155338650996\\
340	0.413448480737169\\
345	0.415078232920823\\
350	0.416862666681002\\
355	0.419120943032069\\
360	0.422152610111294\\
365	0.426175145772563\\
370	0.431145262122722\\
375	0.436422994432006\\
380	0.440389580856546\\
385	0.442607675577536\\
390	0.444316613841395\\
395	0.445903449484446\\
400	0.44765086768031\\
405	0.449795233566286\\
410	0.452472879681887\\
415	0.455693260751797\\
420	0.459335509042022\\
425	0.46308227988377\\
430	0.46612525034337\\
};
\addlegendentry{pMOR Grassmann}

\addplot [color=black, draw=none, mark=square*, mark options={solid, fill=black, black}]
  table[row sep=crcr]{%
280	0.369410199794652\\
330	0.408893837391706\\
380	0.440389580856546\\
430	0.46612525034337\\
};
\addlegendentry{Sample Points}

\end{axis}
\end{tikzpicture}%

%% file: tikz/sigma_vs_pmax_sample1.tex
\begin{tikzpicture}[trim axis left,trim axis right]

\begin{axis}[%
width=0.951\figurewidth,
height=\figureheight,
at={(0\figurewidth,0\figureheight)},
scale only axis,
xmin=280,
xmax=430,
xtick={280,430},
minor x tick num=9,
xminorgrids,
xlabel style={font=\color{white!15!black}},
ymin=80,
ymax=95,
ylabel style={font=\color{white!15!black}},
ylabel={Max. LVP [mmHg]},
ylabel style={at={(axis description cs:0.15,.5)},anchor=south},
xticklabels={,,},
axis background/.style={fill=white},
xmajorgrids,
ymajorgrids,
legend style={at={(0.03,0.97)}, anchor=north west, legend cell align=left, align=left, draw=white!15!black},
title style={font=\footnotesize},xlabel style={font=\footnotesize},ylabel style={font=\footnotesize},ticklabel style={font=\footnotesize},line join=round,every axis legend/.code={\let\addlegendentry\relax}
]
\addplot [color=fom]
  table[row sep=crcr]{%
280	83.1009904154952\\
285	83.4213840503183\\
290	83.7359360165049\\
295	84.0456211948194\\
300	84.3498730424087\\
305	84.6491827298874\\
310	84.9434641174014\\
315	85.2331231224666\\
320	85.5184749404688\\
325	85.7992108723065\\
330	86.0759014311325\\
335	86.3483570350275\\
340	86.6167148333281\\
345	86.8814144265842\\
350	87.142505903693\\
355	87.3996571859964\\
360	87.6534423969611\\
365	87.9038062972142\\
370	88.1504073227232\\
375	88.3938263703264\\
380	88.6341847251328\\
385	88.8718588189841\\
390	89.1060294748025\\
395	89.33724867302\\
400	89.565581965988\\
405	89.7915127632295\\
410	90.0142408012635\\
415	90.2340930357819\\
420	90.452073679103\\
425	90.6674653487719\\
430	90.8795761304342\\
};
\addlegendentry{FOM}

\addplot [color=rom_const]
  table[row sep=crcr]{%
280	83.7277778215195\\
295	84.5314641007367\\
310	85.2967014140489\\
325	86.0278080943625\\
340	86.7277194955348\\
355	87.3996769903868\\
370	88.0456184704058\\
385	88.6681202521177\\
400	89.2681405797032\\
415	89.8469824718974\\
430	90.4080881846101\\
};
\addlegendentry{MOR no Interpolation}

\addplot [color=black, draw=none, mark=square*, mark options={solid, fill=black, black}]
  table[row sep=crcr]{%
355	87.3996769903868\\
};
\addlegendentry{Sample Point}

\end{axis}
\end{tikzpicture}%

%% file: tikz/sigma_vs_pmax_sample2.tex
\begin{tikzpicture}[trim axis left,trim axis right]

\begin{axis}[%
width=0.951\figurewidth,
height=\figureheight,
at={(0\figurewidth,0\figureheight)},
scale only axis,
xmin=280,
xmax=430,
xtick={280,430},
minor x tick num=9,
xminorgrids,
xlabel style={font=\color{white!15!black}},
ymin=80,
ymax=95,
ylabel style={font=\color{white!15!black}},
xticklabels={,,},
yticklabels={,,},
axis background/.style={fill=white},
xmajorgrids,
ymajorgrids,
legend style={at={(0.03,0.97)}, anchor=north west, legend cell align=left, align=left, draw=white!15!black},
title style={font=\footnotesize},xlabel style={font=\footnotesize},ylabel style={font=\footnotesize},ticklabel style={font=\footnotesize},line join=round,every axis legend/.code={\let\addlegendentry\relax}
]
\addplot [color=fom]
  table[row sep=crcr]{%
280	83.1009904154952\\
285	83.4213840503183\\
290	83.7359360165049\\
295	84.0456211948194\\
300	84.3498730424087\\
305	84.6491827298874\\
310	84.9434641174014\\
315	85.2331231224666\\
320	85.5184749404688\\
325	85.7992108723065\\
330	86.0759014311325\\
335	86.3483570350275\\
340	86.6167148333281\\
345	86.8814144265842\\
350	87.142505903693\\
355	87.3996571859964\\
360	87.6534423969611\\
365	87.9038062972142\\
370	88.1504073227232\\
375	88.3938263703264\\
380	88.6341847251328\\
385	88.8718588189841\\
390	89.1060294748025\\
395	89.33724867302\\
400	89.565581965988\\
405	89.7915127632295\\
410	90.0142408012635\\
415	90.2340930357819\\
420	90.452073679103\\
425	90.6674653487719\\
430	90.8795761304342\\
};
\addlegendentry{FOM}

\addplot [color=rom_cos, dashed]
  table[row sep=crcr]{%
280	83.1009917797674\\
295	84.0395611877149\\
310	84.9338760958861\\
325	85.7882284319633\\
340	86.6055438713556\\
355	87.3892347066597\\
370	88.1414030795\\
385	88.8648095804066\\
400	89.5608146413528\\
415	90.2315914755339\\
430	90.8795778242993\\
};
\addlegendentry{pMOR Concatenation of snapshots}

\addplot [color=rom_grassmann]
  table[row sep=crcr]{%
280	83.1009917797674\\
295	84.1820964791302\\
310	85.0744194161285\\
325	85.8223440345506\\
340	86.5302220806142\\
355	87.2386961753053\\
370	87.9625113056257\\
385	88.7067653720629\\
400	89.4691485146515\\
415	90.2219413611573\\
430	90.8795778242993\\
};
\addlegendentry{pMOR Grassmann}

\addplot [color=black, draw=none, mark=square*, mark options={solid, fill=black, black}]
  table[row sep=crcr]{%
280	83.1009917797674\\
430	90.8795778242993\\
};
\addlegendentry{Sample Points}

\end{axis}
\end{tikzpicture}%

%% file: tikz/sigma_vs_pmax_sample4.tex
\begin{tikzpicture}[trim axis left, trim axis right]

\begin{axis}[%
width=0.951\figurewidth,
height=\figureheight,
at={(0\figurewidth,0\figureheight)},
scale only axis,
xmin=280,
xmax=430,
xtick={280, 330, 380, 430},
minor x tick num=9,
xminorgrids,
xlabel style={font=\color{white!15!black}},
ymin=80,
ymax=95,
ylabel style={font=\color{white!15!black}},
xticklabels={,,},
yticklabels={,,}
axis background/.style={fill=white},
xmajorgrids,
ymajorgrids,
legend style={at={(0.03,0.97)}, anchor=north west, legend cell align=left, align=left, draw=white!15!black},
title style={font=\footnotesize},xlabel style={font=\footnotesize},ylabel style={font=\footnotesize},ticklabel style={font=\footnotesize},line join=round,every axis legend/.code={\let\addlegendentry\relax}
]
\addplot [color=fom]
  table[row sep=crcr]{%
280	83.1009904154952\\
285	83.4213840503183\\
290	83.7359360165049\\
295	84.0456211948194\\
300	84.3498730424087\\
305	84.6491827298874\\
310	84.9434641174014\\
315	85.2331231224666\\
320	85.5184749404688\\
325	85.7992108723065\\
330	86.0759014311325\\
335	86.3483570350275\\
340	86.6167148333281\\
345	86.8814144265842\\
350	87.142505903693\\
355	87.3996571859964\\
360	87.6534423969611\\
365	87.9038062972142\\
370	88.1504073227232\\
375	88.3938263703264\\
380	88.6341847251328\\
385	88.8718588189841\\
390	89.1060294748025\\
395	89.33724867302\\
400	89.565581965988\\
405	89.7915127632295\\
410	90.0142408012635\\
415	90.2340930357819\\
420	90.452073679103\\
425	90.6674653487719\\
430	90.8795761304342\\
};
\addlegendentry{FOM}

\addplot [color=rom_cos, dashed]
  table[row sep=crcr]{%
280	83.1009917797674\\
285	83.4206962477498\\
290	83.7346808271216\\
295	84.0441042453877\\
300	84.3482209522176\\
305	84.6475582856846\\
310	84.9418828655298\\
315	85.2317743664656\\
320	85.5175579745736\\
325	85.7986895648082\\
330	86.0759210022639\\
335	86.3477901673225\\
340	86.6157457590732\\
345	86.8802203193205\\
350	87.1412589801672\\
355	87.3983779077299\\
360	87.652253317076\\
365	87.9028325180149\\
370	88.1496099546086\\
375	88.3933728226493\\
380	88.634191444811\\
385	88.8714542357666\\
390	89.1053302307571\\
395	89.3363039916661\\
400	89.564582682422\\
405	89.7904700590935\\
410	90.0131876258088\\
415	90.2331895288052\\
420	90.451412318619\\
425	90.6670887884464\\
430	90.8795778242993\\
};
\addlegendentry{pMOR Concatenation of snapshots}

\addplot [color=rom_grassmann]
  table[row sep=crcr]{%
280	83.1009917797674\\
285	83.4331360709509\\
290	83.7678297111037\\
295	84.0768951268651\\
300	84.3652225652087\\
305	84.6473156827168\\
310	84.9300114352777\\
315	85.2163517817826\\
320	85.5064326428584\\
325	85.7973477228139\\
330	86.0759210022639\\
335	86.3463954978846\\
340	86.6100829612228\\
345	86.855623830546\\
350	87.091064301451\\
355	87.3247932623839\\
360	87.5666058413601\\
365	87.8238352847398\\
370	88.0980630179314\\
375	88.3797196548215\\
380	88.634191444811\\
385	88.8712282291829\\
390	89.1123554827554\\
395	89.3507107004775\\
400	89.5836580528154\\
405	89.8112848048998\\
410	90.035106586633\\
415	90.2554242934594\\
420	90.4702709589807\\
425	90.6788260267506\\
430	90.8795778242993\\
};
\addlegendentry{pMOR Grassmann}

\addplot [color=black, draw=none, mark=square*, mark options={solid, fill=black, black}]
  table[row sep=crcr]{%
280	83.1009917797674\\
330	86.0759210022639\\
380	88.634191444811\\
430	90.8795778242993\\
};
\addlegendentry{Sample Points}

\end{axis}
\end{tikzpicture}%

%% file: tikz/sigma_vs_avpd_max_sample1.tex
\begin{tikzpicture}[trim axis left,trim axis right]

\begin{axis}[%
width=0.951\figurewidth,
height=\figureheight,
at={(0\figurewidth,0\figureheight)},
scale only axis,
xmin=280,
xmax=430,
xtick={280,430},
extra x ticks={355},
minor x tick num=9,
xminorgrids,
xlabel style={font=\color{white!15!black}},
xlabel={$\text{Contractility }\sigma\text{ [kPa]}$},
ymin=9,
ymax=12,
ylabel style={font=\color{white!15!black}},
ylabel={Max. LAVPD [mm]},
ylabel style={at={(axis description cs:0.15,.5)},anchor=south},
axis background/.style={fill=white},
xmajorgrids,
ymajorgrids,
legend style={at={(0.03,0.97)}, anchor=north west, legend cell align=left, align=left, draw=white!15!black},
title style={font=\footnotesize},xlabel style={font=\footnotesize},ylabel style={font=\footnotesize},ticklabel style={font=\footnotesize},line join=round,every axis legend/.code={\let\addlegendentry\relax}
]
\addplot [color=fom]
  table[row sep=crcr]{%
280	9.46292589436094\\
285	9.54540541097568\\
290	9.62666833559362\\
295	9.70674504956379\\
300	9.78566752543436\\
305	9.86346198792152\\
310	9.94015927788245\\
315	10.0157886800801\\
320	10.0903782398474\\
325	10.1639562759113\\
330	10.236544734213\\
335	10.3081680343837\\
340	10.3788499975887\\
345	10.4486129301594\\
350	10.5174799515081\\
355	10.5854727560976\\
360	10.6526106853579\\
365	10.7189146461857\\
370	10.7844048451872\\
375	10.8490985901012\\
380	10.9130140310905\\
385	10.9761693402966\\
390	11.0385824184553\\
395	11.1002685371258\\
400	11.1612438703143\\
405	11.221525213261\\
410	11.2811277145565\\
415	11.3400650582398\\
420	11.398334637521\\
425	11.4559254528825\\
430	11.512887416598\\
};
\addlegendentry{FOM}

\addplot [color=rom_const]
  table[row sep=crcr]{%
280	9.12190520139414\\
295	9.43292064724264\\
310	9.73339630485576\\
325	10.0248043799074\\
340	10.3084629622195\\
355	10.5854729309471\\
370	10.8567262200711\\
385	11.1228637138551\\
400	11.3840265781323\\
415	11.6406606552866\\
430	11.8928495898085\\
};
\addlegendentry{MOR no Interpolation}

\addplot [color=black, draw=none, mark=square*, mark options={solid, fill=black, black}]
  table[row sep=crcr]{%
355	10.5854729309471\\
};
\addlegendentry{Sample Point}

\end{axis}
\end{tikzpicture}%

%% file: tikz/sigma_vs_avpd_max_sample2.tex
\begin{tikzpicture}[trim axis left,trim axis right]

\begin{axis}[%
width=0.951\figurewidth,
height=\figureheight,
at={(0\figurewidth,0\figureheight)},
scale only axis,
xmin=280,
xmax=430,
xtick={280,430},
extra x ticks={355},
minor x tick num=9,
xminorgrids,
xlabel style={font=\color{white!15!black}},
xlabel={$\text{Contractility }\sigma\text{ [kPa]}$},
ymin=9,
ymax=12,
ylabel style={font=\color{white!15!black}},
yticklabels={,,}
axis background/.style={fill=white},
xmajorgrids,
ymajorgrids,
legend style={at={(0.03,0.97)}, anchor=north west, legend cell align=left, align=left, draw=white!15!black},
title style={font=\footnotesize},xlabel style={font=\footnotesize},ylabel style={font=\footnotesize},ticklabel style={font=\footnotesize},line join=round,every axis legend/.code={\let\addlegendentry\relax}
]
\addplot [color=fom]
  table[row sep=crcr]{%
280	9.46292589436094\\
285	9.54540541097568\\
290	9.62666833559362\\
295	9.70674504956379\\
300	9.78566752543436\\
305	9.86346198792152\\
310	9.94015927788245\\
315	10.0157886800801\\
320	10.0903782398474\\
325	10.1639562759113\\
330	10.236544734213\\
335	10.3081680343837\\
340	10.3788499975887\\
345	10.4486129301594\\
350	10.5174799515081\\
355	10.5854727560976\\
360	10.6526106853579\\
365	10.7189146461857\\
370	10.7844048451872\\
375	10.8490985901012\\
380	10.9130140310905\\
385	10.9761693402966\\
390	11.0385824184553\\
395	11.1002685371258\\
400	11.1612438703143\\
405	11.221525213261\\
410	11.2811277145565\\
415	11.3400650582398\\
420	11.398334637521\\
425	11.4559254528825\\
430	11.512887416598\\
};
\addlegendentry{FOM}

\addplot [color=rom_cos, dashed]
  table[row sep=crcr]{%
280	9.46292348057265\\
295	9.70846105524436\\
310	9.94268952844407\\
325	10.1660032824517\\
340	10.3798805127346\\
355	10.5852876999297\\
370	10.7829874392781\\
385	10.9739450113742\\
400	11.1587511807151\\
415	11.3381592588021\\
430	11.5128907600774\\
};
\addlegendentry{pMOR Concatenation of snapshots}

\addplot [color=rom_grassmann]
  table[row sep=crcr]{%
280	9.46292348057265\\
295	9.69232719772668\\
310	9.88918632460435\\
325	10.0803795229261\\
340	10.2812410829515\\
355	10.4981444683024\\
370	10.7287290689318\\
385	10.9652120789321\\
400	11.1917168673155\\
415	11.3750043046487\\
430	11.5128907600774\\
};
\addlegendentry{pMOR Grassmann}

\addplot [color=black, draw=none, mark=square*, mark options={solid, fill=black, black}]
  table[row sep=crcr]{%
280	9.46292348057265\\
430	11.5128907600774\\
};
\addlegendentry{Sample Points}

\end{axis}
\end{tikzpicture}%

%% file: tikz/sigma_vs_avpd_max_sample4.tex
\begin{tikzpicture}[trim axis left, trim axis right]

\begin{axis}[%
width=0.951\figurewidth,
height=\figureheight,
at={(0\figurewidth,0\figureheight)},
scale only axis,
xmin=280,
xmax=430,
xtick={280, 330, 380, 430},
minor x tick num=9,
xminorgrids,
xlabel style={font=\color{white!15!black}},
xlabel={$\text{Contractility }\sigma\text{ [kPa]}$},
ymin=9,
ymax=12,
ylabel style={font=\color{white!15!black}},
yticklabels={,,}
axis background/.style={fill=white},
xmajorgrids,
ymajorgrids,
legend style={at={(0.03,0.97)}, anchor=north west, legend cell align=left, align=left, draw=white!15!black},
title style={font=\footnotesize},xlabel style={font=\footnotesize},ylabel style={font=\footnotesize},ticklabel style={font=\footnotesize},line join=round,every axis legend/.code={\let\addlegendentry\relax}
]
\addplot [color=fom]
  table[row sep=crcr]{%
280	9.46292589436094\\
285	9.54540541097568\\
290	9.62666833559362\\
295	9.70674504956379\\
300	9.78566752543436\\
305	9.86346198792152\\
310	9.94015927788245\\
315	10.0157886800801\\
320	10.0903782398474\\
325	10.1639562759113\\
330	10.236544734213\\
335	10.3081680343837\\
340	10.3788499975887\\
345	10.4486129301594\\
350	10.5174799515081\\
355	10.5854727560976\\
360	10.6526106853579\\
365	10.7189146461857\\
370	10.7844048451872\\
375	10.8490985901012\\
380	10.9130140310905\\
385	10.9761693402966\\
390	11.0385824184553\\
395	11.1002685371258\\
400	11.1612438703143\\
405	11.221525213261\\
410	11.2811277145565\\
415	11.3400650582398\\
420	11.398334637521\\
425	11.4559254528825\\
430	11.512887416598\\
};
\addlegendentry{FOM}

\addplot [color=rom_cos, dashed]
  table[row sep=crcr]{%
280	9.46292348057265\\
285	9.54546097769415\\
290	9.62672118783256\\
295	9.70676687792983\\
300	9.78564195471065\\
305	9.86338079483488\\
310	9.94001781656373\\
315	10.0156000758906\\
320	10.0902017871967\\
325	10.1638017118177\\
330	10.2365448252771\\
335	10.308255102955\\
340	10.3789746945297\\
345	10.4487382606472\\
350	10.5175766553235\\
355	10.5855246243506\\
360	10.6526200676526\\
365	10.7188872230588\\
370	10.7843590232875\\
375	10.8490428252226\\
380	10.9130152132255\\
385	10.9762338968782\\
390	11.0386933420609\\
395	11.1003869342789\\
400	11.1613571693693\\
405	11.2216159389774\\
410	11.2811815390499\\
415	11.3400688234693\\
420	11.3983031450094\\
425	11.4558907537768\\
430	11.5128907600774\\
};
\addlegendentry{pMOR Concatenation of snapshots}

\addplot [color=rom_grassmann]
  table[row sep=crcr]{%
280	9.46292348057265\\
285	9.53505880312545\\
290	9.58830789583759\\
295	9.64580506278218\\
300	9.71821746119737\\
305	9.80434746365947\\
310	9.89824691581092\\
315	9.99313180415379\\
320	10.0821874569716\\
325	10.1611223424926\\
330	10.2365448252771\\
335	10.3212675991226\\
340	10.4040816503032\\
345	10.4853066375004\\
350	10.5636449599163\\
355	10.6372753584519\\
360	10.7031154833874\\
365	10.7589094890994\\
370	10.8061016221668\\
375	10.8525220755892\\
380	10.9130152132255\\
385	10.9864821002063\\
390	11.0582589734406\\
395	11.1286613396075\\
400	11.1967664502622\\
405	11.2608604978711\\
410	11.319337285446\\
415	11.3713836327272\\
420	11.4172917259213\\
425	11.4601677169626\\
430	11.5128907600774\\
};
\addlegendentry{pMOR Grassmann}

\addplot [color=black, draw=none, mark=square*, mark options={solid, fill=black, black}]
  table[row sep=crcr]{%
280	9.46292348057265\\
330	10.2365448252771\\
380	10.9130152132255\\
430	11.5128907600774\\
};
\addlegendentry{Sample Points}

\end{axis}
\end{tikzpicture}%

%% file: tikz/legend_scalar.tex
\begin{tikzpicture}

\begin{axis}[%
width=\figurewidth,
height=0.876\figureheight,
at={(0\figurewidth,0\figureheight)},
hide axis,
scale only axis,
xmin=10,
xmax=100,
ymin=0,
ymax=0.4,
legend columns=-1,
legend style={draw=none,column sep=1ex, font=\footnotesize,cells={align=center}}
]
\addlegendimage{color=fom,thick}
\addlegendentry{\fom{}}
\addlegendimage{color=rom_const,thick}
\addlegendentry{\rom{}\\constant}
\addlegendimage{color=rom_grassmann,thick}
\addlegendentry{\prom{}\\Grassmann}
\addlegendimage{color=rom_cos, dashed,thick}
\addlegendentry{\prom{}\\CoS}
\addlegendimage{only marks, draw=none, mark=square*, mark options={solid, fill=black, black}}
\addlegendentry{Sample points}

\end{axis}
\end{tikzpicture}%

%% file: tikz/eva_cost.tex
\definecolor{mycolor1}{rgb}{0.00000,0.44700,0.74100}%
\definecolor{mycolor2}{rgb}{0.85000,0.32500,0.09800}%
\begin{tikzpicture}

\begin{axis}[%
width=\figurewidth,
height=0.992\figureheight,
at={(0\figurewidth,0\figureheight)},
scale only axis,
xmin=0,
xmax=10,
xlabel style={font=\color{white!15!black}},
xlabel={Iteration $i$ [-]},
ymode=log,
ymin=1e-10,
ymax=1,
yminorticks=true,
ylabel style={font=\color{white!15!black}},
ylabel={Objective function $S^i/S^0$},
axis background/.style={fill=white},
xmajorgrids,
ymajorgrids,
yminorgrids,
legend style={legend cell align=left, align=left, draw=white!15!black},
title style={font=\footnotesize},xlabel style={font=\footnotesize},ylabel style={font=\footnotesize},ticklabel style={font=\footnotesize},line join=round,every axis legend/.code={\let\addlegendentry\relax},ylabel style={at={(axis description cs:0.1,.5)},anchor=south},
]
\addplot [color=fom]
  table[row sep=crcr]{%
0	1\\
1	0.171776453109024\\
2	0.0129671102218598\\
3	0.00498677494437403\\
4	0.0241361170267885\\
5	6.36774813717931e-05\\
6	1.80486471542915e-05\\
7	9.76665393281705e-06\\
8	2.33423592667262e-07\\
9	3.43584858733626e-10\\
10	3.02700317195571e-10\\
11	3.02567235646702e-10\\
12	3.02567175323517e-10\\
13	3.02567176202936e-10\\
};
\addlegendentry{FOM}

\addplot [color=rom300]
  table[row sep=crcr]{%
0	1\\
1	0.183775166340097\\
2	0.0094804107713761\\
3	0.00356542228649867\\
4	0.0131446369920804\\
5	0.000320794923816256\\
6	4.23035750761714e-05\\
7	1.91898425892132e-06\\
8	7.01176326137768e-07\\
9	5.74113221657472e-08\\
10	2.48195922482883e-08\\
};
\addlegendentry{ROM300}

\addplot [color=black, forget plot]
  table[row sep=crcr]{%
0	1e-05\\
15	1e-05\\
};
\end{axis}
\end{tikzpicture}%

%% file: tikz/eva_grad-norm.tex
\definecolor{mycolor1}{rgb}{0.00000,0.44700,0.74100}%
\definecolor{mycolor2}{rgb}{0.85000,0.32500,0.09800}%
\begin{tikzpicture}

\begin{axis}[%
width=\figurewidth,
height=0.992\figureheight,
at={(0\figurewidth,0\figureheight)},
scale only axis,
xmin=0,
xmax=10,
xlabel style={font=\color{white!15!black}},
xlabel={Iteration $i$ [-]},
ymode=log,
ymin=1e-8,
ymax=1,
yminorticks=true,
ylabel style={font=\color{white!15!black}},
ylabel={Gradient $||\nabla S^i||/||\nabla S^0||$},
axis background/.style={fill=white},
xmajorgrids,
ymajorgrids,
yminorgrids,
legend style={at={(0.03,0.03)}, anchor=south west, legend cell align=left, align=left, draw=white!15!black},
title style={font=\footnotesize},xlabel style={font=\footnotesize},ylabel style={font=\footnotesize},ticklabel style={font=\footnotesize},line join=round,every axis legend/.code={\let\addlegendentry\relax},ylabel style={at={(axis description cs:0.1,.5)},anchor=south},
]
\addplot [color=fom]
  table[row sep=crcr]{%
0	1\\
1	0.395860235072325\\
2	0.0325726710503235\\
3	0.0644273887897039\\
4	0.253011133838555\\
5	0.00469130857738507\\
6	0.000365921567808111\\
7	0.00462666665809898\\
8	0.000874558062511056\\
9	9.26875778844971e-06\\
10	6.02502920866891e-07\\
11	1.09341460593892e-08\\
12	4.82172434993173e-10\\
13	3.24461478159482e-10\\
};
\addlegendentry{FOM}

\addplot [color=rom300]
  table[row sep=crcr]{%
0	1\\
1	0.41542146544801\\
2	0.0483906826418413\\
3	0.0400489096421111\\
4	0.185475622785423\\
5	0.0163195773760349\\
6	0.00320849108846318\\
7	0.000828705689884469\\
8	0.000834710309990251\\
9	0.000276374296681891\\
10	0.000156237509881898\\
};
\addlegendentry{ROM300}

%\addplot [color=black, dashed]
%  table[row sep=crcr]{%
%0	1\\
%1	0.25\\
%2	0.0625\\
%3	0.015625\\
%4	0.00390625\\
%5	0.0009765625\\
%6	0.000244140625\\
%7	6.103515625e-05\\
%8	1.52587890625e-05\\
%9	3.814697265625e-06\\
%10	9.5367431640625e-07\\
%11	2.38418579101562e-07\\
%12	5.96046447753906e-08\\
%13	1.49011611938477e-08\\
%14	3.72529029846191e-09\\
%15	9.31322574615479e-10\\
%};
%\addlegendentry{quadratic convergence}

\end{axis}
\end{tikzpicture}%

%% file: tikz/legend_inv_ana_performance.tex
\begin{tikzpicture}

\begin{axis}[%
width=\figurewidth,
height=0.876\figureheight,
at={(0\figurewidth,0\figureheight)},
hide axis,
scale only axis,
xmin=10,
xmax=100,
ymin=0,
ymax=0.4,
legend columns=-1,
legend style={draw=none,column sep=1ex, font=\footnotesize}
]
\addlegendimage{color=fom,thick}
\addlegendentry{\fom{}}
\addlegendimage{color=rom300,thick}
\addlegendentry{\prom{300}}
%\addlegendimage{color=start,thick}
%\addlegendentry{Start}
%\addlegendimage{color=black,thick,dashed}
%\addlegendentry{Ground truth}

\end{axis}
\end{tikzpicture}%

%% file: tikz/legend_inv_ana.tex
\begin{tikzpicture}

\begin{axis}[%
width=\figurewidth,
height=0.876\figureheight,
at={(0\figurewidth,0\figureheight)},
hide axis,
scale only axis,
xmin=10,
xmax=100,
ymin=0,
ymax=0.4,
legend columns=-1,
legend style={draw=none,column sep=1ex, font=\footnotesize}
]
%\addlegendimage{color=fom,thick}
%\addlegendentry{\fom{}}
\addlegendimage{color=start,thick}
\addlegendentry{Initial state}
\addlegendimage{color=converged,thick}
\addlegendentry{Converged solution (\prom{300} and \fom{})}
\addlegendimage{color=black,thick,dashed}
\addlegendentry{Ground truth}

\end{axis}
\end{tikzpicture}%

%% file: tikz/eva_params_FOM.tex
\definecolor{mycolor1}{rgb}{0.00000,0.44700,0.74100}%
\definecolor{mycolor2}{rgb}{0.85000,0.32500,0.09800}%
\definecolor{mycolor3}{rgb}{0.92900,0.69400,0.12500}%
\definecolor{mycolor4}{rgb}{0.49400,0.18400,0.55600}%
\definecolor{mycolor5}{rgb}{0.46600,0.67400,0.18800}%
\begin{tikzpicture}

\begin{axis}[%
width=\figurewidth,
height=0.986\figureheight,
at={(0\figurewidth,0\figureheight)},
scale only axis,
xmin=0,
xmax=10,
xtick={ 0,  2,  4,  6,  7,  8, 10, 12, 14},
xlabel style={font=\color{white!15!black}},
xlabel={Iteration $i$ [-]},
ymin=0.4,
ymax=2.1,
ytick={0.5,   1, 1.5,   2},
ylabel style={font=\color{white!15!black}},
ylabel={Parameter $\mu_p^i/\mu_p^0$ [-]},
axis background/.style={fill=white},
xmajorgrids,
ymajorgrids,
legend style={legend cell align=left, align=left, draw=white!15!black},
title style={font=\footnotesize},xlabel style={font=\footnotesize},ylabel style={font=\footnotesize},ticklabel style={font=\footnotesize},line join=round,every axis legend/.code={\let\addlegendentry\relax},ylabel style={at={(axis description cs:0.1,.5)},anchor=south},
]
\addplot [color=mycolor1]
  table[row sep=crcr]{%
0	1\\
1	0.920151450523606\\
2	1.05299447514016\\
3	1.15253568925147\\
4	1.28154742479808\\
5	1.35370642555033\\
6	1.36393695389862\\
7	1.39292539201469\\
8	1.3997853477673\\
9	1.39996511149575\\
10	1.39999552126705\\
11	1.39999676351034\\
12	1.39999676421008\\
13	1.39999675233289\\
};
\addlegendentry{data1}

\addplot [color=mycolor2]
  table[row sep=crcr]{%
0	1\\
1	1.40388904230014\\
2	1.75925233079791\\
3	1.2544417264123\\
4	0.702495138778377\\
5	0.719330464504844\\
6	0.716514611212688\\
7	0.67336256948629\\
8	0.666860165907586\\
9	0.666708484962327\\
10	0.666673012906218\\
11	0.6666714459225\\
12	0.666671444952038\\
13	0.666671458317415\\
};
\addlegendentry{data2}

\addplot [color=mycolor3]
  table[row sep=crcr]{%
0	1\\
1	1.20836078878401\\
2	1.49555342976446\\
3	1.76519194447294\\
4	1.99357110175799\\
5	2.0460383905404\\
6	2.06426007205717\\
7	2.01755092297224\\
8	2.00150482976372\\
9	2.00012060971705\\
10	2.00000567606296\\
11	2.00000175238936\\
12	2.00000179679733\\
13	2.00000186988244\\
};
\addlegendentry{data3}

\addplot [color=mycolor4]
  table[row sep=crcr]{%
0	1\\
1	0.812469187549186\\
2	0.777117233313406\\
3	0.764506591889724\\
4	0.722863433020911\\
5	0.707574163826876\\
6	0.708739065701164\\
7	0.703997689671992\\
8	0.702737846422152\\
9	0.702860062480397\\
10	0.702857960539859\\
11	0.702857826428858\\
12	0.70285782405207\\
13	0.702857825072073\\
};
\addlegendentry{data4}

\addplot [color=mycolor5]
  table[row sep=crcr]{%
0	1\\
1	0.866426164807479\\
2	0.844178944178452\\
3	0.843710216435299\\
4	0.841214986772855\\
5	0.841350163450278\\
6	0.840986244749811\\
7	0.837781021375087\\
8	0.836773979057972\\
9	0.836674212908225\\
10	0.83666717558307\\
11	0.836666887786716\\
12	0.836666888605604\\
13	0.836666892502672\\
};
\addlegendentry{data5}

\addplot [color=mycolor1, dashed]
  table[row sep=crcr]{%
0	1.4\\
20	1.4\\
};
\addlegendentry{data6}

\addplot [color=mycolor2, dashed]
  table[row sep=crcr]{%
0	0.666666666666667\\
20	0.666666666666667\\
};
\addlegendentry{data7}

\addplot [color=mycolor3, dashed]
  table[row sep=crcr]{%
0	2\\
20	2\\
};
\addlegendentry{data8}

\addplot [color=mycolor4, dashed]
  table[row sep=crcr]{%
0	0.702857142857143\\
20	0.702857142857143\\
};
\addlegendentry{data9}

\addplot [color=mycolor5, dashed]
  table[row sep=crcr]{%
0	0.836666666666667\\
20	0.836666666666667\\
};
\addlegendentry{data10}

\addplot [color=black]
  table[row sep=crcr]{%
7	0\\
7	3\\
};
\addlegendentry{data11}

\end{axis}
\end{tikzpicture}%

%% file: tikz/eva_params_ROM300.tex
\definecolor{mycolor1}{rgb}{0.00000,0.44700,0.74100}%
\definecolor{mycolor2}{rgb}{0.85000,0.32500,0.09800}%
\definecolor{mycolor3}{rgb}{0.92900,0.69400,0.12500}%
\definecolor{mycolor4}{rgb}{0.49400,0.18400,0.55600}%
\definecolor{mycolor5}{rgb}{0.46600,0.67400,0.18800}%
\begin{tikzpicture}

\begin{axis}[%
width=\figurewidth,
height=0.986\figureheight,
at={(0\figurewidth,0\figureheight)},
scale only axis,
xmin=0,
xmax=10,
xtick={ 0,  2,  4,  6,  7,  8, 10, 12, 14},
xlabel style={font=\color{white!15!black}},
xlabel={Iteration $i$ [-]},
ymin=0.4,
ymax=2.1,
ytick={0.5,   1, 1.5,   2},
ylabel style={font=\color{white!15!black}},
ylabel={Parameter $\mu_p^i/\mu_p^0$ [-]},
axis background/.style={fill=white},
xmajorgrids,
ymajorgrids,
legend style={legend cell align=left, align=left, draw=white!15!black},
title style={font=\footnotesize},xlabel style={font=\footnotesize},ylabel style={font=\footnotesize},ticklabel style={font=\footnotesize},line join=round,every axis legend/.code={\let\addlegendentry\relax},ylabel style={at={(axis description cs:0.1,.5)},anchor=south},
]
\addplot [color=mycolor1]
  table[row sep=crcr]{%
0	1\\
1	0.91072794480301\\
2	1.08914194239008\\
3	1.16173862897128\\
4	1.28179929800659\\
5	1.35062495134185\\
6	1.36518532225588\\
7	1.38972566658539\\
8	1.39905726883719\\
9	1.39906277217099\\
10	1.4001637993696\\
};
\addlegendentry{data1}

\addplot [color=mycolor2]
  table[row sep=crcr]{%
0	1\\
1	1.48143973369435\\
2	1.70705996750714\\
3	1.33538047799974\\
4	0.789169932404523\\
5	0.732844243746513\\
6	0.721282254139458\\
7	0.679882033486153\\
8	0.669083985952354\\
9	0.6676885905958\\
10	0.66653942063611\\
};
\addlegendentry{data2}

\addplot [color=mycolor3]
  table[row sep=crcr]{%
0	1\\
1	1.21733410914159\\
2	1.48201483198261\\
3	1.83424918573892\\
4	1.47677527708949\\
5	1.79832666818332\\
6	1.92952183528838\\
7	2.00685746392287\\
8	2.01014805629726\\
9	2.00583090660776\\
10	2.00277037646441\\
};
\addlegendentry{data3}

\addplot [color=mycolor4]
  table[row sep=crcr]{%
0	1\\
1	0.821403511432602\\
2	0.780023806083564\\
3	0.771617096518964\\
4	0.74067279741267\\
5	0.714434746183449\\
6	0.710764589612645\\
7	0.704559159264671\\
8	0.703251668072206\\
9	0.702940219708602\\
10	0.702850309748355\\
};
\addlegendentry{data4}

\addplot [color=mycolor5]
  table[row sep=crcr]{%
0	1\\
1	0.863019070131957\\
2	0.843779613538911\\
3	0.848135651413324\\
4	0.816924476560408\\
5	0.831733310682634\\
6	0.834613352262243\\
7	0.837565219628234\\
8	0.836914764411192\\
9	0.837009967282856\\
10	0.836791184050304\\
};
\addlegendentry{data5}

\addplot [color=mycolor1, dashed]
  table[row sep=crcr]{%
0	1.4\\
20	1.4\\
};
\addlegendentry{data6}

\addplot [color=mycolor2, dashed]
  table[row sep=crcr]{%
0	0.666666666666667\\
20	0.666666666666667\\
};
\addlegendentry{data7}

\addplot [color=mycolor3, dashed]
  table[row sep=crcr]{%
0	2\\
20	2\\
};
\addlegendentry{data8}

\addplot [color=mycolor4, dashed]
  table[row sep=crcr]{%
0	0.702857142857143\\
20	0.702857142857143\\
};
\addlegendentry{data9}

\addplot [color=mycolor5, dashed]
  table[row sep=crcr]{%
0	0.836666666666667\\
20	0.836666666666667\\
};
\addlegendentry{data10}

\addplot [color=black]
  table[row sep=crcr]{%
7	0\\
7	3\\
};
\addlegendentry{data11}

\end{axis}
\end{tikzpicture}%

%% file: tikz/legend_params.tex
\begin{tikzpicture}
\begin{axis}[%
width=\figurewidth,
height=0.876\figureheight,
at={(0\figurewidth,0\figureheight)},
hide axis,
scale only axis,
xmin=10,
xmax=100,
ymin=0,
ymax=0.4,
legend columns=-1,
legend style={draw=none,column sep=1ex, font=\footnotesize,cells={align=center}}
]
\addlegendimage{color=p_sigma,thick}
\addlegendentry{$\sigma$}
\addlegendimage{color=p_act_max,thick}
\addlegendentry{$\alpha_{\text{max}}$}
\addlegendimage{color=p_act_min,thick}
\addlegendentry{$\alpha_{\text{min}}$}
\addlegendimage{color=p_t_act,thick}
\addlegendentry{$t_{\text{syst}}$}
\addlegendimage{color=p_t_deact,thick}
\addlegendentry{$t_{\text{dias}}$}
\end{axis}
\end{tikzpicture}